\documentstyle[preprint,tighten,aps]{revtex}
\newcommand{\ewxy}[2]{\setlength{\epsfxsize}{#2}\epsfbox[30 60 540 570]{#1}}

\newcommand{\delv}{{\bf \nabla}}
\newcommand{\delvt}{{\bf \tilde{\nabla}}}

\newcommand{\delfour}{{\Delta^{(4)}}}
\newcommand{\delsq}{\Delta^{(2)}}

\newcommand{\vev}[1]{\langle #1 \rangle}
\newcommand{\Mbz}{{(a_sM_0)}}

\newcommand{\half}{{\mbox{$\frac{1}{2}$}}}
\newcommand{\order}{{\cal O}}

\newcommand{\Ev}{{\bf \tilde{E}}}
\newcommand{\Bv}{{\bf \tilde{B}}}

\newcommand{\sigmav}{\mbox{\boldmath$\sigma$}}

\newcommand{\ainvsp}{$a^{-1}\;$}

\newcommand{\tsig}{${\it T}_{signal}$}
\newcommand{\tsigsp}{${\it T}_{signal} \;$}
\newcommand{\ppri}{p^\prime}
\newcommand{\pprimu}{p^{\prime\mu}}

\newcommand{\act}{{\cal S}}
\newcommand{\be}{\begin{equation}}
\newcommand{\ee}{\end{equation}}
\newcommand{\nl}{\nonumber \\}

\begin{document}
\preprint{\begin{minipage}{2in}\begin{flushright}
 OHSTPY-HEP-T-00-021 \\ GUPTA/01/01/01 \end{flushright}
  \end{minipage}}
\input epsf
\draft

\title{ {\bf Comparison Studies of Finite Momentum Correlators  \\
on Anisotropic and Isotropic Lattices}}

\author{
  S.Collins$^{a*}$, C.T.H.Davies$^a$\footnote{UKQCD Collaboration},
 J.Hein$^{b*}$\footnote{Present address: Dept. of Physics \& Astronomy,
 University of Edinburgh, Edinburgh EH9 3JZ. }
  R.R.Horgan$^{c*}$, \\
G.P.Lepage$^b$, J.Shigemitsu$^d$.\\[.4cm]
\small $^a$Department of Physics \& Astronomy, University of Glasgow, 
 Glasgow, G12 8QQ, UK. \\[.2cm]
\small $^b$Newman Laboratory of Nuclear Studies, Cornell University, 
 Ithaca, NY 14853, USA. \\[.2cm]
\small $^c$D.A.M.T.P. , CMS, Wilberforce Road, Cambridge, England CB3 0WA, 
UK. \\[.2cm]
\small $^d$Physics Department, The Ohio State University,
 Columbus, OH 43210, USA.
\\ }

%\date{ DRAFT VERSION --- 19th January 2001}

\maketitle

\begin{abstract}
\noindent
 We study hadronic two- and three-point correlators relevant for 
heavy to light pseudoscalar meson semi-leptonic decays, 
using Symanzik improved glue, D234 light quark and NRQCD 
heavy quark actions. 
Detailed comparisons are made between simulations on anisotropic 
and isotropic lattices involving 
finite momentum hadrons.  We find evidence that having an
anisotropy helps in extracting better signals at higher momenta.
Initial results for the 
form factors $f_+(q^2)$ and $f_0(q^2)$  are presented with tree-level 
matching of the lattice heavy-light currents.

\end{abstract}

\vspace{.2in}

\newpage

\section{Introduction}

The use of anisotropic lattices with a finer grid in the temporal 
direction ($a_t < a_s$) has been advocated 
for studies of correlation functions
 that produce usable signals only over a limited 
time range, \tsig, in physical units \cite{gpltsukuba}. 
 Even when \tsigsp is reasonably large 
(e.g. $> 1fm$), if one is working on coarse isotropic lattices with lattice
spacing $a > 0.2 fm$ only 
a few data points will lie within \tsigsp and fitting data becomes 
problematic. One is then naturally led to anisotropic lattices in order 
to retain the advantages of working with only a small number of 
lattice sites in  spatial directions and nevertheless have enough 
temporal resolution within \tsig.
Many of the recent examples of successful 
employment of anisotropic lattices have been in simulations on very 
coarse spatial lattices.  These include investigations
 of the glueball spectrum \cite{mikecolin} and NRQCD studies 
of heavy hybrid states \cite{manke,ianron} and of quarkonium fine structure 
\cite{cppacs}. 
 Anisotropic 
lattices have also been useful in finite temperature studies 
\cite{karsch,sakai} and 
have been employed as an alternate approach to simulations of heavy 
quarks \cite{klassen1,columbia}.
  In several of these examples one turns to anisotropic 
lattices for two reasons: to be able to extract a better signal, and to 
avoid discretization errors coming from large $a_t  M$ or $a_t E$.

\vspace{.1in}
\noindent
In this article we investigate the extent to which anisotropic lattices 
could also be useful 
in simulations of finite momentum hadrons.  We have 
in mind, for instance, heavy meson semi-leptonic decays such as 
$B \rightarrow \pi \, (\rho) \, l\nu$ \cite{semi}.
In order to be able to cover the full 
 kinematic range of interest to experimentalists
and map out the $q^2$ dependence of form factors ($q^\mu = p^\mu
 - \pprimu$ with 
$p^\mu=$ momentum of the decaying B meson and $\pprimu=$ the 
pion momentum) one 
needs to calculate matrix elements between hadrons with large momenta.  
In the B meson rest frame, for instance,  one would ideally like to 
simulate pions with momenta all the way from $\ppri\equiv |\vec{\ppri}|
 = 0$ to 
approximately $\ppri = M_B/2$.  Correlators for hadrons with finite 
momenta are much noisier than in the zero momentum case, this being 
particularly true for light hadrons such as the pion.  Hence one is 
dealing with a situation where \tsigsp 
is shrinking rapidly as the momentum increases.  Even for 
reasonably fine lattices (e.g. $a_t=a_s \sim 0.1 fm$) once \tsigsp falls 
significantly below $1 fm$ one might consider going to anisotropic lattices.

\vspace{.1in}
\noindent
To make the above statements more explicit, and come up with concrete 
examples that could guide us in the future,  we have studied heavy meson 
 semi-leptonic decays on both isotropic and anisotropic lattices 
with comparable coarse
 spatial lattice spacings around $a_s \approx 0.25 - 0.29 fm$. 
Simulations of semi-leptonic decays require good control over both 
two- and three-point correlators. 
We use identical operators and smearing functions on the isotropic and 
anisotropic lattices.  This way one creates 
 similar signals on the initial timeslices and can  watch 
how they propagate in time on the two lattices.  We then compare 
the ease with which physics is extracted on the anisotropic versus 
the isotropic lattice.  We find that at higher momenta anisotropic simulations
 are superior in providing more reliable signals.

\vspace{.1in}
\noindent
Extracting a good signal is just one of the 
challenges one faces in lattice
 studies of semileptonic decays involving hadrons with high momenta.
 One must, for instance, also control 
large $ap$ discretization errors.  The second objective of the present 
simulations was to check how successful highly improved quark and glue actions 
are in removing lattice artifacts.
Working with Symanzik improved gauge actions \cite{weisz}, 
D234 light quark \cite{alford} and $O(a^2)$ 
improved NRQCD heavy quark \cite{cornell} actions we have
studied dispersion relations, speed of light renormalizations and also 
heavy meson decay constants at finite momenta.  We find continuum-like 
behaviour within $5$\% up to about $p\sim 1.2$GeV.
  Even for $p \sim 1.5$GeV, which on our coarse lattices corresponds 
to $a_sp \sim 2$, deviations are 10\% or less in many cases. 
Although additional tests are clearly still called for, we are very encouraged 
by these findings.  
In the future we plan to go to lattices finer than those used 
in the present initial study.  Our experience to date 
indicates it may not be necessary to go very much finer to get 
phenomenologically meaningful results.
 With highly improved actions, simulations of hadrons 
with momenta as high as $p \sim 1.5-2.0$GeV are possible once 
$a_s < 0.2 fm$. It is not necessary to have $a_sp < 1$.  We also note that 
the findings of the present work should carry over from the 
actions used here to other highly improved actions.  For instance, the most 
highly improved light quark action with better chiral properties than the D234 
action would be the staggered action \cite{gpltsukuba,staggered}. 
 One could also consider using heavy 
clover \cite{fermilab} rather than NRQCD heavy quarks.  
It will be worthwhile investigating these alternative 
options in the future.

\vspace{.1in}
\noindent
In the next section we introduce the highly improved gauge and quark actions 
used in our study and discuss simulation parameters.  In the current 
exploratory study, we have not tried to tune quark masses very accurately.  
On both the isotropic and anisotropic lattices we work with one 
light quark mass slightly heavier than the $strange$ quark.  On the anisotropic
lattice we accummulated results for two heavy quark masses, one around the 
$bottom$ quark and the other close to the $charm$ quark.  On the isotropic 
lattice only one heavy quark mass near the $bottom$ quark was used.
Section III concentrates on two-point function results.  
We compare effective masses on isotropic and anisotropic lattices and discuss 
dispersion relations and ratios of decay constant matrix elements for 
heavy mesons with and without spatial momenta.  In section IV we present 
results for three-point functions relevant for pseudoscalar $\rightarrow$ 
pseudoscalar semileptonic decays.  Again comparisons are made between signals 
on isotropic and anisotropic lattices.  The formfactors 
 $f_+(q^2)$ and $f_0(q^2)$ are extracted.

\section{Gauge and Quark Actions and Simulation Parameters }

\subsection{Gauge Actions}
We use the standard Symanzik improved isotropic gauge
action including square and six-link rectangular loops \cite{weisz}.
\be
\label{isoglue}
\act^{(iso)}_G = - \beta \sum_{x,\,\mu > \nu}  \left\{
\frac{5}{3} \frac{P_{\mu \nu}}{u_L^4} 
- \frac{1}{12} \frac{R_{\mu \nu}}{u_L^6} 
- \frac{1}{12} \frac{R_{\nu \mu}}{u_L^6} \right\} ,
\ee
with
\begin{eqnarray}
P_{\mu \nu} &=& \frac{1}{N_c}  Real \left( Tr \{ U_\mu(x)
 U_\nu(x+a_\mu) U^\dagger_\mu(x+a_\nu) U^\dagger_\nu(x) \} \right) ,  \\
R_{\mu \nu} &=& \frac{1}{N_c}  Real \left( Tr \{ U_\mu(x)
 U_\mu(x+a_\mu) U_\nu(x+2a_\mu) 
U^\dagger_\mu(x+a_\mu + a_\nu) U^\dagger_\mu(x+a_\nu) U^\dagger_\nu(x) \}
 \right)  .
\end{eqnarray}
$\beta = 2N_c/g^2$ and 
$u_L$ is the tadpole improvement ``$u_0$'' factor, for which we use the
Landau-link definition in this article \cite{lepmac}.  
For each $\beta$, $u_L$ must 
be determined iteratively via simulations.  One could also consider 
using perturbative expressions for $u_L$.  
On anisotropic lattices one can drop rectangles that extend over two 
links in the time direction and one has, 
\begin{eqnarray}
\act^{(aniso)}_G &=& - \beta
 \sum_{x,\,s > s^\prime} \frac{1}{\chi_0} \left\{
\frac{5}{3} \frac{P_{ss^\prime}}{u_s^4} 
- \frac{1}{12} \frac{R_{ss^\prime}}{u_s^6} 
- \frac{1}{12} \frac{R_{s^\prime s}}{u_s^6} \right\} \nl
 & & - \beta \;\; \sum_{x,s} \chi_0 \left\{
\frac{4}{3} \frac{P_{st}}{u_s^2 u_t^2} 
- \frac{1}{12} \frac{R_{st}}{u_s^4u_t^2} \right\} .
\label{anisoglue}
\end{eqnarray}
The variables $s$ and $s^\prime$ run only over spatial directions and 
one must now distinguish between temporal and spatial Landau-link tadpole 
improvement factors, $u_t$ and $u_s$.  
$\chi_0$ is the bare anisotropy.  It differs from the true or renormalized 
anisotropy,
\be
\chi \equiv a_s/a_t,
\ee
once quantum corrections are taken into account.
Just as one fixes lattice spacings through some experimental
input in conventional lattice calculations,
  when working on anisotropic lattices one must, in addition,
determine the ratio of the spatial and temporal lattice spacings, $a_s/a_t$, 
via some physics requirement.  This leads to the renormalized anisotropy 
$\chi$.  In the absence of lattice artifacts, it should not matter 
which physical quantity is used to fix $\chi$.  Conversely the dependence 
of $\chi$ on how it was determined provides a measure of discretization 
errors in the lattice system.
 In reference \cite{ron} the renormalized anisotropy was calculated for 
the action $\act^{(aniso)}_G$ of eq.(\ref{anisoglue}) using both the 
torelon dispersion relation and the sideways potential method for 
several values of $\beta$ and $\chi_0$.  Agreement was found between 
the two determinations within 3-4\%.  We will be using those results 
in this article.

\vspace{.1in}
\noindent
An alternate and equivalent procedure would be to write,
\begin{eqnarray}
\act^{(aniso)}_G &=& - \beta
 \sum_{x,\,s > s^\prime} \frac{\eta}{\chi} \left\{
\frac{5}{3} \frac{P_{ss^\prime}}{u_s^4} 
- \frac{1}{12} \frac{R_{ss^\prime}}{u_s^6} 
- \frac{1}{12} \frac{R_{s^\prime s}}{u_s^6} \right\} \nl
 & & - \beta \;\; \sum_{x,s} \frac{\chi}{\eta} \left\{
\frac{4}{3} \frac{P_{st}}{u_s^2 u_t^2} 
- \frac{1}{12} \frac{R_{st}}{u_s^4u_t^2} \right\} ,
\label{anisoglue2}
\end{eqnarray}
($\eta \equiv \chi/\chi_0$) 
and adjust $\eta$ at fixed $\chi$ to satisfy some physical criterion.

\subsection{Light Quark Actions}
The isotropic D234 quark action is given by \cite{alford},
\begin{eqnarray}
\label{sd234a}
\act^{(iso)}_{D234} &=& a^4  \sum_x \overline{\Psi}_c 
\left\{  \gamma_t \frac{1}{a} (\nabla_t - \frac{1}{6} 
C_{3t} \nabla_t^{(3)}) + 
\frac{C_0}{a} \vec{\gamma} \cdot (\vec{\nabla} -
 \frac{1}{6} 
C_{3} \vec\nabla^{(3)}) + m_0 \right. \nl
 &  & - \frac{r a}{2} \left [ \frac{1}{a^2} ( \nabla_t^{(2)} 
- \frac{1}{12} C_{4t} \nabla_t^{(4)} )
+  \frac{1}{a^2} \sum_{j=1}^3 ( \nabla_j^{(2)} 
- \frac{1}{12} C_{4} \nabla_j^{(4)} ) \right ] \nl
  & & - r a \left. \frac{C_F}{4} 
\frac{i \sigma_{\mu \nu} \tilde{F}^{\mu \nu}} {a^2} 
\right\} 
\Psi_c    \\
\label{sd234b}
 &=&  \sum_x \overline{\Psi}
\left\{  \gamma_t (\nabla_t - \frac{1}{6} 
C_{3t} \nabla_t^{(3)}) + 
C_0 \vec{\gamma} \cdot (\vec{\nabla} -
 \frac{1}{6} 
C_{3} \vec\nabla^{(3)}) + a m_0 \right. \nl
 &  & - \frac{r }{2} \left [  ( \nabla_t^{(2)} 
- \frac{1}{12} C_{4t} \nabla_t^{(4)} )
+   \sum_{j=1}^3 ( \nabla_j^{(2)} 
- \frac{1}{12} C_{4} \nabla_j^{(4)} ) \right ] \nl
  & & - r \left. \frac{C_F}{4} 
i \sigma_{\mu \nu} \tilde{F}^{\mu \nu} 
\right\} 
\Psi .
\end{eqnarray}
The quark fields $\Psi_c$ and the dimensionless lattice 
fields $\Psi$ are related through
\be
\label{psiscale}
\Psi = a^{3/2}  \Psi_c .
\ee
Definitions for tadpole improved 
dimensionless covariant derivatives and 
field strength tensors are summarized for instance in the Appendix of 
reference \cite{pert}.  We note here just the relation between unimproved 
$F_{\mu \nu}$ 
and the $O(a^2)$ improved field strength tensors 
 $\tilde{F}_{\mu\nu}$ used in D234 actions.
\begin{eqnarray}
& &\tilde{F}_{\mu\nu}(x) = \frac{5}{3}F_{\mu\nu}(x) \nl
& & \quad \; - \; \frac{1}{6} \left[\,\frac{1}{u_\mu^2}
(\,U_\mu(x)F_{\mu\nu}(x+a_\mu)U^\dagger_\mu(x) + U^\dagger_\mu(x-a_\mu)
F_{\mu\nu}(x-a_\mu)U_\mu(x-a_\mu)\,)  
\; - \;  (\mu \leftrightarrow \nu)\,\right]  \nl
& & \quad \; + \; \frac{1}{6} \, ( \frac{1}{u_\mu^2} + \frac{1}{u_\nu^2} 
- 2 ) \, F_{\mu\nu}(x).
\label{fmunu}
\end{eqnarray}
The last term is needed so that factors of $1/u_\mu$ are correctly removed 
from  contributions to $U  F_{\mu\nu}  U^\dagger$ and $U^\dagger 
 F_{\mu\nu} U$ that end up being 
four link objects rather than six link ones.  Effects of this term are 
nonnegligible on coarse lattices. \\
After tadpole improving the action we set the coefficients, $C_{ti}$, 
$C_i$, $C_0$ and $C_F$ equal to their tree-level value of unity in our 
simulations.  We also work with $r=1$.
 
\vspace{.1in}
\noindent
The anisotropic D234 action is,
\begin{eqnarray}
\label{sd234c}
\act^{(aniso)}_{D234} &=& a_s^3 a_t \sum_x \overline{\Psi}_c 
\left\{  \gamma_t \frac{1}{a_t} \nabla_t 
 + \frac{C_0}{a_s} \vec{\gamma} \cdot (\vec{\nabla} -
 \frac{1}{6} 
C_{3} \vec\nabla^{(3)}) + m_0 \right. \nl
 &  & - \frac{r a_s}{2} \left [ \frac{1}{a_t^2}  
 \nabla_t^{(2)} 
+  \frac{1}{a_s^2} \sum_{j=1}^3 ( \nabla_j^{(2)} 
- \frac{1}{12} C_{4} \nabla_j^{(4)} ) \right ] 
 - r a_s \left. \frac{C_F}{4} 
\frac{i \sigma_{\mu \nu} \tilde{F}^{\mu \nu}} {a_\mu a_\nu} 
\right\} 
\Psi_c    \\
\label{sd234d}
 &=&  \sum_x \overline{\Psi}
\left\{  \gamma_t \nabla_t +
\frac{C_0}{\chi} \vec{\gamma} \cdot (\vec{\nabla} -
 \frac{1}{6} 
C_{3} \vec\nabla^{(3)}) + a_t m_0 \right. \nl
 &  & - \frac{r }{2} \left [ \chi  \nabla_t^{(2)} 
+  \frac{1}{\chi} \sum_{j=1}^3 ( \nabla_j^{(2)} 
- \frac{1}{12} C_{4} \nabla_j^{(4)} ) \right ] 
 - r \left. \frac{C_F}{4} 
i \sigma_{\mu \nu} \tilde{F}^{\mu \nu} \frac{a_s a_t}{a_\mu a_\nu} 
\right\} 
\Psi ,
\end{eqnarray}
where we have used the spatial lattice spacing $a_s$ to rescale the 
quark fields according to eq.(\ref{psiscale}). Note that the renormalised 
anisotropy $\chi = a_s/a_t$ appears in the quark action.  In a quenched 
calculation it is permissible to first fix $\chi$ in the pure glue 
sector and use this value for the 
 ratio of spatial and temporal lattice spacings in the quark action.  
$\act^{(aniso)}_{D234}$ has its own ``speed of light'' renormalization 
term $C_0$.  For fixed $\chi$ this coefficient must be tuned to 
ensure correct dispersion relations in fermionic correlators. 
In the future, especially in unquenched calculations, 
it may be simplest to work with eq.(\ref{anisoglue2}) and eq.(
\ref{sd234d}) at fixed $\chi$ and simultaneously and iteratively adjust 
$\eta$ and $C_0$ using appropriate physics criteria.  Another possibility 
is to use perturbative expressions for $\eta$ and $C_0$ throughout.
In the present article we will use nonperturbatively determined 
$\chi$, $u_L$, $u_s$ and $u_t$ 
from references \cite{alford,ron}
 and a one-loop perturbative estimate for $C_0$ 
from reference \cite{pert}.  Just as in $\act^{(iso)}_{D234}$, we set 
$C_3$, $C_4$ and $C_F$ in $\act^{(aniso)}_{D234}$ equal to unity.

\subsection{Heavy Quark Actions}
It suffices to write down one expression for both the isotropic 
and anisotropic NRQCD actions, the former corresponding simply to 
 $\chi=1$ \cite{cornell,ianron2}.  We work with dimensionless 2-spinor fields 
$\Phi = a_s^{3/2}\Phi_c$ in terms of which
\begin{eqnarray}
 \label{nrqcdact}
&& \act_{NRQCD} =  \nl
&& \sum_x \left\{  \overline{\Phi}_t \Phi_t - 
 \overline{\Phi}_t
\left(1 \!-\!\frac{a_t \delta H}{2}\right)_t
 \left(1\!-\!\frac{a_tH_0}{2n}\right)^{n}_t
 U^\dagger_4
 \left(1\!-\!\frac{a_tH_0}{2n}\right)^{n}_{t-1}
\left(1\!-\!\frac{a_t\delta H}{2}\right)_{t-1} \Phi_{t-1} \right\}.
 \end{eqnarray}
 $H_0$ is the nonrelativistic kinetic energy operator,
 \be
a_t H_0 = - {\delsq\over2\chi\Mbz},
 \ee
and $\delta H$ includes relativistic and finite-lattice-spacing
corrections,
 \begin{eqnarray}
a_t\delta H 
&=& - c_1\,\frac{1}{2\chi\Mbz}\,\sigmav\cdot\Bv \nl
& & + c_2\,\frac{i}{8\Mbz^2}\left(\delv\cdot\Ev - \Ev\cdot\delv\right)
 - c_3\,\frac{1}{8\Mbz^2} \sigmav\cdot(\delvt\times\Ev - \Ev\times\delvt)\nl
& & - c_4\,\frac{(\delsq)^2}{8\chi\Mbz^3} 
  + c_5\,\frac{\delfour}{24\chi\Mbz}  - c_6\,\frac{(\delsq)^2}
{16n\chi^2\Mbz^2} .
\label{deltaH}
\end{eqnarray}
All derivatives are tadpole improved and,
\be
\delsq = \sum_{j=1}^3\nabla_j^{(2)}, \qquad \qquad \delfour = \sum_{j=1}^3
\nabla_j^{(4)}
\ee
\be
\tilde{\nabla}_k = \nabla_k - \frac{1}{6}\nabla_k^{(3)}
\ee
The dimensionless Euclidean electric and magnetic fields are,
\be
\tilde{E}_k = \tilde{F}_{k4}, \qquad \qquad \tilde{B}_k =
-\half \epsilon_{ijk}\tilde{F}_{ij}
\ee
$\nabla_k$, $\nabla_k^{(j)},$ $j=2,3,4$ and $\tilde{F}_{\mu \nu}$ are 
the same as in the light quark actions of the previous subsection.  In 
our simulations we again set all $c_i = 1$.

\subsection{Simulation Parameters}
Simulations have been carried out on $8^3 \times 20$ isotropic and 
$8^3 \times 48$ ($\chi  = 5.3$) anisotropic lattices using 
ensembles of 200 configurations each.  Details are summarized in Table I.
The $\beta$, $u_0$ and $a^{-1}$ values  are 
taken from reference \cite{alford} for the isotropic and from reference 
\cite{ron} 
for the anisotropic lattices, respectively. The latter reference also 
provided $\chi_0$ and $\chi$.  None of the lattice spacing determinations 
are very precise.  For the anisotropic lattice we have taken the 
4.503GeV quoted in \cite{ron}
 for $a_t^{-1}$ from $\Upsilon$ S-P splittings and 
 reduced it by 22\%, approximately the amount by which light hadron 
\ainvsp values differ from those fixed by $\Upsilon$ splittings in 
quenched calculations. This is also consistent with \cite{ron}'s finding for 
$a^{-1}$ from the string tension.  In heavy-light physics one expects 
light hadron \ainvsp values to be more appropriate.
The ``speed of light'' renormalization coefficient $C_0$ was estimated 
from the one-loop perturbative result for $\chi=5.3$, 
$ C_0 = 1 - 0.45 \alpha_s$ \cite{pert}, using $\alpha_s \approx 0.4$.

\vspace{.1in}
\noindent
The bare light quark mass has been adjusted so that one has the same 
pseudoscalar-to-vector ratio, P/V, on the isotropic and anisotropic 
lattices.  We found that this required approximately a factor of 
$3/2$ more BiCGstab iterations 
to create light quark propagators 
on the anisotropic lattice compared to on the isotropic one
 (actual numbers are given in Table I).
We believe the reason is critical slowing down due to an increase in the
condition number with $a^{-1}_t$.
Although our pion is still quite heavy ($\sim 840$MeV), on the anisotropic 
lattice we did, on a few occasions, encounter problems with 
``exceptional'' configurations.  In the process of accumulating an ensemble of
200 configurations, 5 had to be skipped.
 We will see below that the anisotropic 
pion propagators are noisier than the isotropic ones for the same pion 
mass.  Other mesons such as the $\rho$ and $B$, which
 use the same light quark propagators are not affected, however.
Although this aspect of simulations on anisotropic lattices is worrisome and 
needs further study, we do not, at the moment, believe that this 
implies anisotropic lattices will never be useful for light quarks. 
The problem has been exaggerated in the present calculations since we 
are on very coarse spatial lattices and $\chi = 5.3$ is a large 
anisotropy.  In future, more realistic, simulations we plan to use
 finer lattices and work with moderate anisotropies such as $\chi = 2 - 3$. 
We expect to be able to go to lighter pions there without encountering 
problems.  We mention that the tadpole improvement adopted in eq.(\ref{fmunu})
leads to a larger effective ``$c_{SW}$'' than has been used 
in the past, and it is known that 
large $c_{SW}$, such as the nonperturbative $c_{SW}$, leads to problems 
 on coarse lattices.  This phenomenon appears to set 
in at heavier quark masses when $\chi >1$.  There were no hints of 
problems with exceptional configurations on our coarse isotropic lattice 
for the light quark masses considered.

\section{ Results from Two Point Functions }
We calculated two-point correlators for the light-light pseudoscalar 
and vector, the $B_s$ and $B_s^\ast$ and, on the anisotropic lattice, also 
for the $D_s$ and $D_s^\ast$ mesons.  We will generically call these the ``pion'',
 ``rho'', ``B meson'' and ``D meson'' channels respectively. 
Current sinks corresponding to the 
temporal and spatial components of the heavy-light axial vector currents 
were also considered.  We employed gauge invariant smearings, separately 
for the light and heavy quarks, of the form
\be
( 1 + c_{sm} \delsq)^l \;\delta^{(3)}(\vec{x}-\vec{x}_0).
\ee
For light quarks we used $l = 10$ and $c_{sm} = 1/12$.  We have two 
smearings for the heavy quarks both with $c_{sm}=1/24$ but with 
different $l$ values, $l_1 = 2$ and $l_2 = 10$.  We found that 
$l_2$ works better for zero and low momentum $B$ mesons and $l_1$ gives
better signals for higher momentum correlators.  We accumulated data 
at 7 spatial momenta,  (0,0,0), (0,0,1), (0,1,1), (1,1,1),
 (0,0,2), (0,2,2) and 
(0,0,3) in units of $2\pi/aL$, averaging over all equivalent momenta.  
In the following 
subsections we will compare effective masses, present fit results for 
a small number of states and dispersion relations, and look at 
ratios of matrix elements with current sinks.

\subsection{Effective Masses }
We start by comparing raw data for B meson correlators on the anisotropic 
and isotropic lattices. 
 Figs. 1. and 2. show fractional errors of the correlators 
versus physical time for several momenta.  
One sees that, where there is overlap, the fluctuations of B correlators 
are identical on the two lattices.  
Effective mass plots are presented in Figs. 3 to 5 for the B meson 
for three representative momenta. Each figure
shows results in lattice units on the left and results in physical units, 
using central values of the scales from Table I, on the right.
 We label time slices such that sources are at $t=0$.  
By looking at the anisotropic data one finds  plateaus starting at
 about $t=0.3$fm - $0.4$fm for all momenta.  This must also be the case 
for the isotropic data since the original signals are essentially the same. 
 However, 
based only on information from the isotropic data,
 this is not always obvious, especially
for the higher momenta.  There is also an unfortunate upward fluctuation at 
around $t=0.7$fm which makes fitting the isotropic data more difficult. 
One agonizes over whether to fit starting before or after the hump.
This illustrates some of the immediate advantages of anisotropic lattices.
Once a plateau has set in it is more easily recognized, and one is also less 
sensitive to one or two points fluctuating up or down, since there are 
enough other points around.

\vspace{.1in}
\noindent
Effective mass plots for pions and rhos are given in Figs. 6 - 9 and 
corresponding correlator fractional errors in Figs. 10 and 11. 
In contrast to the B meson case, one 
sees that fluctuations are considerably larger for the pion correlators on 
anisotropic lattices.  This indicates that some configurations are close 
to being ``exceptional''. 
The rho correlators show no enhanced 
errors relative to the isotropic lattice.  
Despite the larger fluctuations, it is not difficult to 
 fit the anisotropic pion data
and fluctuations cancel to some degree upon calculating 
physical quantities such as dispersion 
relations or semi-leptonic formfactors.  In Fig. 12 we show fitted energies 
for the anisotropic lattice pion and rho at momentum (1,1,1) versus 
$t_{min}/a_t$ from single cosh fits.   The pion energies have larger errors 
but look otherwise normal.   In future simulations 
one will have to monitor pion correlators carefully.

\vspace{.1in}
\noindent
Meson correlators were calculated up to the center of the lattice in the 
time direction. 
We find that for zero and low momenta 
\tsigsp covers the entire $T/2$ region
(1.24fm for the anisotropic and 2.25fm for the isotropic lattice).
However as the momentum increases \tsigsp starts to shrink.  For the 
B meson this occurs only at the highest two momenta (0,2,2) and (0,0,3).  
 For the pion and the rho, \tsigsp shrinks below $\sim 1fm$  
once one goes beyond momentum (1,1,1).  It then becomes much harder to 
get reliable information from the isotropic lattice.
For instance, it should be obvious from 
 Fig. 9 that trying to fit the isotropic data is considerably more 
frustrating than carrying out fits to the anisotropic data.   Fig. 13
shows fitted energies versus $t_{min}/a_t$ for the isotropic rho and pion 
with momentum (1,1,1).  This should be compared with  the corresponding 
anisotropic results in Fig. 12.
With the isotropic data, picking $t_{min}$ is extremely tricky.  In 
 past work on isotropic lattices we often adopted a criterion whereby if 
$t_{min}^{(0)}$ is the smallest $t_{min}$ which gives an 
acceptable $Q$-value ($>0.1$) 
then our preferred choice for $t_{min}$ in a single exponential (cosh) fit 
to a single correlator 
would be $t_{min}^{(0)} + 2 a_t$ \cite{scri,joachim}. 
 This is a fairly conservative criterion 
leading to larger statistical errors than if one chooses a smaller 
$t_{min}$.  We will call fits using this criterion ``B-fits''.  In the present 
calculation, B-fits would dictate $t_{min}/a_t = 3$ or $
4$ for most of our fits 
to isotropic data.  We will abide by this $t_{min}$
 for fitting B meson correlators.
However, for the rho and the pion we will be less conservative and use 
$t_{min}/a_t = 2$, which is equal to 
$0.5$fm in physical units.  $Q$-values are always acceptable already for 
$t_{min}/a_t = 2$.  
We call this latter 
choice ``A-fits''.  Based only on the 
isotropic data, one would be hard pressed to argue why A-fits should 
be preferred over B-fits.
 In the present comparison study, however, 
we have additional information from the anisotropic lattice that tells us 
that plateaus have set in by $t \sim 0.3 - 0.4$fm, i.e. before $t=0.5$fm, 
and hence A-fits should be fine.  We should also 
note that correlated fits were used throughout our analysis.

\vspace{.1in}
\noindent
The results presented in this subsection demonstrate that for the 
pion and rho correlators there is a clear advantage to anisotropic 
simulations starting with momenta around (1,1,1).  For B correlators 
benefits start around (0,2,2).  Below we will present physics results 
extracted from the isotropic and anisotropic lattices, concentrating on 
momentum dependent quantities.  For the B meson on both lattices and 
the pion and rho on the anisotropic lattice we can go to the 
highest momentum (0,0,3).  On the isotropic lattice, pion and rho results can
only be extracted up to momentum (0,0,2) using A-fits.  
With B-fits, only momenta up to (1,1,1) can be reached and errors are
larger than with A-fits.

\vspace{.1in}
\noindent
Another lesson to be drawn from the present exercise is the importance of 
good smearings. 
We can get away here with single exponential 
(cosh) fits, because our smearings are reasonable and plateaus set in 
well within \tsig.  In the case of poorer smearings it would still be 
 advantageous to be on an anisotropic lattice with many 
data points within \tsig, since that would facilitate multi-exponential 
fits.  Going to lattices with longer time extent in physical units in 
search of a plateau will not work for high momentum correlators since 
\tsigsp stops well before the end of the lattice.  It is much more 
profitable to explore better smearings and/or increase the number of 
points within \tsig.

\subsection{ Dispersion Relation}
A convenient way to investigate how well lattice simulations are reproducing 
 relativistic dispersion relations is to consider the following 
quantity \cite{alford},
\be
\label{cp}
C(p) \equiv \sqrt{\frac{E^2(p) - E^2(0)}{p^2}} ,
\ee
where $p \equiv |\vec{p}|$ and $E(p)$ is the total energy of the particle. 
In a relativistic theory $C(p)=1$ for all $p$.  
$C(p)$ is shown for the pion and the rho in Figs. 14 and 15.  One 
sees that on both the isotropic and anisotropic lattices, continuum 
behavior is observed to better than  $\sim5$\% for momenta of
up to 1.2GeV at the minimum and, in the case 
of the rho, even beyond that.  Hence, it appears one can simulate
 particles with 
momentum as high as $p \approx 1.5/a_s$ to $2.0/a_s$ without introducing large 
discretization errors.
 One notices slightly better behavior on isotropic 
 compared to on anisotropic lattices  
 for the smaller momentum points.  The anisotropic results are more 
sensitive to the tuning of $C_0$, for which we have used the one-loop 
estimate.  
We have studied the effect of this one-loop correction by calculating 
anisotropic light quark propagators with $C_0=1$ on a subset of 50  
configurations (at the same time adjusting $a_tm_0$ to get the same 
pion mass). The fancy squares in Fig.16 show results for $C(p)$ for 
$C_0=1$, i.e. without the one-loop correction.  They are compared with 
results from 50 configurations using $C_0=0.82$.  For completeness we also 
include corresponding points from Figs.14 and 15, which are based on the 
full set of 200 configurations with $C_0=0.82$. 
 Although  effects from the one-loop 
correction are not large, nevertheless, they are crucial for getting the 
$\sim2$\% to $\sim5$\% agreement with relativistic behavior. 
 Even if we had attempted 
to tune $C_0$ nonperturbatively, it would have been difficult to do much 
better simultaneously for both pions and rhos,
 than our perturbative estimate.  
The errors on $C(p)$ for anisotropic pions (from
 a bootstrap analysis) are smaller than one might have expected based 
on the large fluctuations noted above in pion correlators.  We are
 evidently observing 
large cancellations between fluctuations in the 
 zero and nonzero momentum correlators.

\vspace{.1in}
\noindent
The isotropic results in Figs. 14 and 15 are based on A-fits, defined above.  
In Fig. 17 we compare with 
results using B-fits. Results from the two different 
fits are consistent.  But, as expected, errors are much larger for B-fits 
and one cannot go beyond the three lowest momenta using them. 
 Unless specified 
otherwise, for the rest of this article we will use A-fits for 
pions and rhos on isotropic lattices.  The isotropic results in Figs. 14
and 15 are also consistent with findings in reference \cite{alford}, 
where comparisons 
were made with clover quarks at similar lattice spacings.  In those 
studies $C(p)$, at lattice spacing $a=0.25fm$,  
was consistent with unity within errors up to about $p=1.2$GeV for 
the D234 quark action and exhibited 10\% $\sim$ 20\% deviations from 
unity for the clover quark action.

\vspace{.1in}
\noindent
$C(p)$ is not applicable for heavy-light mesons involving NRQCD heavy 
quarks.  The NRQCD action omits the heavy quark rest mass and meson 
correlators fall off with an energy $E_{sim}(p)$ that differs from the 
total energy $E(p)$.  However one has,
\be
\delta E(p) \equiv E_{sim}(p) - E_{sim}(0) = E(p) - E(0),
\ee
and one can define the  ``kinetic'' mass, $M_{kin}=M_2$
through,
\be
\label{mkin}
M_{kin} = (p^2 - \delta E^2(p)) / (2 \,\delta E(p))  .
\ee
If $E_{sim}(p)$ has the correct momentum dependence, then $M_{kin}$ 
should be independent of the momentum used in the RHS of eq.(\ref{mkin}).
Fig.18 shows $M_{kin}$ for several mesons versus the momentum of the 
correlator from which it was extracted.  On the anisotropic lattice we have 
data for two heavy-light mesons, both the $B_s$ and the $D_s$.  The 
latter meson uses NRQCD charm quarks. 
 One sees that for all mesons, results for $M_{kin}$ are 
independent of momentum within errors
 up to about $p\sim 1.5$GeV.  For the pion and the 
rho we can also compare $M_{kin}$ with the rest mass $M_1\equiv E(0)$.  The 
deviation of $M_{kin}$ from $M_1$ reflects the deviation of
 $C(p)$ from unity in Figs.14 and 15 and, like the latter, is very
small.

\subsection{ Decay Constant Ratios}
Heavy meson semi-leptonic decay formfactors, which will be 
the focus of the next section, 
require matrix elements of heavy-light currents between hadronic states 
with and without momentum.  It is worthwhile considering first simpler 
 matrix elements  of currents between 
mesons and the vacuum and studying their momentum dependence. Such 
matrix elements are relevant for meson leptonic decays.  
Starting from the usual definition of the decay constant (in Euclidean space),
\be \label{deffb}
   \langle \, 0 \,| \, A_{\mu} \,|\,B,\,\vec{p} \, \rangle = \tilde{p}_{\mu} 
f_{B} ,
\ee
($\tilde{p}_\mu = (iE,\vec{p})$) one can form the ratio,
\be\label{ratio}
 \frac{\langle 0 | A_0 | PS \;,\; \vec{p} \rangle /\sqrt{E(p)}}
{\langle 0 | A_0 | PS \;,\; \vec{p}=0 \rangle /\sqrt{M_{PS}}}
= \frac{\sqrt{E(p)}}{\sqrt{M_{PS}}} .
\ee
This ratio was studied recently on finer lattices using a less 
improved action \cite{fbscale}  (similar calculations were done several years 
ago with relativistic heavy fermions in \cite{simone}).  
In the present article we will use the following current for 
the temporal component of the axial vector current,
\be
\label{a0}
A_0 \rightarrow J_{A_0}^{(0)} =   \overline{\Psi} \,\gamma_5 \gamma_0\, Q ,
\ee
where $\Psi$ is the light quark field and 
the heavy quark 4-spinor $Q$ has the 
NRQCD 2-spinor $\Phi$ as 
the upper two components and zero for the lower two components. 
The superscript $(0)$ signifies that $J_{A_0}^{(0)}$ is the 
zeroth order term in an $1/M$ expansion for the axial vector current. 
The LHS of eq.(\ref{ratio}) is replaced by,
\be\label{ratio0}
R^{(0)}(p) \equiv \frac{\langle 0 | J_{A_0}^{(0)} |  
\vec{p} \rangle /\sqrt{E(p)}} 
{\langle 0 | J_{A_0}^{(0)} |  \vec{p}=0 \rangle /\sqrt{M_{PS}}} .
\ee
In reference \cite{fbscale} 
 $1/M$ current and one-loop matching corrections were 
included in the ratio and seen to have only a small effect relative 
to using just the zeroth order current for heavy quark masses around the 
$b$-quark mass.
 Matching calculations for 
the actions of the present article have not been carried out yet so 
we are forced to use the simple ratio $R^{(0)}(p)$ here. 
Figs. 19 and 20 show $R^{(0)}$ for the $B_s$ and $D_s$ leptonic 
decays, compared with the expected continuum behavior of the RHS 
of eq.({\ref{ratio}).  One sees good agreement for most of the 
momentum range studied.  Only at the highest momentum ($> 1.5 $GeV) 
does one see $\sim15$\% deviations for $B_s$ mesons.  One should be able
to reduce these errors below $\sim10$\% by going to 
slightly finer lattices.  Since we do not include higher order (in $p/M$)
 current corrections 
one expects agreement with full continuum QCD behavior to be 
worse for the $D_s$ meson, especially at  higher momenta \cite{ak}. 
 The better agreement found in 
Fig. 20 as compared to in Fig. 19 is, hence, fortuitous.

\vspace{.1in}

\subsection{ Zero Momentum Spectrum }
In the process of studying momentum dependence of meson correlators we 
have also accumulated some zero momentum spectrum results.  They are 
summarized in Table II.
The first errors are statistical and the second represent errors coming 
from uncertainties in \ainvsp which we take to be roughly 10\% not 
including quenching effects.  
In this exploratory study we will not 
try to estimate other systematic errors.  One sees that 
the ``pion'' and the ``rho'' masses are very close to each other 
on the isotropic and anisotropic lattices. 
  The isotropic $B_s$ meson 
is about 250 MeV too heavy and the anisotropic one about 600 MeV too 
light compared to experiment.  On the other hand, 
the uncertainty coming from \ainvsp is at the 500-600 MeV level.

\vspace{.1in}
\noindent
It is amusing that the heavy-light hyperfine splittings, which have been 
the bane of lattice heavy-light spectroscopy in the past, 
agree so well here 
with experiment for both the $B_s$ and $D_s$ mesons.  One should not make 
too much out of this, however,
 until results at other lattice spacings have been obtained.  
At the moment we have no estimate of the size of scaling or unquenching 
corrections to this quantity.  If one corrects approximately for 
the incorrect heavy quark mass in our simulations and
 multiplies by a factor of
[calculated meson mass]/[experimental meson mass], then the entries in 
Table II for
the $B_s^\ast - B_s$ hyperfine splitting, are modified to 48MeV or 40MeV 
for isotropic and anisotropic lattices respectively.  Similarly 
the $D_s^\ast - D_s$ splitting becomes 148MeV.  All these numbers 
are much larger than
 and in better agreement with experiment than in previous 
lattice estimates \cite{joachim,arifa,lewis,boyle,fermi,jlqcd}. 
 One difference between the present and previous 
calculations lies in the last term in eq.(\ref{fmunu}) which enhances 
$\tilde{F}_{\mu\nu}$.  It is hard to imagine, however, that this term alone  
can be the whole story.
It will be interesting to see what happens after 
systematic errors have been investigated more thoroughly and various 
corrections to the present calculation have been incorporated.

\section{Results from Three Point Functions}

For pseudoscalar to pseudoscalar semi-leptonic decays one is interested in 
matrix elements of the heavy-light vector current $V^\mu$. The matrix 
elements are then parametrized in terms of form factors $f_+$ and $f_-$ 
(or $f_1$ and $f_0$).
\be
\label{vmu}
\vev{\pi(\ppri)|V^\mu|B(p)} = f_+(q^2) (p^\mu + \pprimu) + f_-(q^2) 
(p^\mu - \pprimu) ,
\ee
where $q^2 = (p^\mu - \pprimu)^2$.  We will write formulae for 
$B$ decays but they apply also to $D$ decays. 
$f_1$ and $f_0$ are defined as :
\be
\label{fp0}
f_0 = f_+ + \frac{q^2}{(M_{B}^2 - M_\pi^2)}\, f_- \qquad , \qquad f_1 = f_+
\ee

\vspace{.1in}
\noindent
Our main goal in this section will be to compare the quality of signals for 
matrix elements such as those in eq.(\ref{vmu}) between isotropic and 
anisotropic lattices.
Since the light quarks in this study 
are still heavier than the strange quark, 
the physical situation we are simulating 
 will be closest to  $B_s$ or $D_s$ decays into Kaons.

\subsection{Current Matrix Elements}
In order to extract the matrix element of eq.(\ref{vmu}) one starts 
from the following three-point function, 
$G^{(3)}_\mu(\vec{p},\vec{\ppri},t_B,t)$, on the lattice (for technical 
reasons it is more convenient to consider the time reversed matrix 
element)
\be
\label{threepnt1}
G_\mu^{(3)}(\vec{p},\vec{\ppri},t_B,t) = \sum_{\vec{x}}\sum_{\vec{y}}
e^{-i\vec{p}\cdot\vec{x}} e^{i(\vec{p} - \vec{\ppri})\cdot \vec{y} }
\vev{0|\Phi_{B}(t_B,\vec{x})\,V^L_\mu(t,\vec{y})\, \Phi^\dagger_\pi(0)
|0} .
\ee
$\Phi_\pi^\dagger$ and $\Phi_B^\dagger$ are interpolating operators 
used to create the pion or $B$ meson respectively.  $V^L_\mu$ is the 
dimensionless Euclidean space 
lattice heavy-light vector current.  It will be defined more 
precisely below.  It is related to the continuum Minkowski space $V^\mu$ 
through,
\be
V^L_\mu = a_s^3 \sqrt{Z^{(0)}_q}\,\xi(\mu)\, V^\mu .
\ee
 $Z_q^{(0)}$ is the tree-level 
wave function renormalization for lattice light quark actions.  It is 
discussed for the isotropic and anisotropic D234 actions in the Appendix. 
$\xi(\mu)$ is the conversion factor between Euclidean and Minkowski space 
quark bilinear currents which is necessary due to the different $\gamma$-matrix
conventions in the two spaces.
$\xi(0)=1$ and $\xi(k)=-i$, $k=1,2,3$. 
 $t_B$ denotes the time 
slice at which the $B$ meson operator is inserted. 
In the simulations $t_B$ is kept fixed and we vary $t$, the timeslice 
of the current insertion, between 0 and $t_B$.   Physics is extracted from 
those timeslices where the corresponding two-point correlators are  
dominated by the ground state and where $e^{-E_\pi (T-t)}$ can be ignored 
relative to $e^{-E_\pi t}$, $T$ being the time extent of the lattice.
  If these 
conditions are satisfied the three-point correlator (\ref{threepnt1}) becomes,
\be
\label{threepnt2}
G_\mu^{(3)}(\vec{p},\vec{\ppri},t_B,t) 
\rightarrow \frac{\vev{0|\Phi_{B}|B(\vec{p})} 
\,\vev{B(\vec{p})|V^L_\mu|\pi(\vec{\ppri})} \,\vev{\pi(\vec{\ppri})|
\Phi_\pi^\dagger|0}} {(2E_{B}a_s^3) \;\;(2E_\pi a_s^3)}
\,e^{-E^{sim}_{B}(t_B-t)} \, e^{-E_\pi t} .
\ee
The exponential factors in eq.(\ref{threepnt2}) can be removed by 
dividing with the appropriate two-point functions.
\begin{eqnarray}
G^{(2)}_B(\vec{p},t) &=& \sum_{\vec{x}}e^{-i\vec{p}\cdot\vec{x}}
\vev{0|\Phi_B(t,\vec{x})\,\Phi_B^\dagger(0)|0}  \nl
&\rightarrow& 
\frac{|\vev{B(\vec{p})|\Phi_B^\dagger|0}|^2}{(2E_Ba_s^3)} \,e^{-E^{sim}_Bt}
\equiv \zeta_{BB}e^{-E^{sim}_Bt} ,  \\
\label{pioncorr}
G^{(2)}_\pi(\vec{\ppri},t) &\rightarrow&
\frac{|\vev{\pi(\vec{\ppri})|\Phi_\pi^\dagger|0}|^2}{(2E_\pi a_s^3)} \,
[e^{-E_\pi t} + e^{-E_\pi(T-t)}] 
\approx \zeta_{\pi\pi}\,e^{-E_\pi t} .
\end{eqnarray}
The matrix element of the continuum current $V^\mu$ can now be obtained from,
\begin{eqnarray}
\label{c3}
\Gamma^{(3)}_\mu(\vec{p},\vec{\ppri},t_B,t) &\equiv&
\xi^\ast(\mu) \, 
\frac{G^{(3)}_\mu(\vec{p},\vec{\ppri},t_B,t)}{G_B^{(2)}(\vec{p},t_B-t) \,
G^{(2)}_\pi(\vec{\ppri},t)} \sqrt{\zeta_{BB} \zeta_{\pi\pi}} \\
\label{vtilde}
&\rightarrow& \sqrt{Z^{(0)}_q}\, \frac{\vev{B(\vec{p})|V^\mu|\pi(\vec{\ppri})}}
{2\sqrt{E_B E_\pi}} \equiv \vev{\tilde{V}^\mu} .
\end{eqnarray}

\vspace{.1in}
\noindent
Eqs. (\ref{c3}) and (\ref{vtilde}) relate
 the three- and two-point functions evaluated 
in our simulations to the continuum matrix elements of eq.(\ref{vmu}). 
We need to specify now the lattice current $V^L_\mu$ that enters into 
the three-point functions. It is defined  in 
terms of Euclidean space $\gamma$-matrices $\gamma_\mu = 
\xi(\mu)\,\gamma^\mu_{(Mink.)}$.
Since we use the NRQCD formulation of heavy quarks, 
 $V_\mu^L$ becomes an expansion in $1/M$, 
the inverse of the heavy quark mass. After matching to continuum QCD one has,
\be
V_\mu^L =  \sum_j C_j^{(V_\mu)}  J_{V_\mu}^{(j)}.
\ee
For $j=0$ one has the zeroth order, $\order((1/M)^0)$, current,
\be
 J_{V_\mu}^{(0)} =   \overline{\Psi} \,\gamma_\mu \, Q ,
\ee
with the fields $\Psi$ and $Q$ defined as in eq.(\ref{a0}).  Higher order 
currents ($j>0$) are listed in reference \cite{pert2}.  
The matching coefficients $C_j^{(V_\mu)}$ are not yet known 
beyond tree-level so we will work with $C_0^{(V_\mu)} = 1$ and all 
other $C_j$'s equal to zero.  One of the $1/M$ current corrections 
$J_{V_\mu}^{(1)}$ has $C_1^{(V_\mu)} = 1$ + $\order(\alpha_s)$ and also
contributes at tree-level.  However, its matrix elements include power 
law terms that will not be cancelled unless a proper one-loop calculation 
has been carried out.  Hence, we do not include $J_{V_\mu}^{(1)}$ 
contributions in the present study.  

\vspace{.1in}
\noindent
We have evaluated 
$\Gamma^{(3)}_\mu(\vec{p},\vec{\ppri},t_B,t)$ 
 of eq.(\ref{c3}) for several pion momenta 
$\vec{\ppri}$ ranging from (0,0,0) to (0,0,3) in units of $2 \pi/(aL)$. 
The $B$ meson momentum was always set equal to zero.   
 On the anisotropic 
lattice we used $t_B/a_t = 24$ and on the isotropic lattice $t_B/a_t = 10$. 
Figs. 21 and 22 show $\Gamma_0^{(3)}$ versus $t$ in physical units for zero 
momentum and for three nonzero momenta. 
 One sees that, 
where there is overlap, anisotropic and isotropic lattices 
give consistent results. 
One is interested, of course, in the 
region where $\Gamma_0^{(3)}$ is independent of $t$  implying that the 
simple $t$-dependence in eq.(\ref{threepnt2}) and the last 
expression in eq.(\ref{pioncorr}) 
is justified.  $\Gamma_0^{(3)}$ can then 
 be identified with the asymptotic matrix element $\vev{\tilde{V}^0}$ of 
eq.(\ref{vtilde}).
 Again the crucial question for the higher momentum 
isotropic results becomes whether one would believe in the presence of 
a plateau if one did not have the comparison anisotropic data.  To illustrate 
this point we show $\Gamma_0^{(3)}$ versus $t$ in lattice units separately 
for isotropic and anisotropic lattices in Figs. 23 to 26.  One might 
still feel comfortable extracting a signal from the (1,1,1) isotropic
data.  One would be hard pressed, however, to claim that a plateau has 
been established at momentum (0,0,2) based solely on the left hand plot 
in Fig.26. 
  Note that \tsig, the 
time range over which statistical errors are under control, 
is about 
$1fm$ for momentum (1,1,1) and has shrunk to about $0.5fm$ for (0,0,2).
We saw in the previous section that individual two-point correlators 
had reached a plateau by $0.3 \sim 0.4 fm$.  If sufficient data points 
could be introduced between $0.3fm$ and $0.5fm$, then one should be able to 
extract meaningful results for $\vev{\tilde{V}^0}$.  Hence, once again one 
sees an advantage to using anisotropic lattices starting with momentum 
(1,1,1).  This is the same situation as with the pion and rho correlators 
described in the previous section, not a surprising finding since 
statistical errors in $\Gamma_\mu^{(3)}$ are dominated by the pion correlator 
and not by the $B$ correlator.

\vspace{.1in}
\noindent
From the region where $\Gamma^{(3)}_\mu$ is independent of $t$, one can 
extract $\vev{\tilde{V}^\mu}$.  We show 
 $\vev{\tilde{V}^0}$ and 
 $\vev{\tilde{V}^k}$ in Fig. 27 as a function of the pion momentum $\ppri$.  
 For points at the largest momenta on the anisotropic lattice 
one could be seeing some discretization effects.  Only a more careful 
analysis involving simulations at several lattice spacings and/or 
further studies with nonzero $B$ meson momenta will be able to shed
more light on this.
Here we are concentrating mainly on whether 
signals can be extracted, postponing scaling studies for the future.
  This is in contrast to the 
two-point correlator studies of the previous section where several
continuum expectations based just on simple Lorentz symmetry 
considerations could be tested.

\subsection{Form Factors $f_+(q^2)$ and $f_0(q^2)$}

From (\ref{vtilde}), (\ref{vmu}) and (\ref{fp0}) one can extract 
  the form factors $f_+(q^2)$ and $f_0(q^2)$.  Isotropic lattice 
results for the $B_s$ meson are shown in Fig. 28 and anisotropic 
lattice results for the $B_s$ and $D_s$ mesons in Figs. 29 and 30. 
The kinematics, including the range in $q^2$ that is covered, 
depends on the meson masses of Table II and differs for the two 
types of lattices.  The errors in the 
three figures are statistical and come from a simultaneous 
bootstrap analysis of the $V^0$ and $V^k$ three-point functions and 
the B (or D) and pion correlators.

\vspace{.1in}
\noindent
Since we have at present only tree-level matching of the heavy-light 
currents and also have not tuned (or extrapolated in) the heavy and light 
quark masses, the above form factor results cannot be applied yet to 
phenomenology.  What is important, however, is that with just 200 
configurations it was possible to obtain form factors for 
a nontrivial range in $q^2$ and that we were able to demonstrate the
advantages of anisotropic lattices in calculations of this kind.  
By increasing statistics in the future 
one should be able to get good data over an even wider range in $q^2$. 
At that point one would also want to go to finer lattices and 
work with improved currents so that finite momentum errors are minimized 
even at the largest momenta for which signals can be obtained.

\section{Summary}
This article investigates the extent to which anisotropic lattices can help 
in extracting better signals from two- and three-point correlators 
involving high momentum hadrons and whether they 
can play an important role in studies of semi-leptonic heavy meson 
decays into light hadrons.
 To address this question we have 
carried out simulations 
of heavy meson semi-leptonic decays, 
in parallel, on isotropic and anisotropic lattices. 
  In order to have 
a meaningful comparison, we work with similar coarse spatial lattice 
spacings and use identical sources and smearings on the two lattices.

\vspace{.1in}
\noindent
We find that it is considerably easier to extract reliable signals from 
anisotropic simulations once the light hadron momentum reaches 
(1,1,1) $2 \pi/(a_s L)$ or higher.  This advantage 
may not be so obvious just by 
comparing Figs. 28 and 29 
and one needs to 
go back to figures such as Fig. 22 to fully appreciate how much the 
anisotropy is helping here.  The 
last point (at the smallest value of $q^2$) in Fig. 28 comes  from 
the isotropic data in the left hand plot of Fig. 22 (see also Fig. 25). 
It is because the first three isotropic points in Fig. 22 
agree with the anisotropic data, that one feels confident about the 
form factor results in Fig. 28.   Without the anisotropic data, one would 
have to allow for a considerable additional systematic error, which
 one might call 
``fitting error'' or ``$t_{min}$ dependence error'', when presenting 
 isotropic results.  
Hence, the main conclusion from the present work is that 
 anisotropic lattices definitely improve signal quality and 
should be considered in semi-leptonic decay studies, especially 
 if a large range in $q^2$ is of interest.

\vspace{.1in}
\noindent
The advantages of anisotropic lattices come at a certain price.  For instance,
such lattices require more sites in the time direction.  Also light quark 
inversions take more iterations in order to get the same pion-to-rho
mass ratio $P/V$.  In the present simulations, the
 cost increase from just these two effects meant a factor 
of $2.4 \times 1.5 = 3.6$  in CPU time.
 Working with anisotropic lattices also 
requires tuning of more parameters. At  a minimum, two additional 
parameters, $\eta$ in the glue action and $C_0$ in the D234 quark action 
must be determined nonperturbatively or perturbatively.  It is important 
that efficient procedures for carrying out such tunings be developed. 
Another drawback in the current anisotropic simulations was the increased 
susceptability to exceptional configurations.  If it were not 
for this problem and the larger fluctuations in pion correlators, 
 the anisotropic results for the form factors 
$f_0(q^2)$ and $f_+(q^2)$ would have been of even higher quality relative to 
those coming from isotropic lattices.  There are several ways one could 
try and ameliorate this problem in the future.  Just going to finer 
lattices and working with more moderate anisotropies, $\chi < 5.3$, 
should help.  One could also explore other light quark actions that 
have better chiral properties than the clover or D234 actions, such 
as highly improved staggered fermions \cite{gpltsukuba,staggered}, 
domain wall fermions \cite{domainwall} or 
the twisted QCD approach \cite{twistqcd}.  

\vspace{.1in}
\noindent
Ultimately one will have to weigh the additional costs associated with 
anisotropic lattices against the likelihood that an isotropic simulation 
with much higher statistics will allow us to approach the high momenta 
we are seeking.  Based on our (and also other peoples') experience to 
date, we do not believe isotropic lattices can be competitive at 
high momenta.  Signal-to-noise of high momentum correlators will 
be decreasing exponentially with $t$.  When $a_t$ is large it will 
be very costly to move even one additional point from the noise 
into \tsig.  In order to do so one will need to reduce errors 
by roughly a factor of $e^{a_t (E(p) - E0)}$, where $E0$ is the ground 
state energy \cite{tasi}.
In other words, one will need an 
increase in statistics by 
 a factor of $N_{stat} \equiv |e^{a_t (E(p) - E0)}|^2$.  Typical numbers 
 for the current isotropic simulation for momenta starting with (1,1,1) 
and higher, would be 
$N_{stat} \approx 4 - 10$.  This means a factor of 16 - 100 if 
one wants to go from just one or two points in \tsigsp 
 to a marginally useful \tsigsp of 3 to 4 points.  This is much 
more than the cost of going to anisotropic lattices.

\vspace{.1in}
\noindent
We mention another unrelated advantage of anisotropic actions.
One difference between 
$\act^{(iso)}_G$  and $\act^{(aniso)}_G$ is the omission in the latter 
of rectangles 
that span two links in the time direction.  Similarly 
$\act^{(iso)}_{D234}$  includes higher time derivatives that are absent in 
 $\act^{(aniso)}_{D234}$.  As a consequence, for the same amount 
of improvement in spatial directions, 
the anisotropic actions 
 suffer less from 
lack of reflection positivity and/or presence of ghosts.
 In addition to being theoretically cleaner this means perturbation 
theory is more straightforward for the anisotropic actions.  One such 
example is the $Z^{(0)}_q$ calculation in the Appendix.

\vspace{.1in}
\noindent
In the course of this study we accummulated a wealth of
two-point correlator data at finite momentum.  This enabled us to 
 compare lattice results for 
dispersion relations and leptonic decays of finite momentum heavy mesons 
 with continuum expectations.  We found that momentum dependent 
discretization errors were under control 
 and less than 10\% up to about $a_s p \approx 2$.  Discretization errors 
at high momenta in 
three-point functions have not been critically assessed to date. 
However, assuming
a situation  similar to the one found with the two-point functions, 
prospects for simulating hadrons in semi-leptonic decays 
with momenta as high as $1.5-2GeV$ look promising. 
 Only slightly finer lattices 
($a_s^{-1} \geq 1GeV$) than those of the present work may be required.
This would have to be
coupled with an anisotropy of $\chi \approx 2.5$ or higher in order to be able 
to extract a signal at those high momenta.  The experience gained in the 
present work will be indispensable when picking optimal simulation 
parameters in the future and going onto more realistic calculations.

\vspace{.2in}
\noindent
\acknowledgements
This work was supported by the DOE under DE-FG02-91ER40690, PPARC under 
PPA/G/0/1998/00559 and by NSF.  Simulations were 
carried out at the Ohio Supercomputer Center and at NERSC.  We thank 
Mark Alford for help in checking the gauge and the D234 inversion codes and 
Peter Boyle for useful discussions on code improvements.

S.C. acknowledges a fellowship from the Royal Society of Edinburgh.
J.S. thanks Cornell University  and the University of Glasgow for 
their hospitality during the initial stages of this project. 
Support from a PPARC Visiting Fellowship PPA/V/S/1997/00666 is 
gratefully acknowledged.

\appendix

\section{Tree-level Wave Function Renormalization for D234 Quark Actions}
In this appendix we sketch the derivation of the tree-level wave function 
renormalisations, $Z^{(0)}_q$, for the light quark actions used in this 
article.  Results for $\act^{(aniso)}_{D234}$ have already appeared 
in reference \cite{pert}.  One starts from the requirement that the propagator 
for a zero momentum quark have the form,
\begin{eqnarray}
\label{def0}
G(t,\vec{p}=0) & = & \int^{\pi/a_t}_{-\pi/a_t} \frac{dp_0}{2 \pi} 
e^{ip_0 t} \overline{G}(p_0,\vec{p}=0)  \nl
  & \equiv &  Z_q e^{-M_1t} \frac{1 + \gamma_0}{2} 
 \; + \; \ldots \;\; .
\end{eqnarray}
 $\overline{G}(p)$ is the momentum 
space propagator and $M_1$ denotes the pole mass.  
 The dots
refer to lattice artifacts and additional multi-particle states that 
could be created by the lattice fermion field operator $\Psi$ beyond the 
single quark state.  Writing 
\be
\frac{1}{a_t} \, \overline{G}(p_0,\vec{p} = 0) =
 \frac{1}{i\gamma_0 A + B }  ,
\ee
and using the complex variable
\be 
z \equiv e^{i a_tp_0} = e^{-a_tE},
\ee
one finds
\be 
G(t,\vec{p}=0)  =  \int^{\pi/a_t}_{-\pi/a_t} \frac{dp_0}{2 \pi} 
e^{ip_0 t} a_t \frac{-i \gamma_0 A + B}{(A^2 + B^2)}
= \int\frac{dz}{2 \pi i z} z^{t/a_t} \frac{-i \gamma_0 A + B}
{(A + i B)(A-iB)}.
\ee
One can show that the pole,
\be
z_1 \equiv e^{-a_tM_1}
\ee 
corresponding to a physical positive 
energy particle obeys $(A-iB)|_{z=z_1} = 0$ or $B|_{z=z_1} = 
-iA|_{z=z_1}$. The contribution to $G(t,0)$ from the residue at this 
physical pole is then given by,
\be
\left [ \frac{z^{t/a_t} (-i \gamma_0 A + B)}{z \, \frac{d}{dz} 
(A^2 + B^2)} \right ] _{z = z_1} = 
\frac{(\gamma_0 + 1)}{2} e^{-M_1t} \,
\left [ \frac{-i}{z (A^\prime - i B^\prime)} \right ] _{z=z_1}.  
\ee
Using
\be
 \left(z \frac{d \,f}{dz} \right)_{z = z_1} = -i \left( \frac{d \,f}{d(a_tp_0)}
\right)_{p_0 = iM_1}.
\ee
and comparing with eq.(\ref{def0}), one finds,
\be
\label{zq}
Z_q = \left [\frac{1}{\frac{d}{d(a_tp_0)}(A-iB)} \right ]_{p_0 = iM_1}.
\ee

\vspace{.1in}
\noindent
For $\act^{(aniso)}_{D234}$ one has at tree-level,
\begin{eqnarray}
A^{(aniso)} &=& sin(a_tp_0)   \\
B^{(aniso)} &=& a_t m + \chi - \chi cos(a_t p_0)  \\
\end{eqnarray}
and
\be
e^{-a_tM_1} = 
\frac{(a_t m + \chi) - \sqrt{(a_t m + \chi)^2 + 1 - \chi^2}}
{\chi-1}
\ee
or equivalently,
\be
e^{a_tM_1} = 
\frac{(a_t m + \chi) + \sqrt{(a_t m + \chi)^2 + 1 - \chi^2}}
{\chi+1}  .
\ee
As explained in reference \cite{pert} 
$m = m_0-m_c$ and $m_c$ is the value of $m_0$ that gives a massless pion.
From eq.(\ref{zq}) one finds,
\be
Z^{(0),aniso}_q = \frac{1}{cosh(a_tM_1) + \chi sinh(a_t M_1)}
= \frac{1}{\sqrt{(a_tm)^2 + 2 (a_tm)\chi + 1}}.
\ee

\vspace{.1in}
\noindent
For the isotropic action $\act^{(iso)}_{D234}$ with higher time derivatives 
 the formulas are more complicated.
\begin{eqnarray}
A^{(iso)} &=& \frac{4}{3} sin(a_t p_0) - \frac{1}{6} sin( 2 a_t p_0)  \\
B^{(iso)} &=& a_t m + \frac{4}{3}(1 - cos(a_tp_0)) 
- \frac{1}{6}sin^2(a_tp_0) .
\end{eqnarray}
Eq({\ref{zq}) leads to
\be
Z^{(0),iso}_q = \frac{1}{\left [ \frac{4}{3} e^{a_tM_1} - 
\frac{1}{3}(cosh^2(a_tM_1) + sinh^2(a_tM_1) + sinh(a_tM_1) cosh(a_tM_1))
\right ] }.
\ee
The tree-level physical pole $z_1 = e^{-a_tM_1}$ is the solution to,
\be
\label{poly}
z^4 - (24 a_tm + 30) z^2 + 32 z -3 = 0
\ee
that evolves smoothly from $z_1=1$ at $a_tm = 0$.
A lengthy closed expression for $z_1$ as a function of $a_tm$ can be obtained 
(using for instance Mathematica), however we do not consider it 
worthwhile reproducing it here.  It is easier to plug in specific values
for $a_t m$ into eq.(\ref{poly}) before solving for the roots. 
At small $a_t m$ the physical pole is well approximated by 
\be
\label{z10}
\frac{1}{z_1} = e^{a_tM_1} = 1 + (a_t m) + \frac{1}{6} (a_tm)^3 
- \frac{5}{24} (a_tm)^4 
\quad + \quad O((a_tm)^5)  .
\ee
For the mass parameter values used in the current simulations, the 
sum of just the first three terms in (\ref{z10}) differs from the 
exact solution to eq.(\ref{poly}) by only 1\%.

\vspace{.1in}
\noindent
In order to get explicit values for $Z_q^{(0)}$, one needs to 
know $a_tm= a_t(m_0-m_c)$.  Since we have data at only one light quark 
mass, we do not have a nonperturbative estimate for $a_tm_c$ based on 
a vanishing pion mass.  We approximate $a_tm_c$ using perturbation 
theory and find \cite{pert}, $a_tm \approx 0.635$ and $a_t m \approx 0.196$ 
respectively for the isotropic and anisotropic lattices.  This leads to 
\begin{eqnarray}
\sqrt{Z^{(0),iso}_q} \;\; &=& \frac{1}{1.2226}  \\
  & &  \nl
\sqrt{Z^{(0),aniso}_q} &=& \frac{1}{1.3286}  
\end{eqnarray}
which are the values used in section IV.

%%%%%%%%%%%%%%%%%%%%%%%%%%%%%%%%%%%%%%%%%%%%%%%%%%%%%%%%%%%%%%%%%%%%%%
%
%                       TABLES
%
%%%%%%%%%%%%%%%%%%%%%%%%%%%%%%%%%%%%%%%%%%%%%%%%%%%%%%%%%%%%%%%%%%%%%

\begin{table}
\caption{Simulation Details.  
  }
\begin{center}
\begin{tabular}{c|cc}
%\hline
    & isotropic& anisotropic\\\hline
 lattice size  &  $8^3 \times 20$    &  $ 8^3 \times 48 $  \\
\# configs     &  200                &   200              \\
$\beta$        &  1.719              &  1.8               \\
Landau link $u_0$   &  0.797 \cite{alford}
         &  $u_s$=0.721  $\;u_t$=0.992 \cite{ron}  \\
$\chi_0$       &                     &  6.0             \\
$\chi = a_s /a_t$ &     1            &  5.3 \cite{ron}      \\
$C_0$          &        1            &    0.82         \\
$a_s^{-1}$     &        0.8(1) GeV   &   0.7(1) GeV    \\
$a_t^{-1}$    &         0.8(1) GeV   &   3.7(4) GeV    \\
$a_t m_0$     &         1.15         &   0.39          \\
$P/V$         &         0.725(5)     &   0.726(6)     \\
typical \# of BiCGstab iters &  140 - 160           &  220 - 270   \\
$a_sM_0$      &         6.5          &  7.0 and 2.0    \\
\end{tabular}
\end{center}
\end{table}

\begin{table}
\caption{Some Spectrum Results.  The first errors are statistical and 
the second are estimates for errors due to uncertainties in the scale.
The hyperfine splittings have not been adjusted for incomplete 
tuning of the heavy quark mass (see text for adjusted numbers).
  }
\begin{center}
\begin{tabular}{c|ccc}
%\hline
    & isotropic& anisotropic & Experiment\\\hline
 light-light  & & &  \\
\hline
 ``pion'' mass &  0.856(3)(86) GeV &   0.840(7)(84) GeV  &  \\
 ``rho'' mass  &  1.179(10)(118) GeV  &  1.158(7)(116) GeV  &  \\
\hline
  heavy-light  & & &  \\
\hline
$B_s$   &  5.65(31)(57) GeV   &  4.80(20)(48) GeV  &  5.369 GeV\\
$D_s$  &          &       2.04(5)(20) GeV  &  1.969 GeV  \\
\hline
$B_s^\ast - B_s$ &  46(3)(5) MeV    &  45(2)(5) MeV  &  47.0(26) MeV \\
$D_s^\ast - D_s$ &    &  143(4)(14) MeV  & 143.8(4) MeV  \\
\end{tabular}
\end{center}
\end{table}

%%%%%%%%%%%%%%%%%%%%%%%%%%%%%%%%%%%%%%%%%%%%%%%%%%%%%%%%%%%%%%%%%%%%%
%%
%%                               FIGURES
%%
%%%%%%%%%%%%%%%%%%%%%%%%%%%%%%%%%%%%%%%%%%%%%%%%%%%%%%%%%%%%%%%%%%%%%

\newpage
\begin{figure}
\centerline{
\ewxy{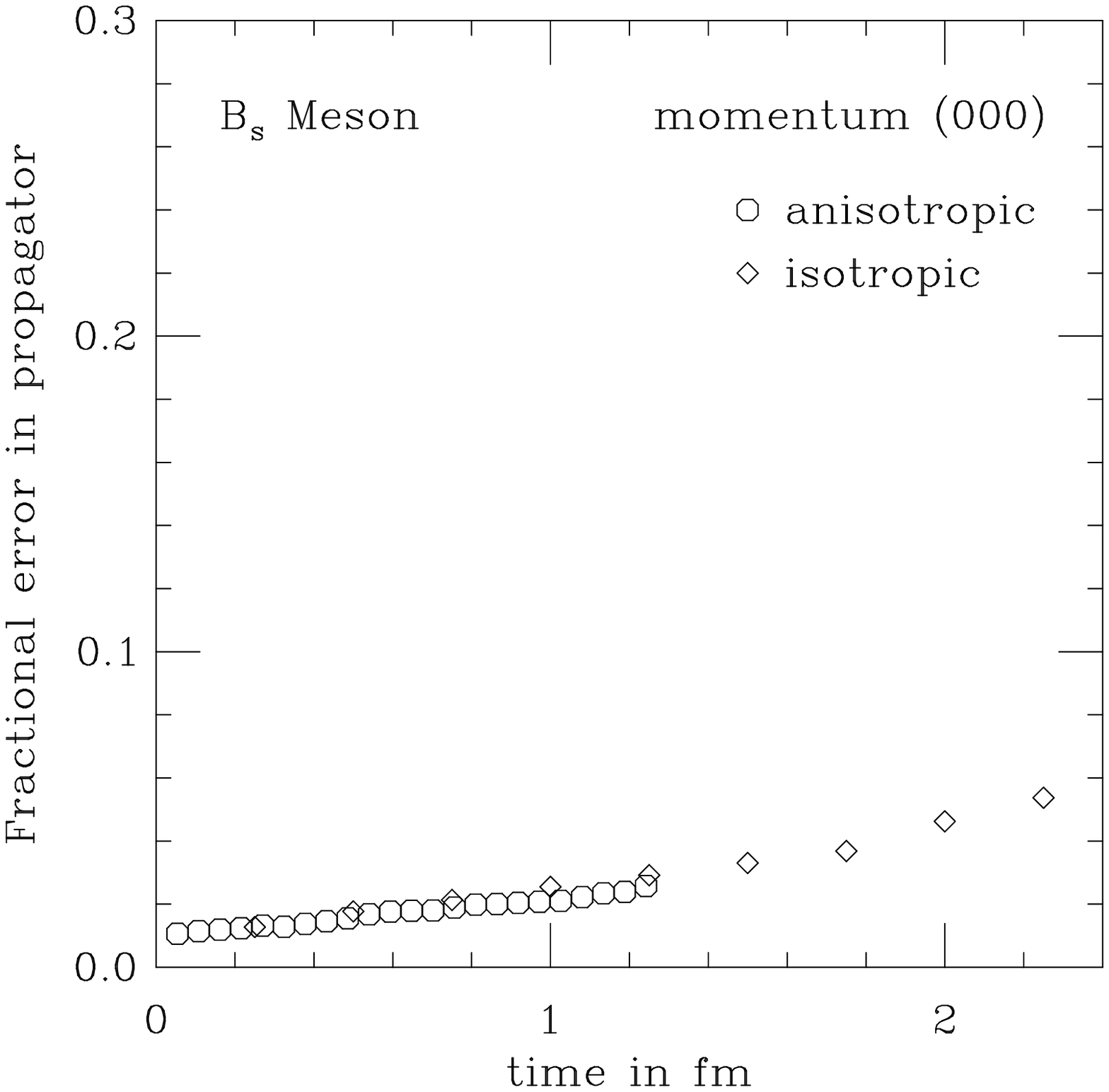}{90mm}
\ewxy{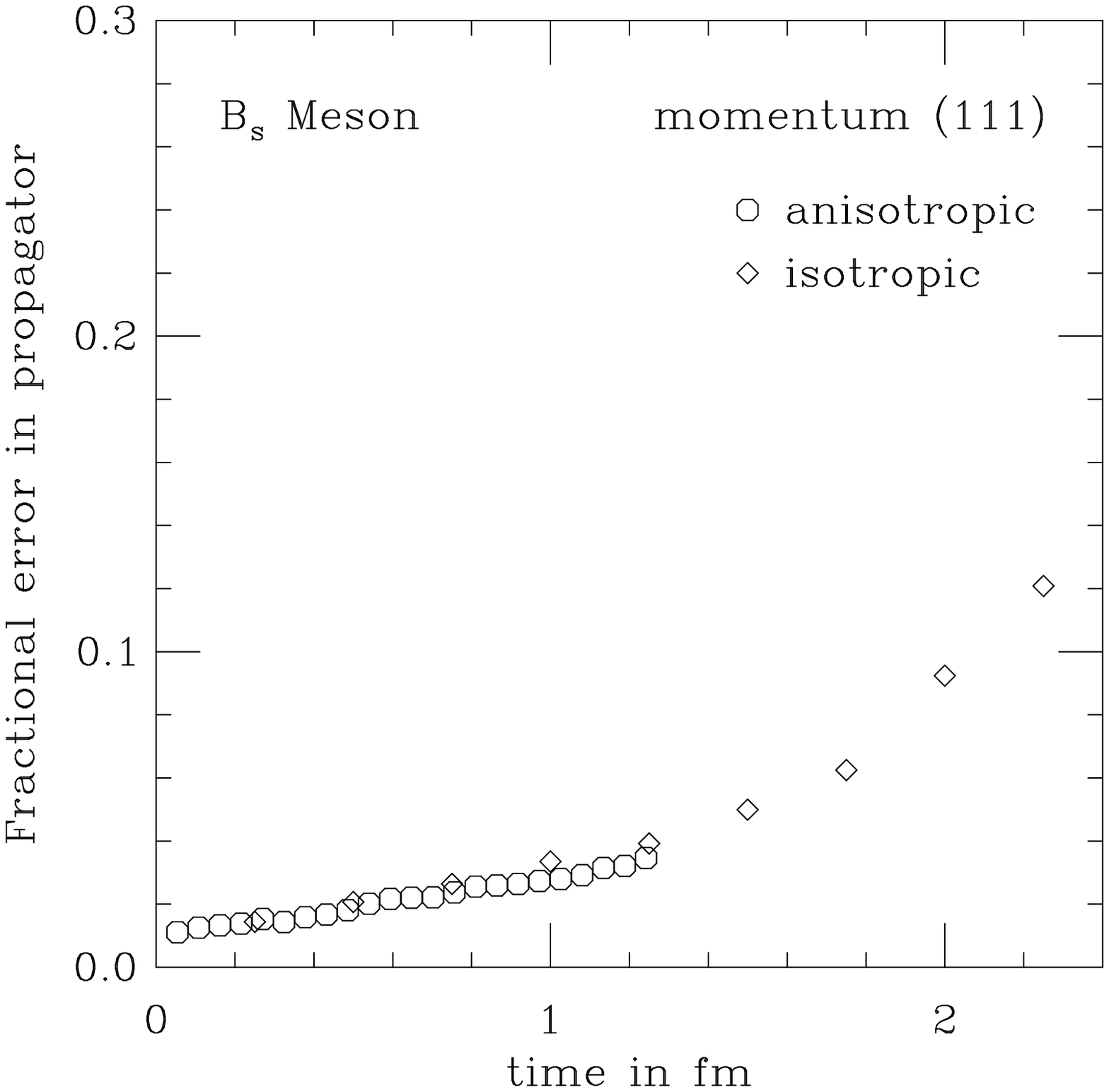}{90mm}
}
%\centerline{(i) \hspace{10cm} (ii)}
\caption{ 
Fractional errors in $B_s$ meson correlators for momentum (0,0,0) and 
(1,1,1).
}
\end{figure}

\begin{figure}
\centerline{
\ewxy{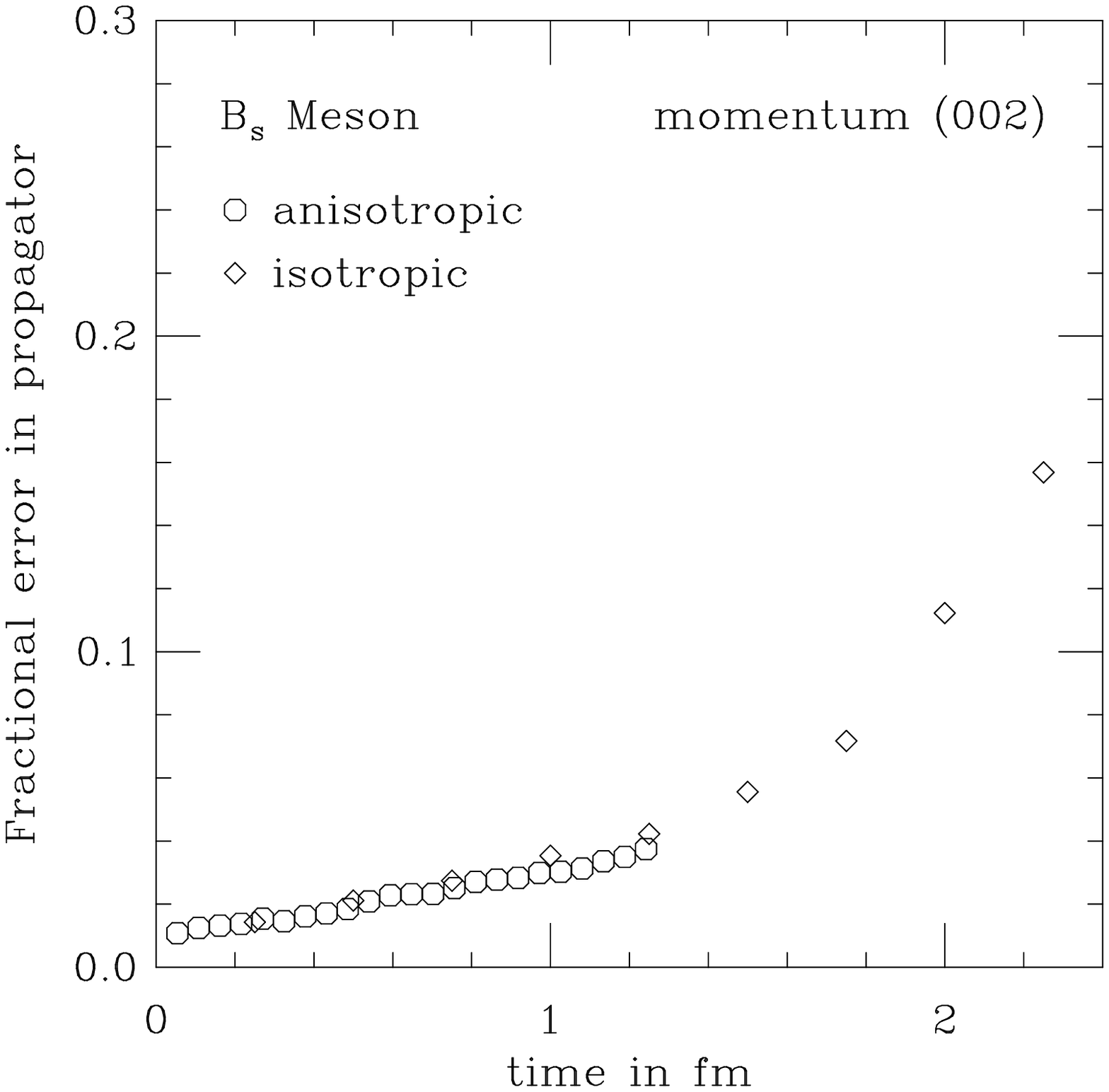}{90mm}
\ewxy{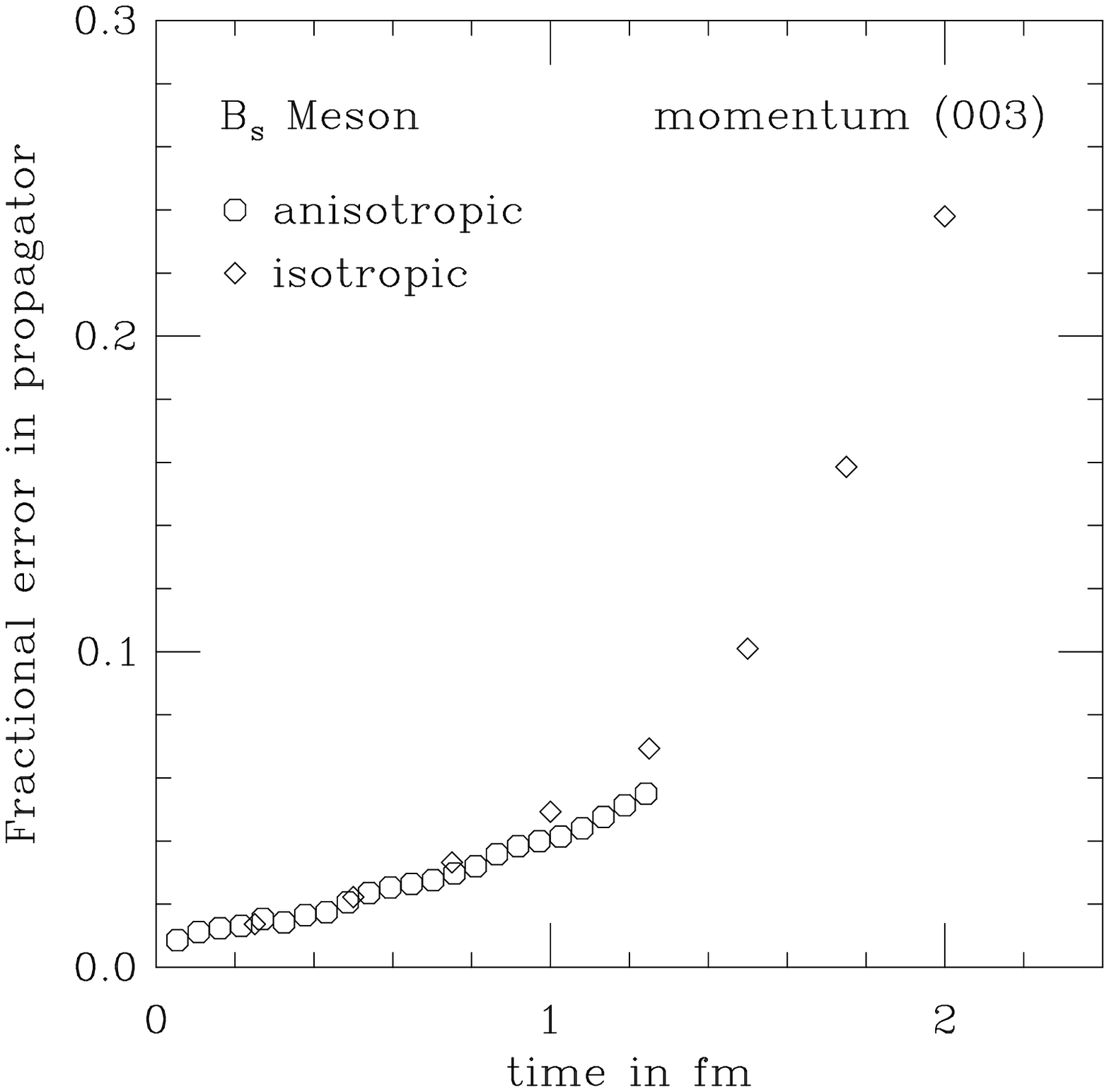}{90mm}
}
%\centerline{(i) \hspace{10cm} (ii)}
\caption{ 
Fractional errors in $B_s$ meson correlators for momentum (0,0,2) and 
(0,0,3).
}
\end{figure}

\newpage
\begin{figure}
\centerline{
\ewxy{ 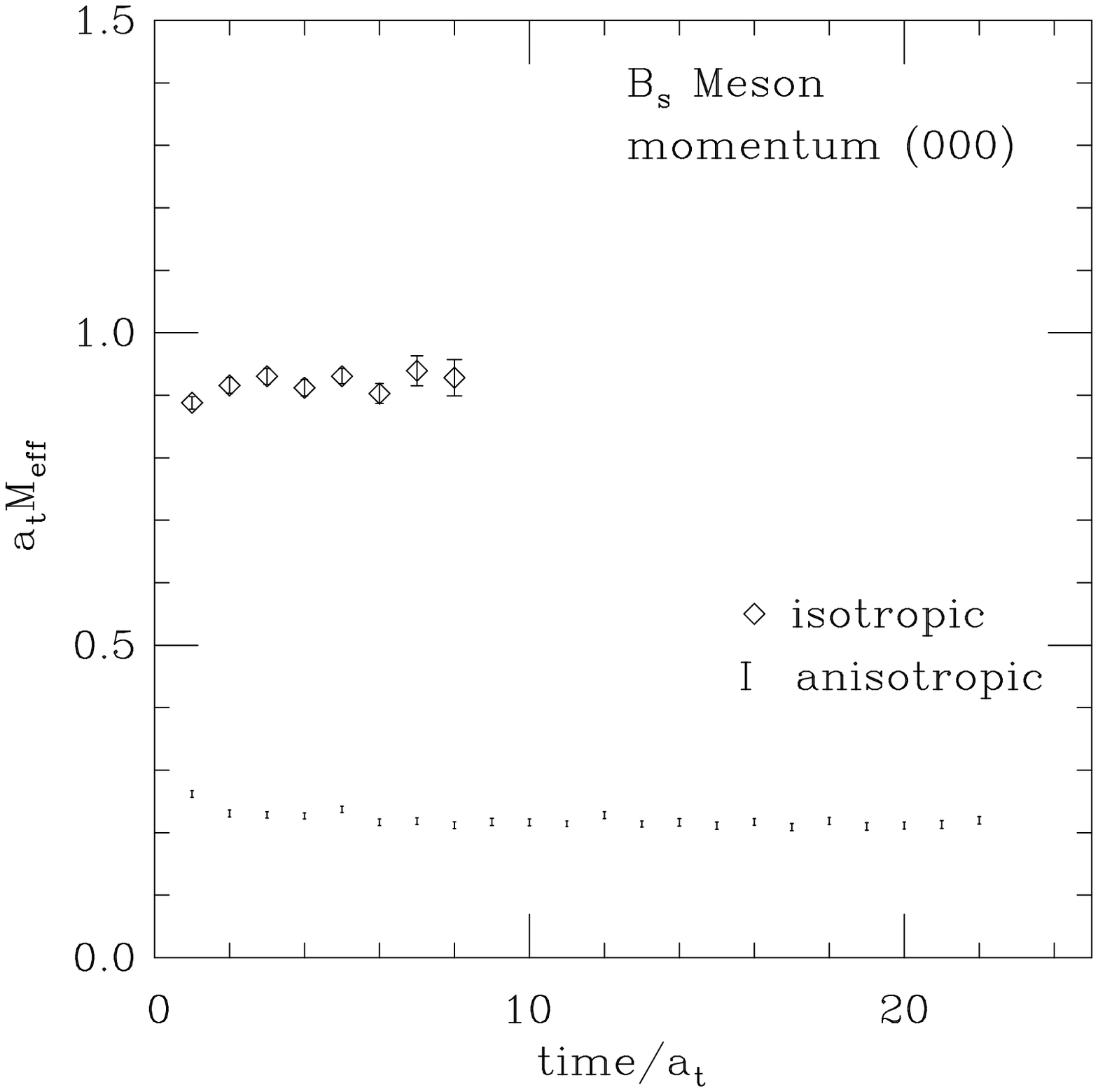}{90mm}
\ewxy{ 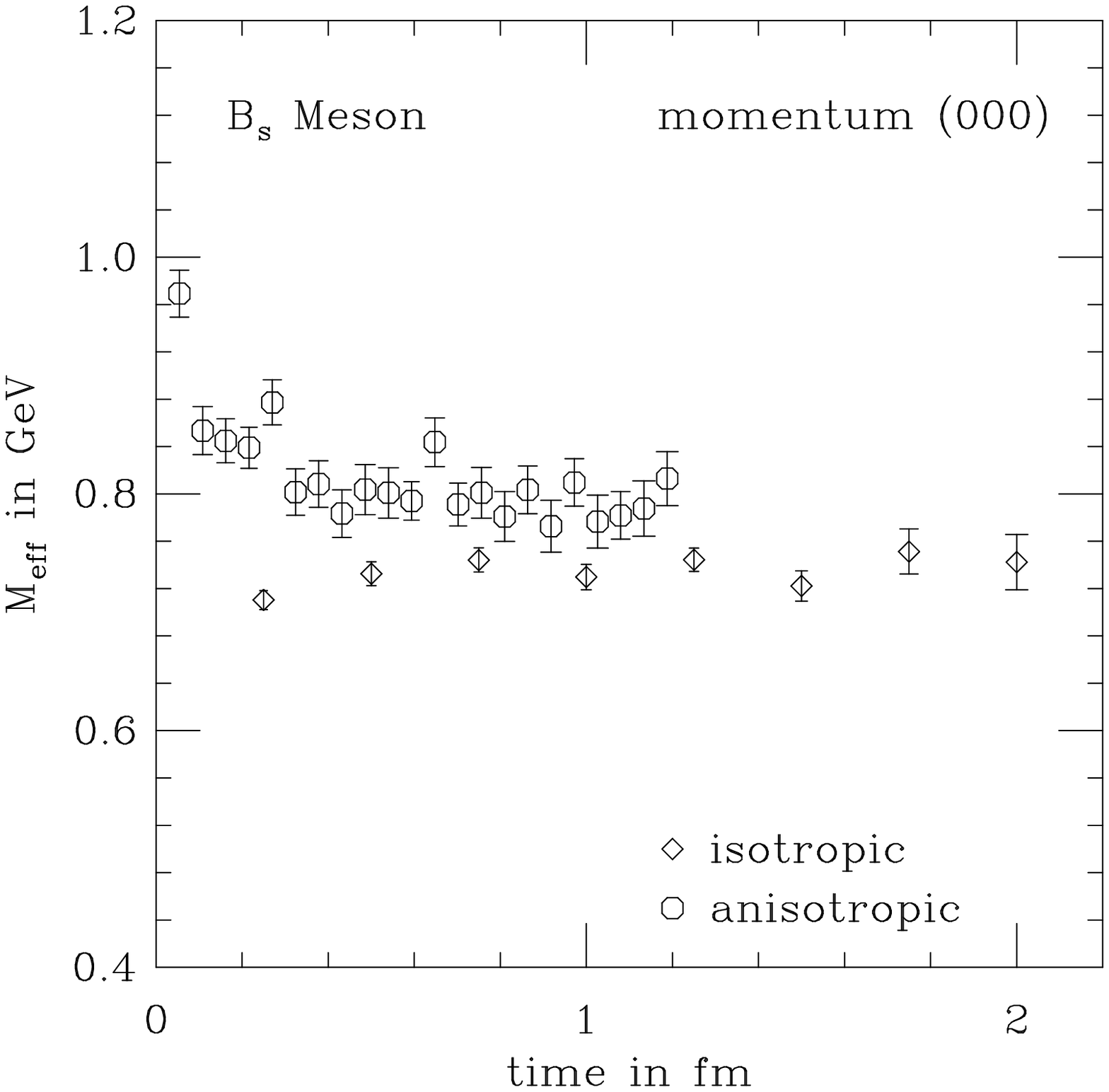}{90mm}
}
\centerline{\hspace{1cm} (i) \hspace{9cm} (ii)}
\caption{ 
Effective masses for the (0,0,0) momentum $B_s$ meson correlators in
 (i) lattice and (ii) physical units.
}
\end{figure}

%\newpage
\begin{figure}
\centerline{
\ewxy{ 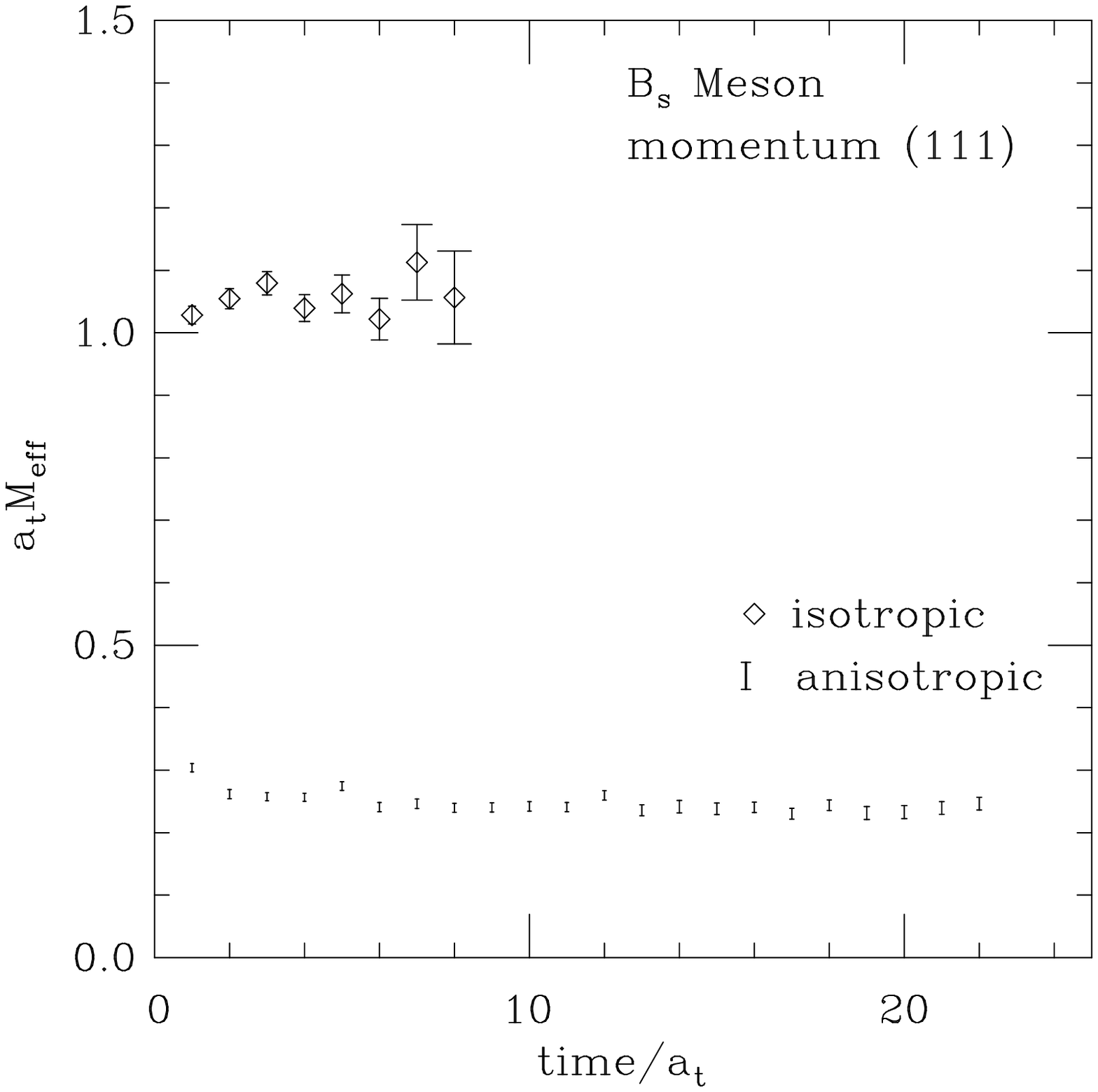}{90mm}
\ewxy{ 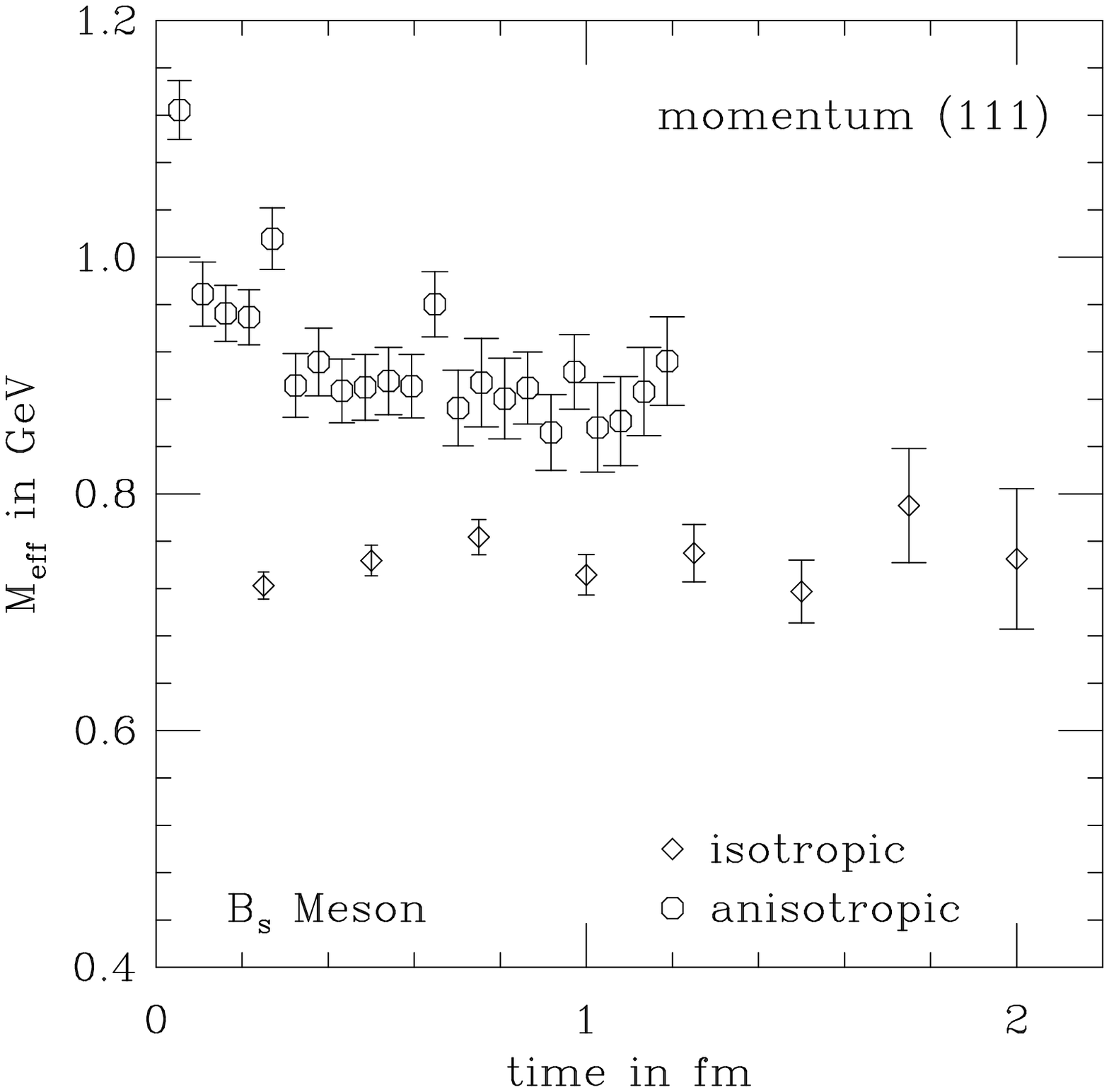}{90mm}
}
\centerline{\hspace{1cm} (i) \hspace{9cm} (ii)}
\caption{ 
Effective masses for the (1,1,1) momentum $B_s$ meson correlators in
 (i) lattice and (ii) physical units.
The isotropic points in (ii) have been shifted down for clarity.
}
\end{figure}

%\newpage
\begin{figure}
\centerline{
\ewxy{ 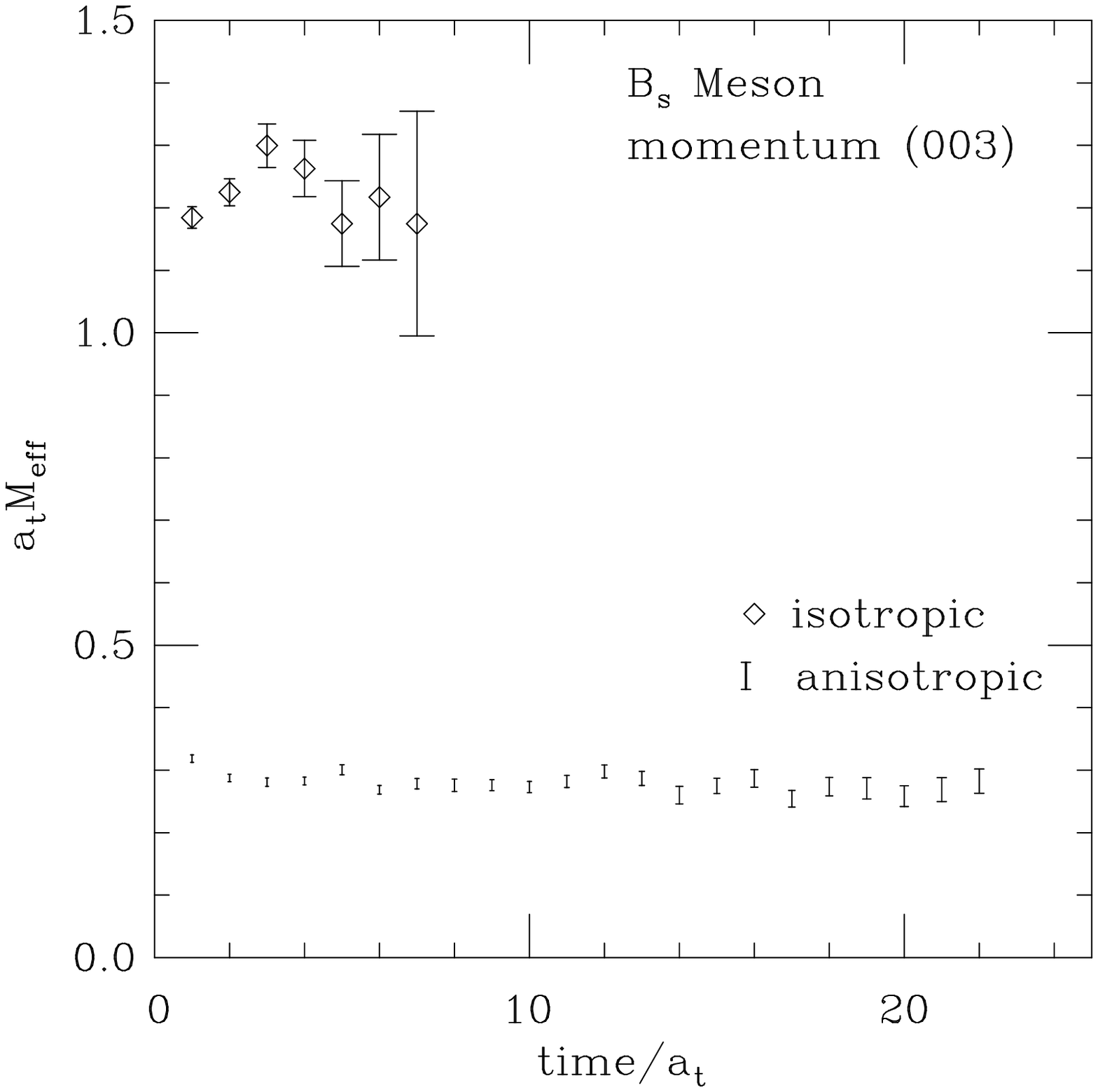}{90mm}
\ewxy{ 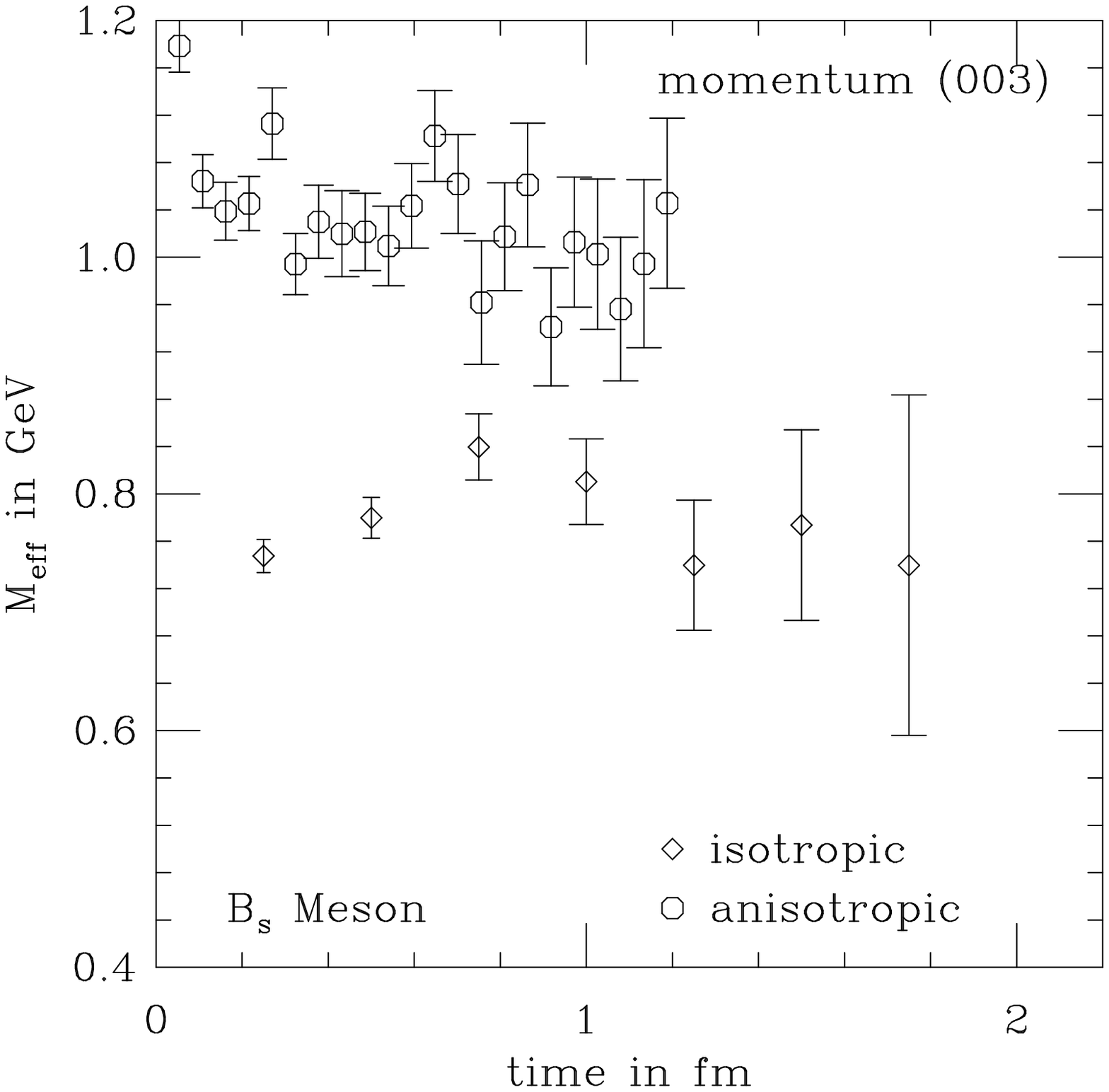}{90mm}
}
\centerline{\hspace{1cm} (i) \hspace{9cm} (ii)}
\caption{ 
Effective masses for the (0,0,3) momentum $B_s$ meson correlators in
 (i) lattice and (ii) physical units.
The isotropic points in (ii) have been shifted down for clarity.
}
\end{figure}

\newpage
\begin{figure}
\centerline{
\ewxy{ 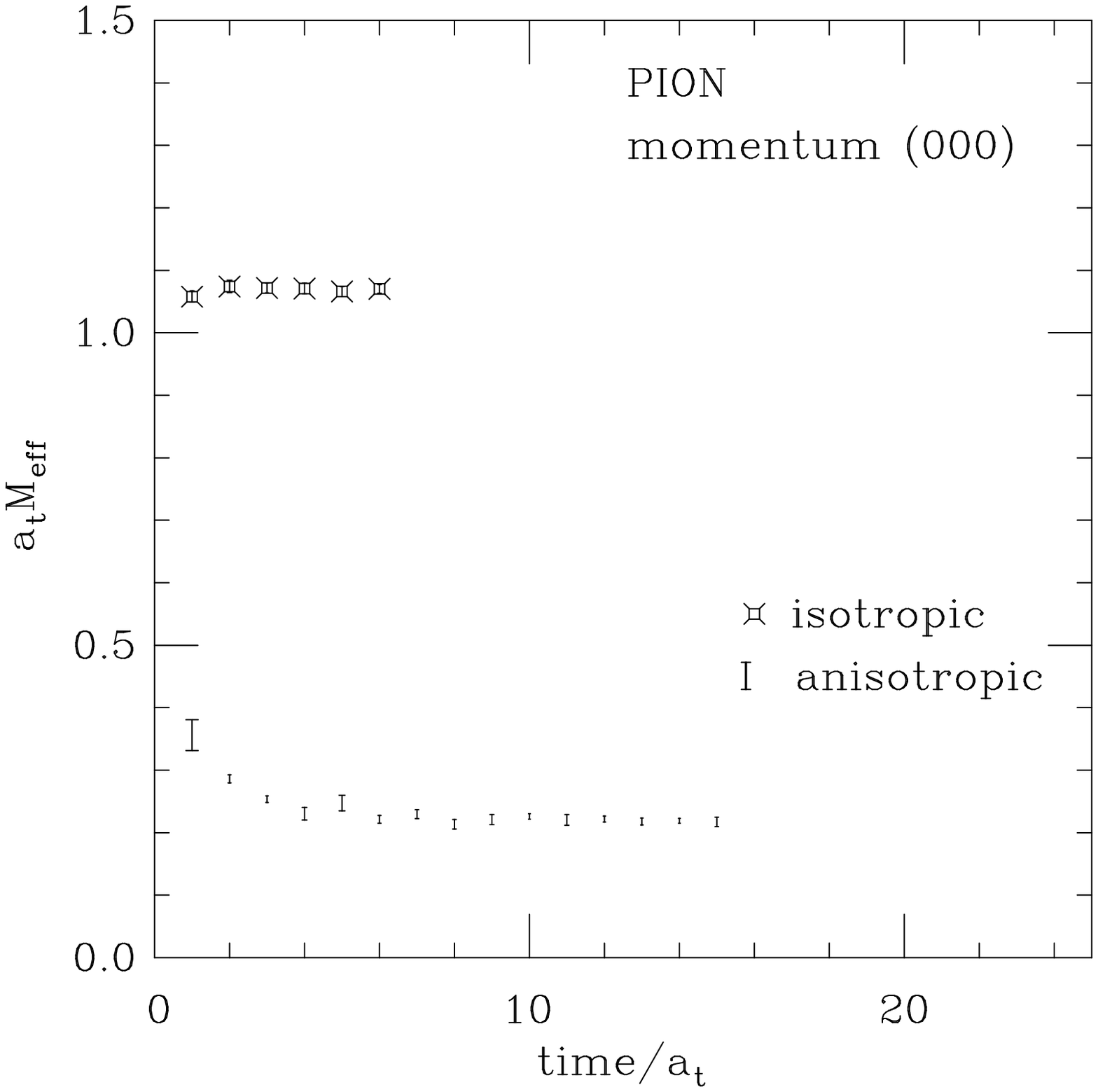}{90mm}
\ewxy{ 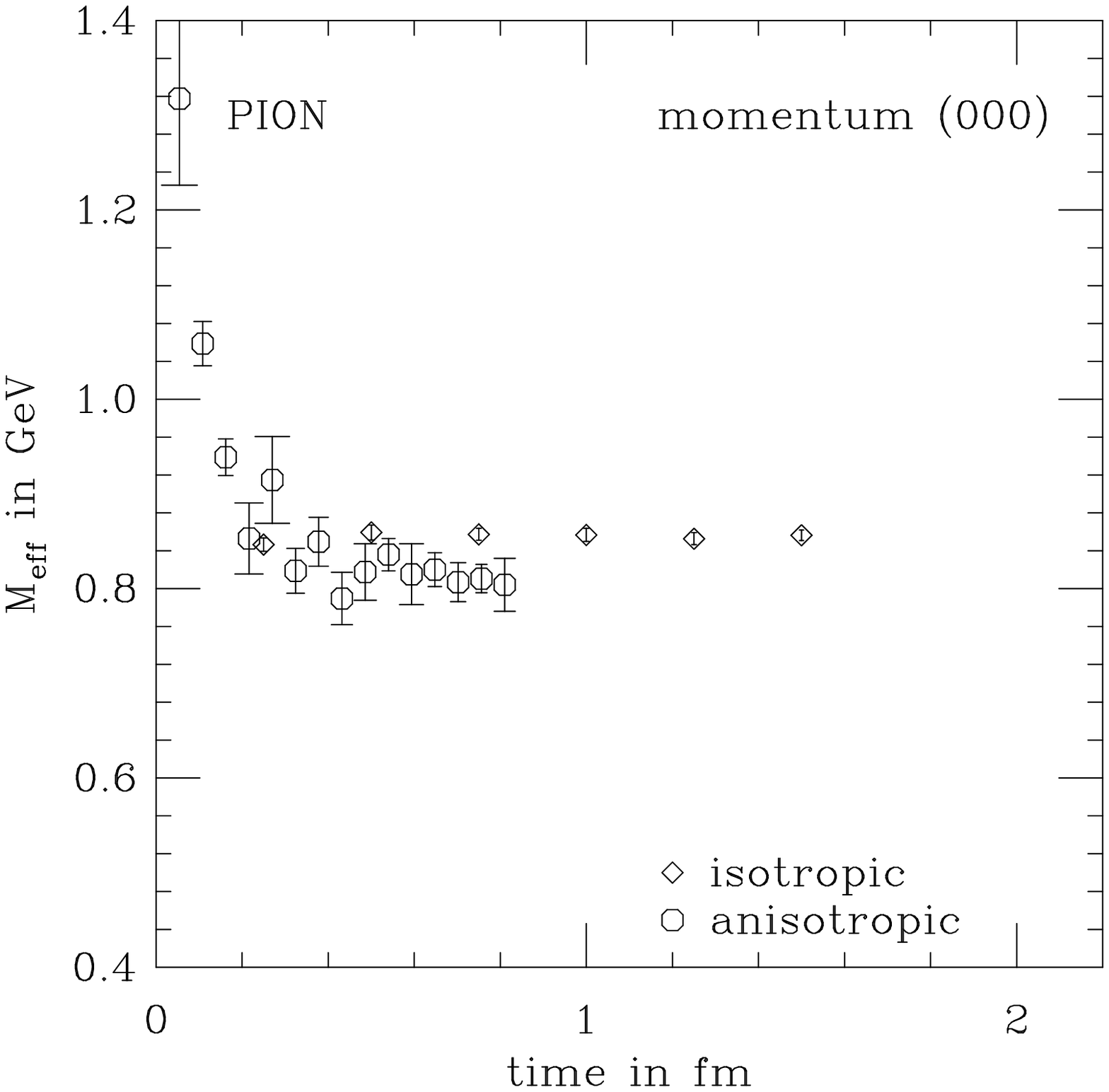}{90mm}
}
\centerline{\hspace{1cm} (i) \hspace{9cm} (ii)}
\caption{ 
Effective masses for the (0,0,0) momentum pion correlators in
 (i) lattice and (ii) physical units.
}
\end{figure}

%\newpage
\begin{figure}
\centerline{
\ewxy{ 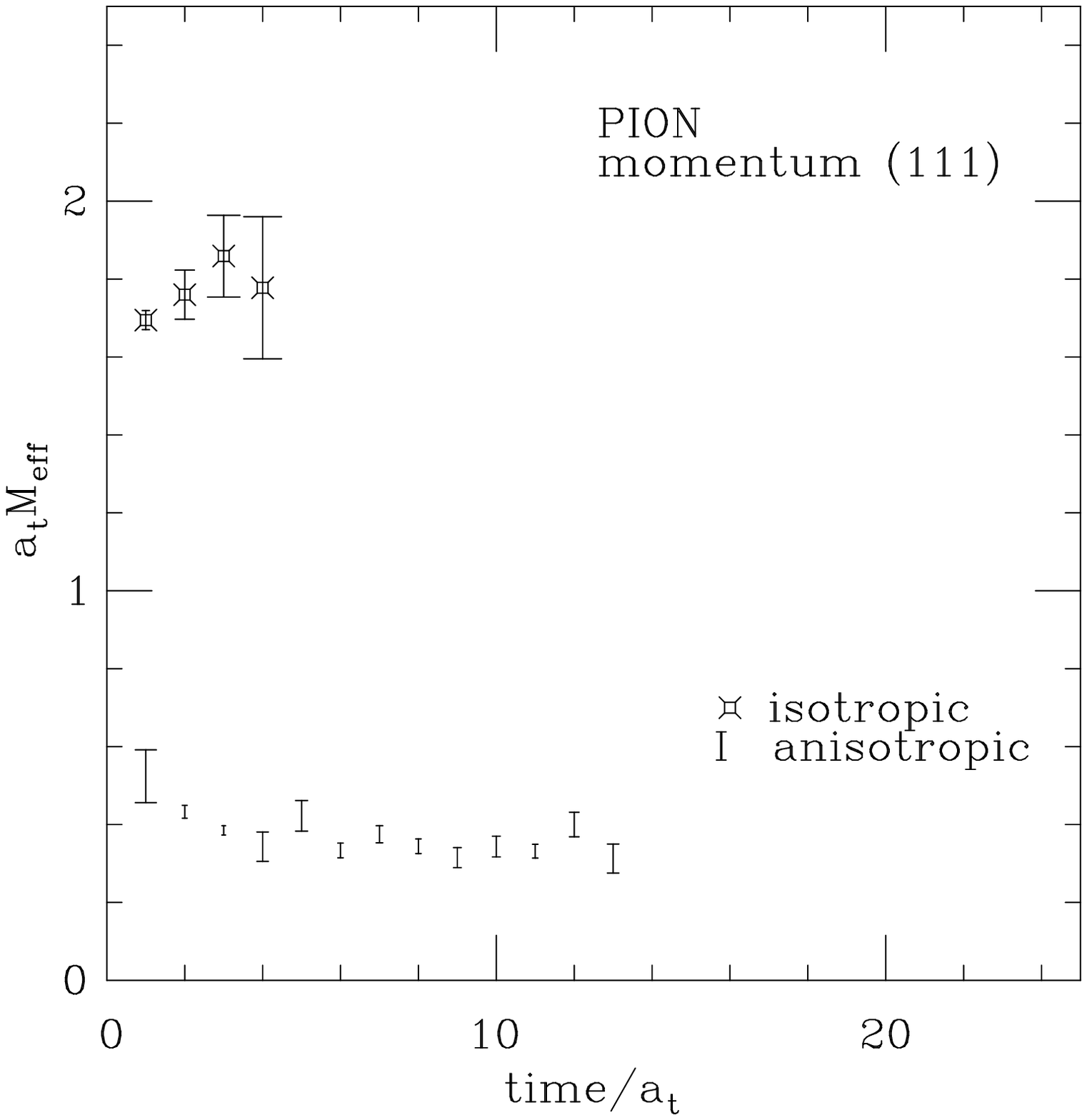}{90mm}
\ewxy{ 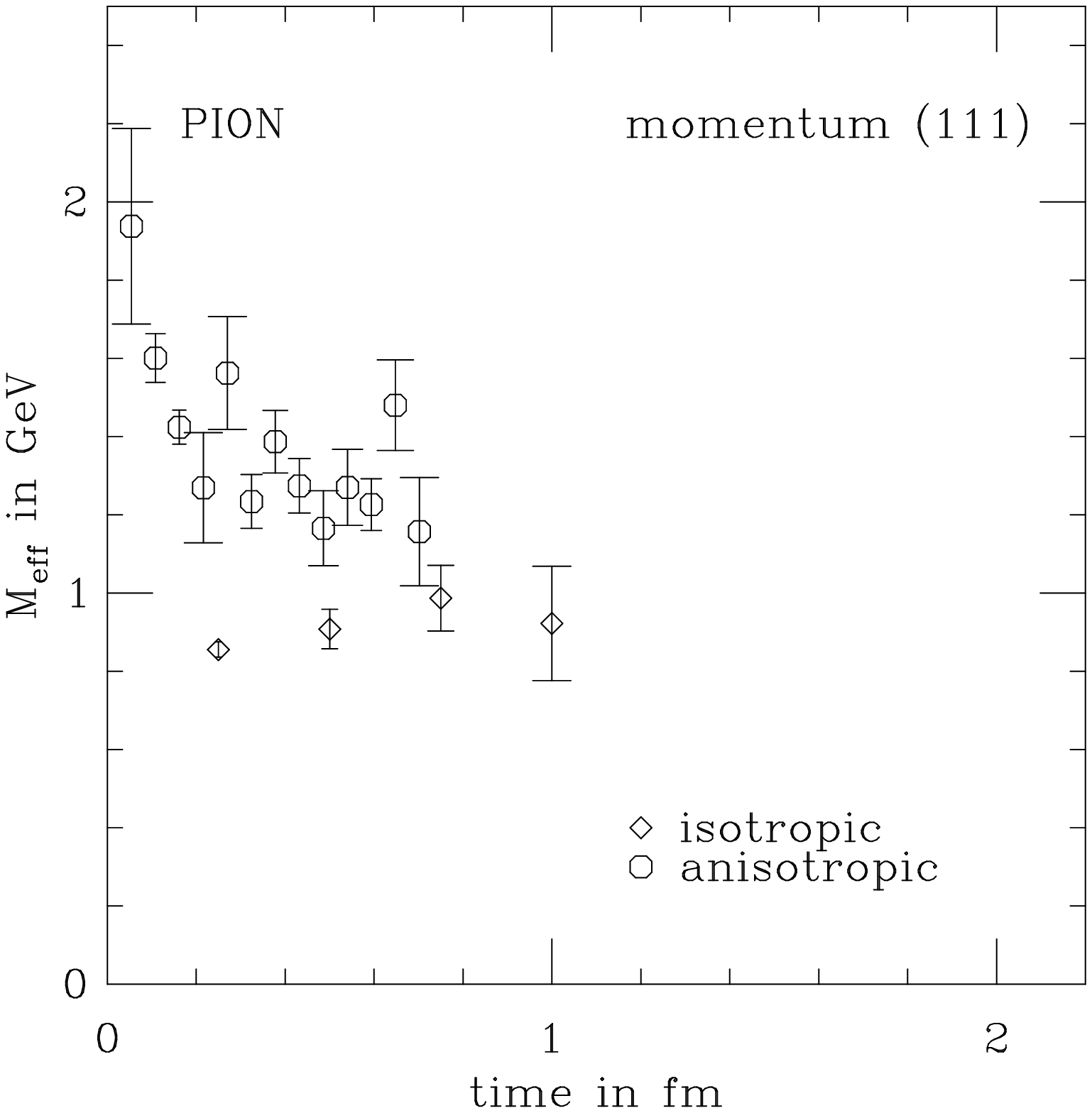}{90mm}
}
\centerline{\hspace{1cm} (i) \hspace{9cm} (ii)}
\caption{ 
Effective masses for the (1,1,1) momentum pion correlators in
 (i) lattice and (ii) physical units.
The isotropic points in (ii) have been shifted down for clarity.
}
\end{figure}

\newpage
\begin{figure}
\centerline{
\ewxy{ 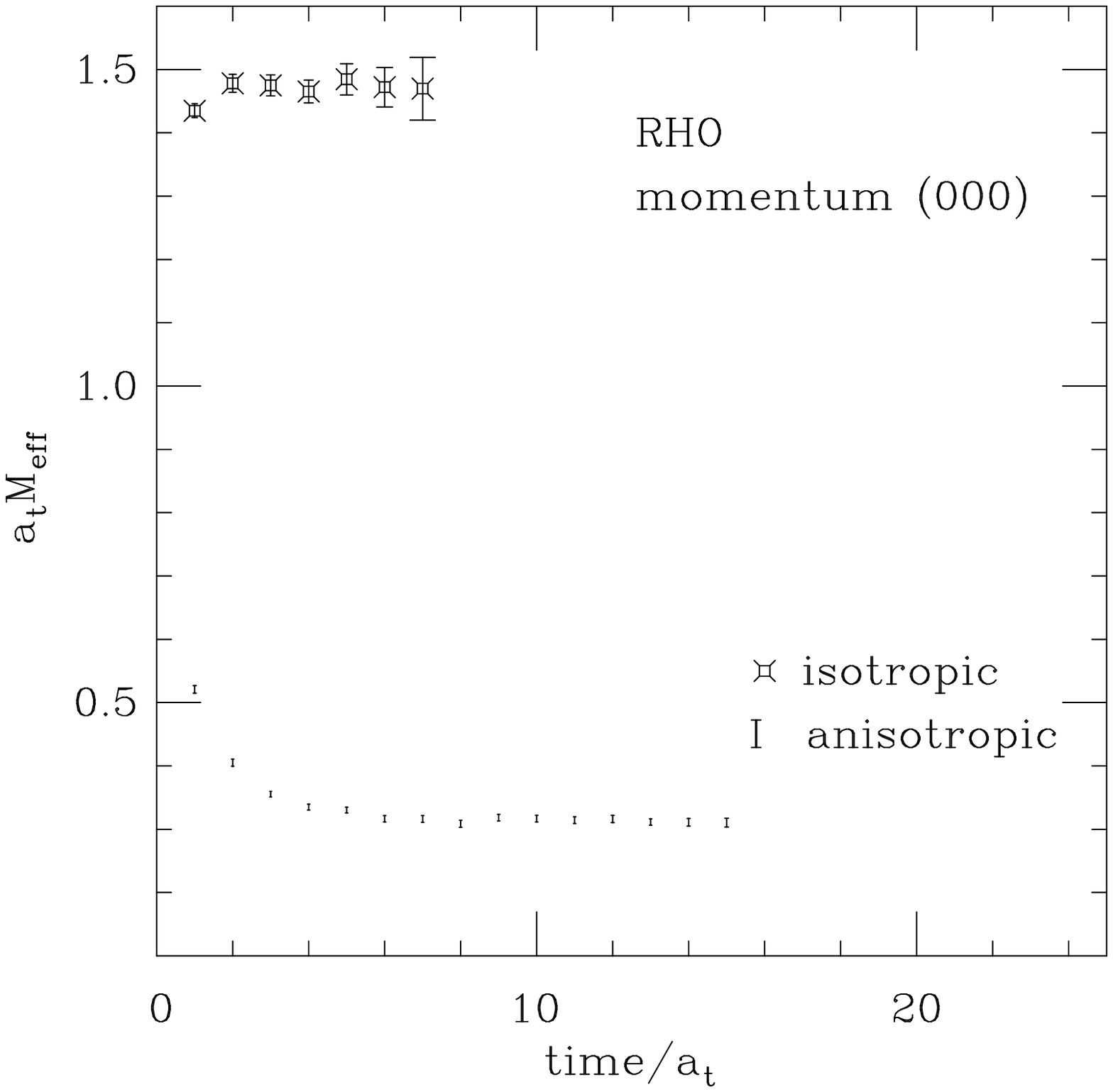}{90mm}
\ewxy{ 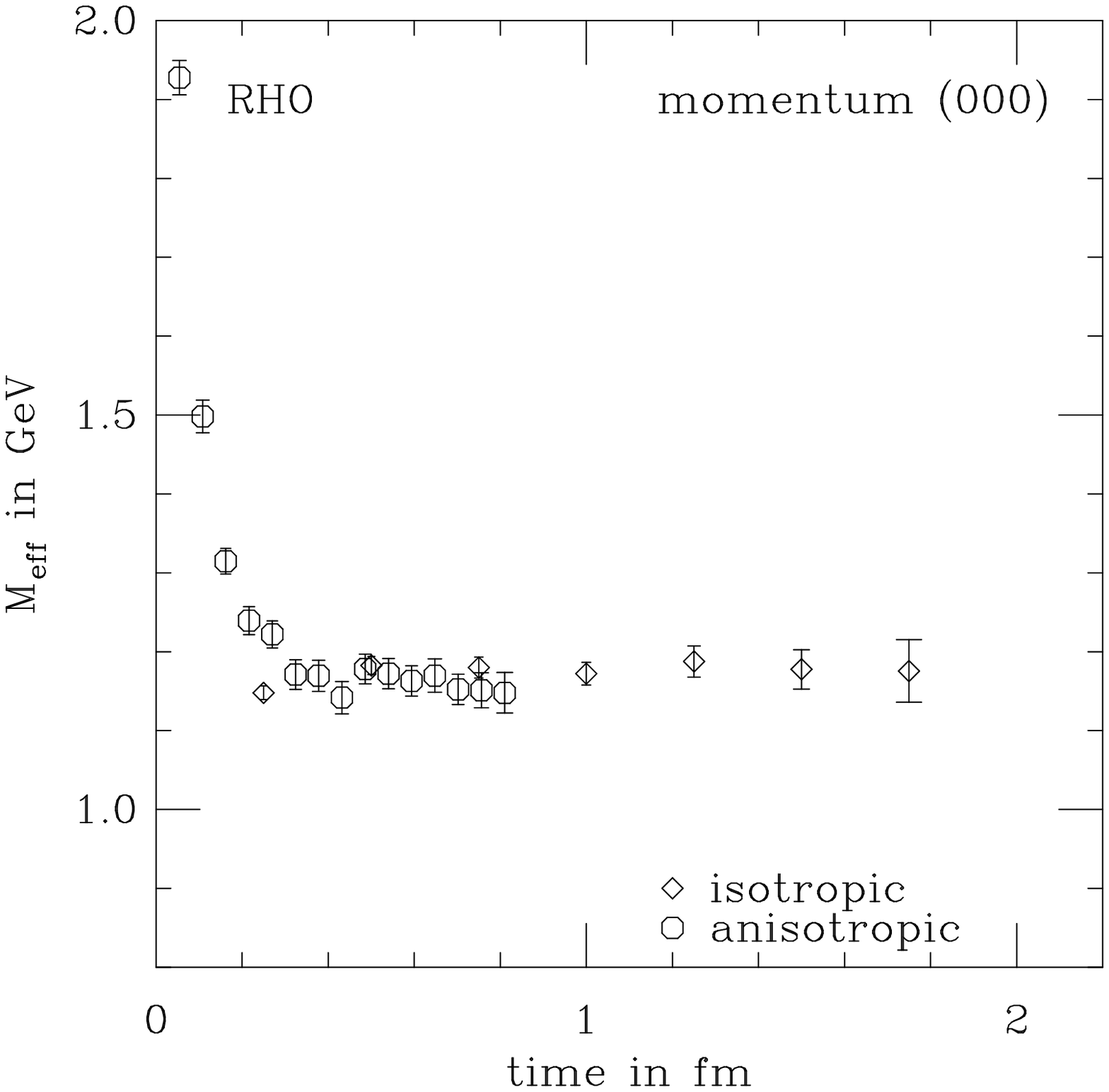}{90mm}
}
\centerline{\hspace{1cm} (i) \hspace{9cm} (ii)}
\caption{ 
Effective masses for the (0,0,0) momentum rho correlators in
 (i) lattice and (ii) physical units.
}
\end{figure}

%\newpage
\begin{figure}
\centerline{
\ewxy{ 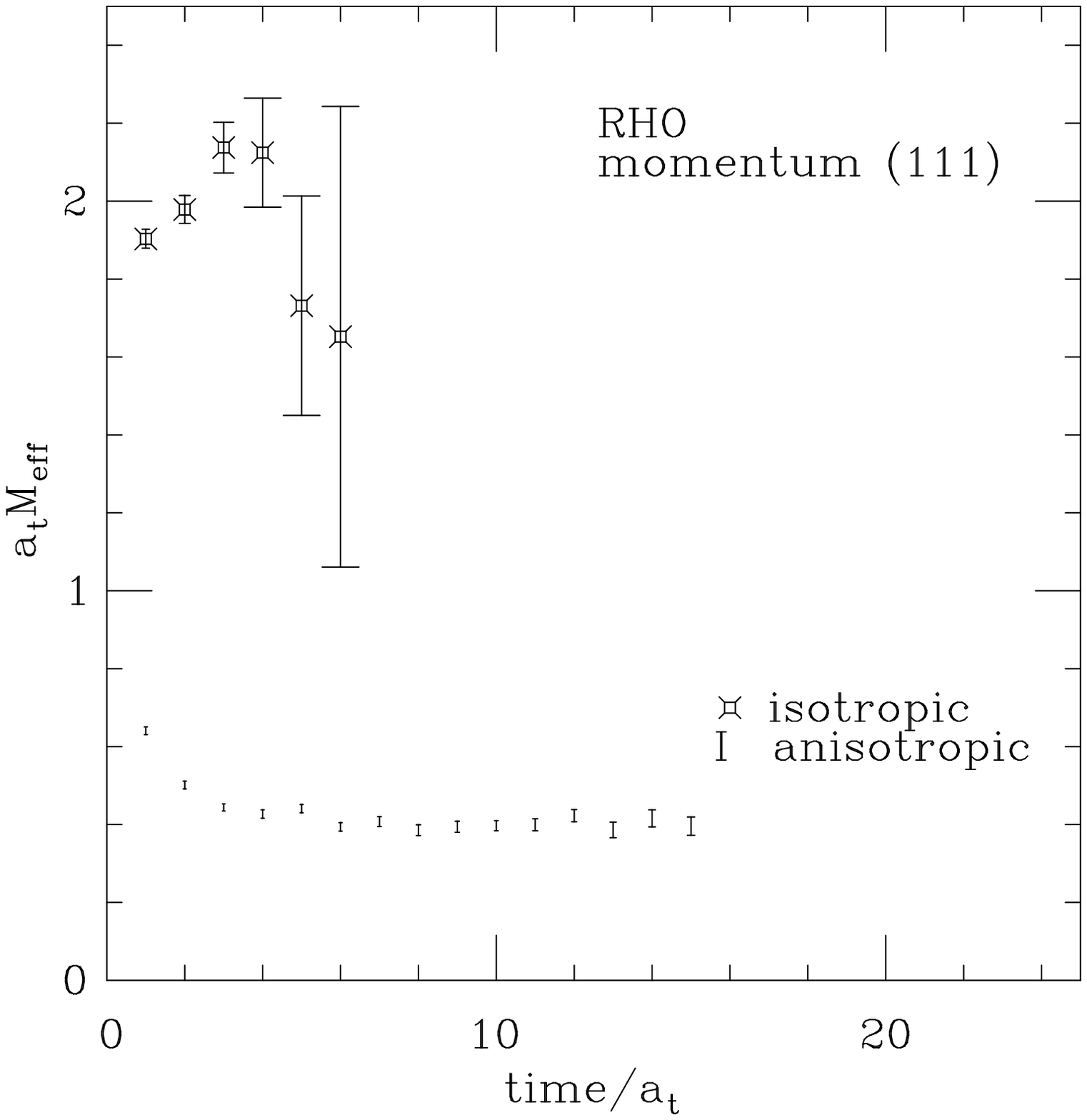}{90mm}
\ewxy{ 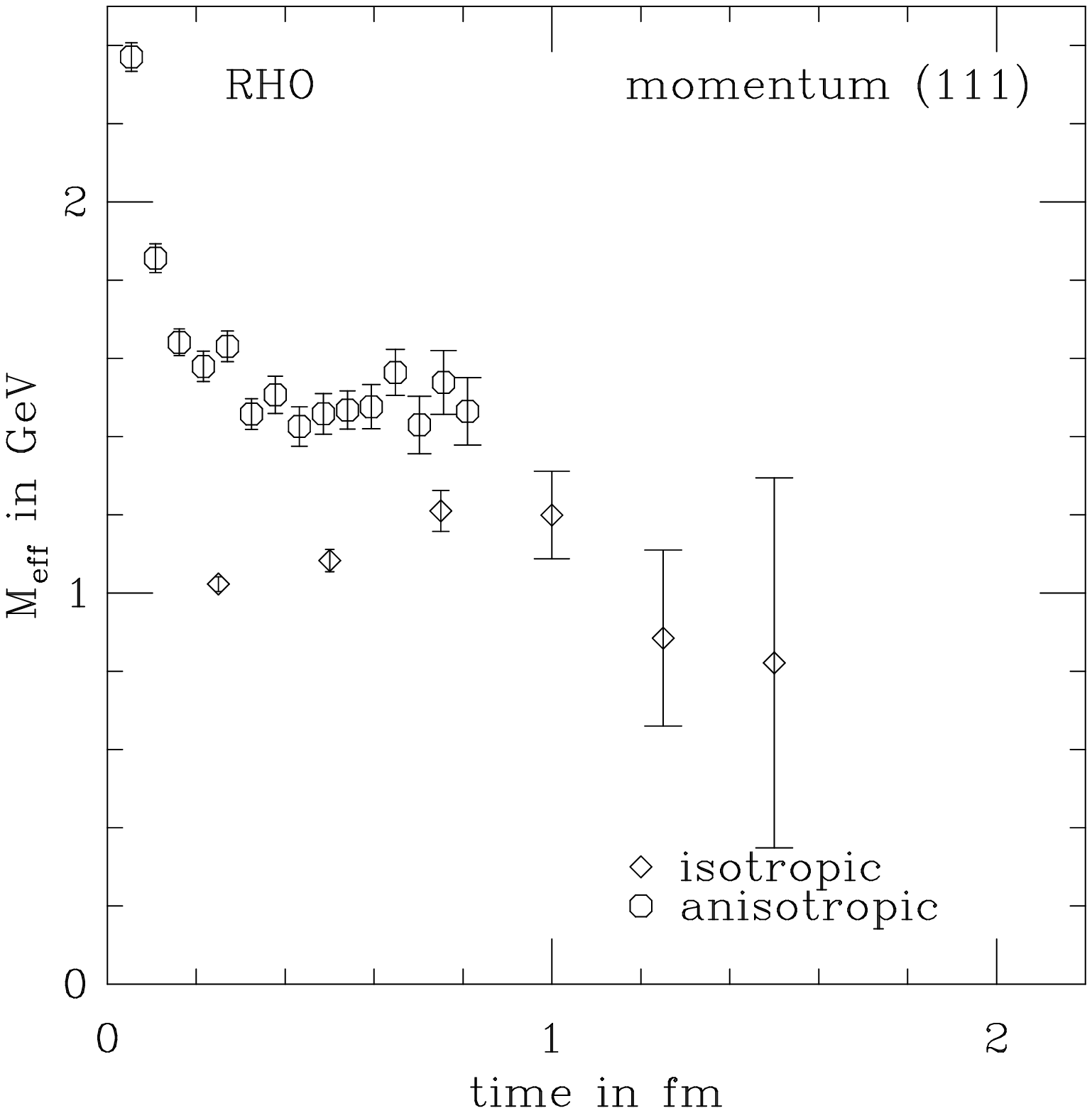}{90mm}
}
\centerline{\hspace{1cm} (i) \hspace{9cm} (ii)}
\caption{ 
Effective masses for the (1,1,1) momentum rho correlators in
 (i) lattice and (ii) physical units.
The isotropic points in (ii) have been shifted down for clarity.
}
\end{figure}

\newpage
\begin{figure}
\centerline{
\ewxy{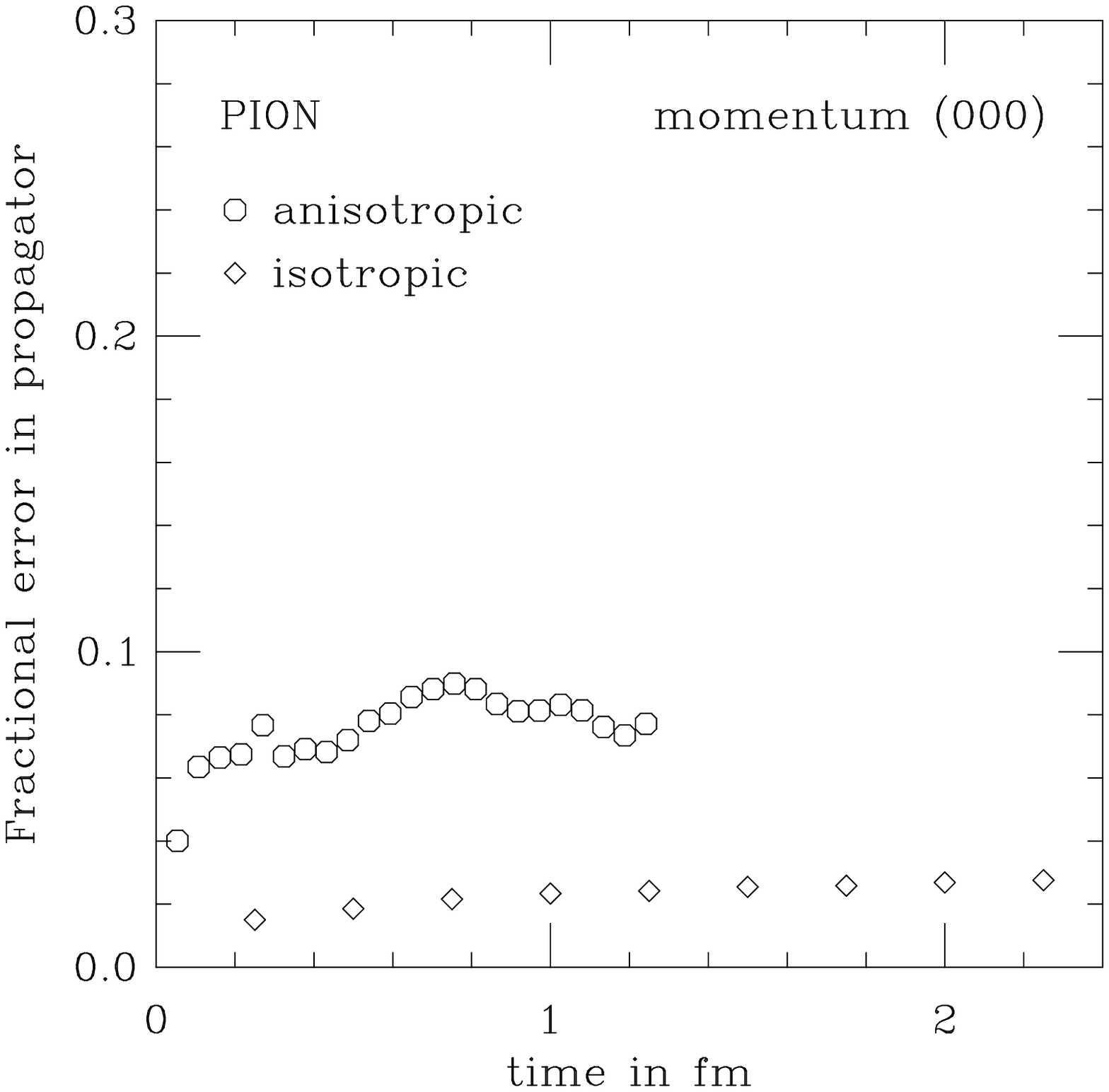}{90mm}
\ewxy{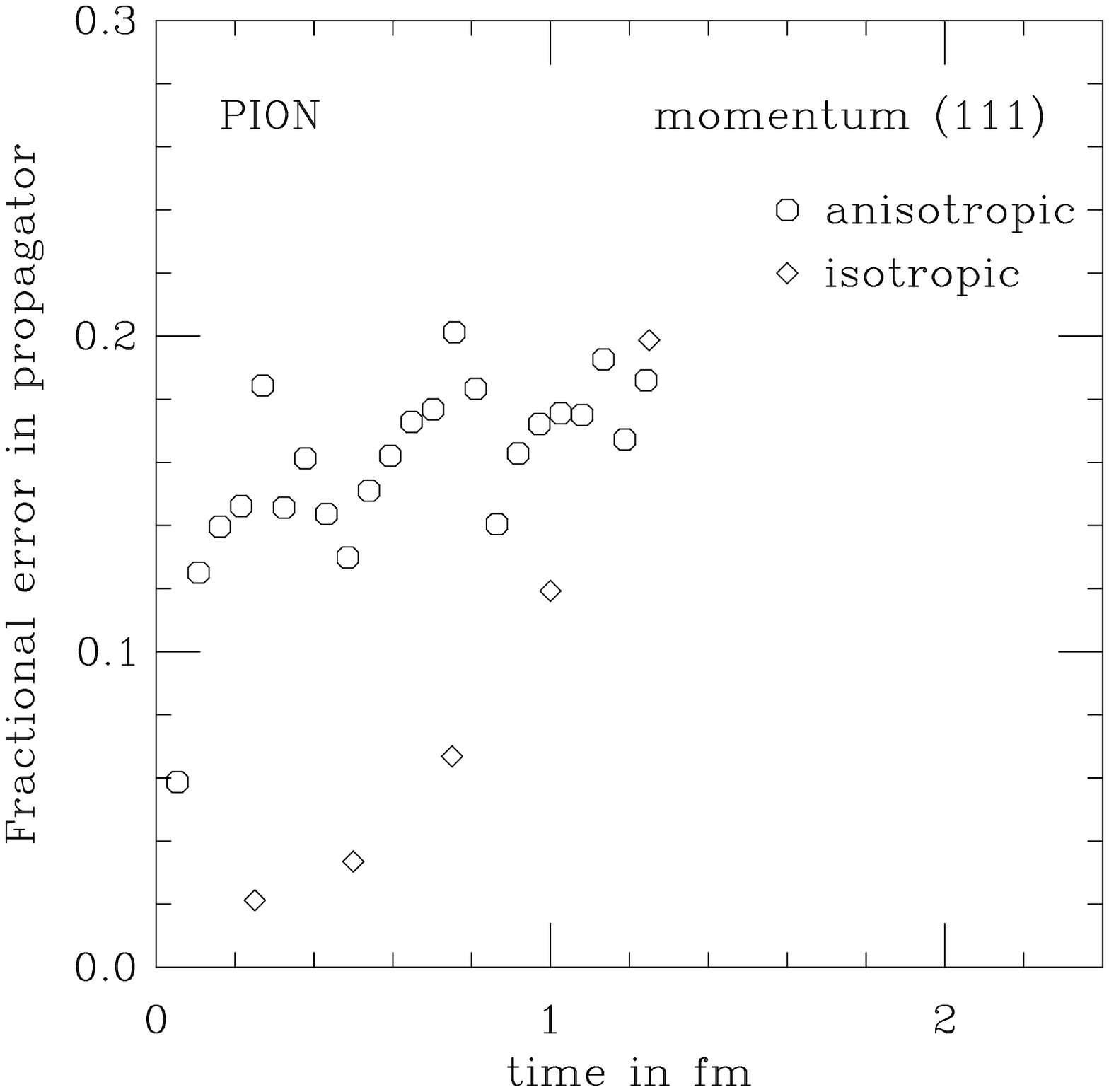}{90mm}
}
%\centerline{(i) \hspace{10cm} (ii)}
\caption{ 
Fractional errors in pion correlators for momentum (0,0,0) and 
(1,1,1).
}
\end{figure}

%\newpage
\begin{figure}
\centerline{
\ewxy{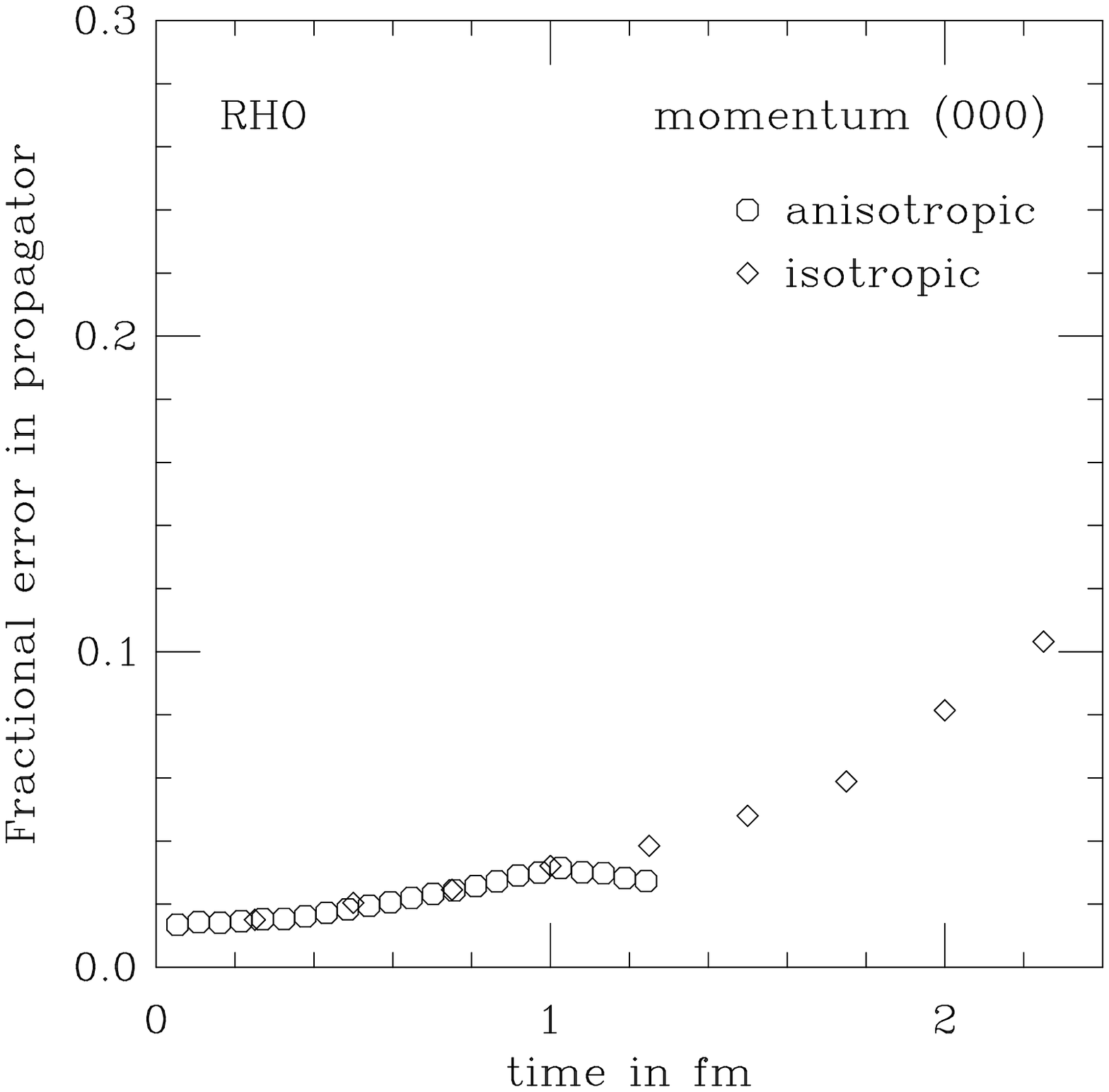}{90mm}
\ewxy{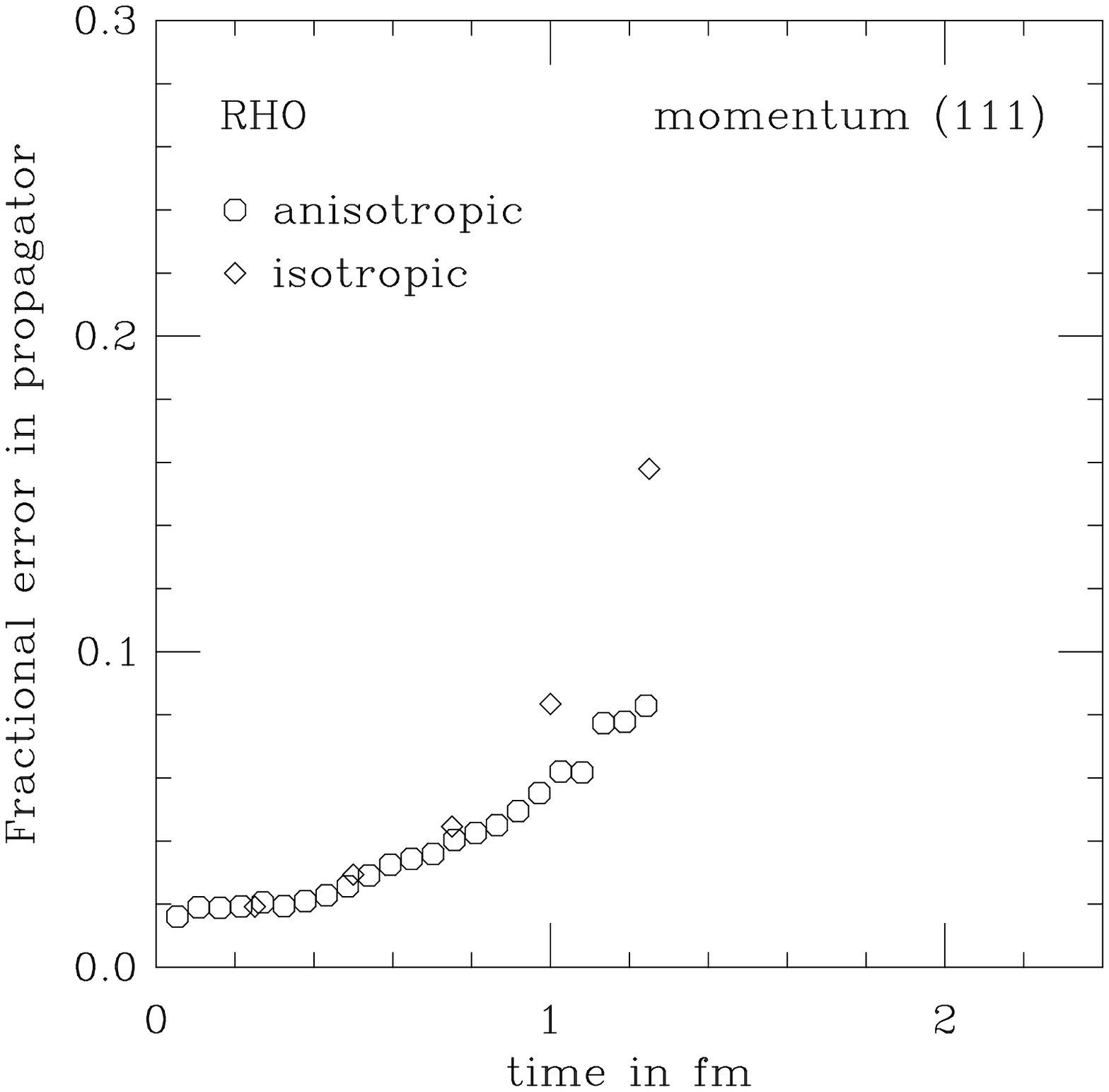}{90mm}
}
%\centerline{(i) \hspace{10cm} (ii)}
\caption{ 
Fractional errors in rho correlators for momentum (0,0,0) and 
(1,1,1).
}
\end{figure}

\newpage
\begin{figure}
\begin{center}
\epsfysize=7.in
\centerline{\epsfbox{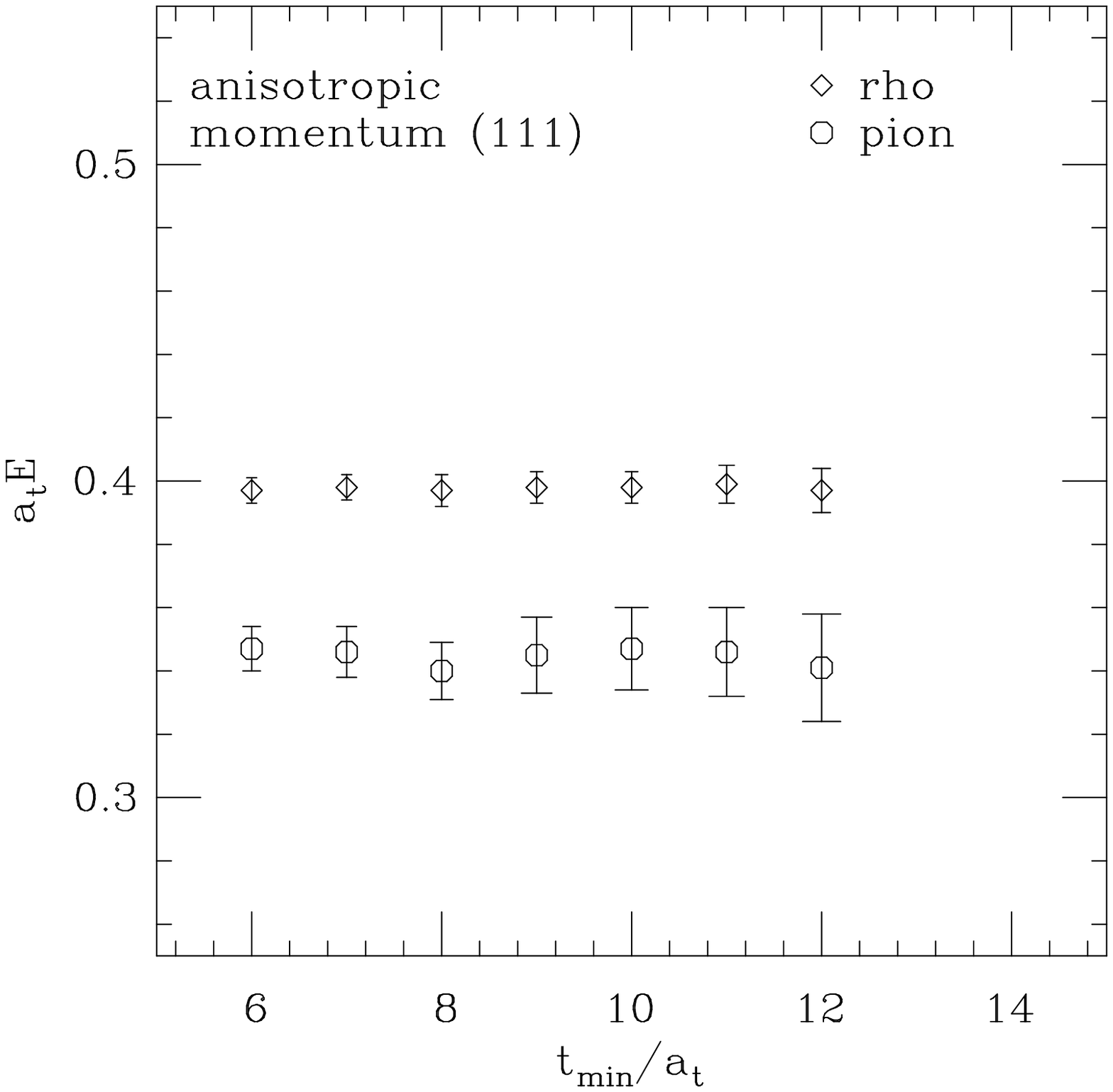}}
\end{center}
\caption{
Fitted energies versus $t_{min}/a_t$ for the pion and the rho for 
momentum (1,1,1) on anisotropic lattices.  Single cosh fits were used with 
$t_{max}/a_t$ fixed at 22.
  Errors are from a bootstrap over 200 ensembles.
  }
\end{figure}

\newpage
\begin{figure}
\begin{center}
\epsfysize=7.in
\centerline{\epsfbox{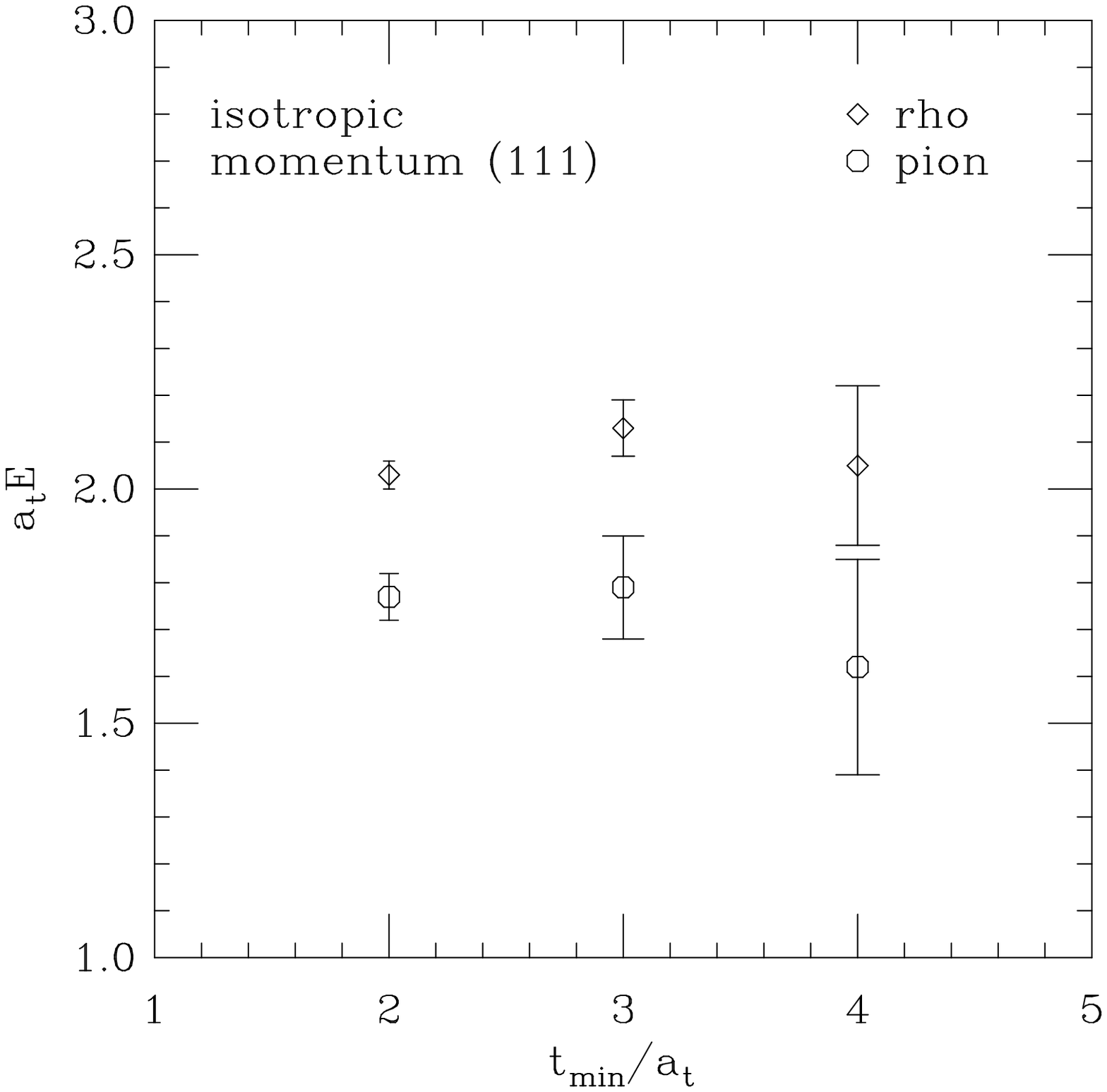}}
\end{center}
\caption{
Fitted energies versus $t_{min}/a_t$ for the pion and the rho for 
momentum (1,1,1) on isotropic lattices.  Single cosh fits were used with 
$t_{max}/a_t$ fixed at 9.
 Errors are from a bootstrap over 200 ensembles.
  }
\end{figure}

\newpage
\begin{figure}
\begin{center}
\epsfysize=7.in
\centerline{\epsfbox{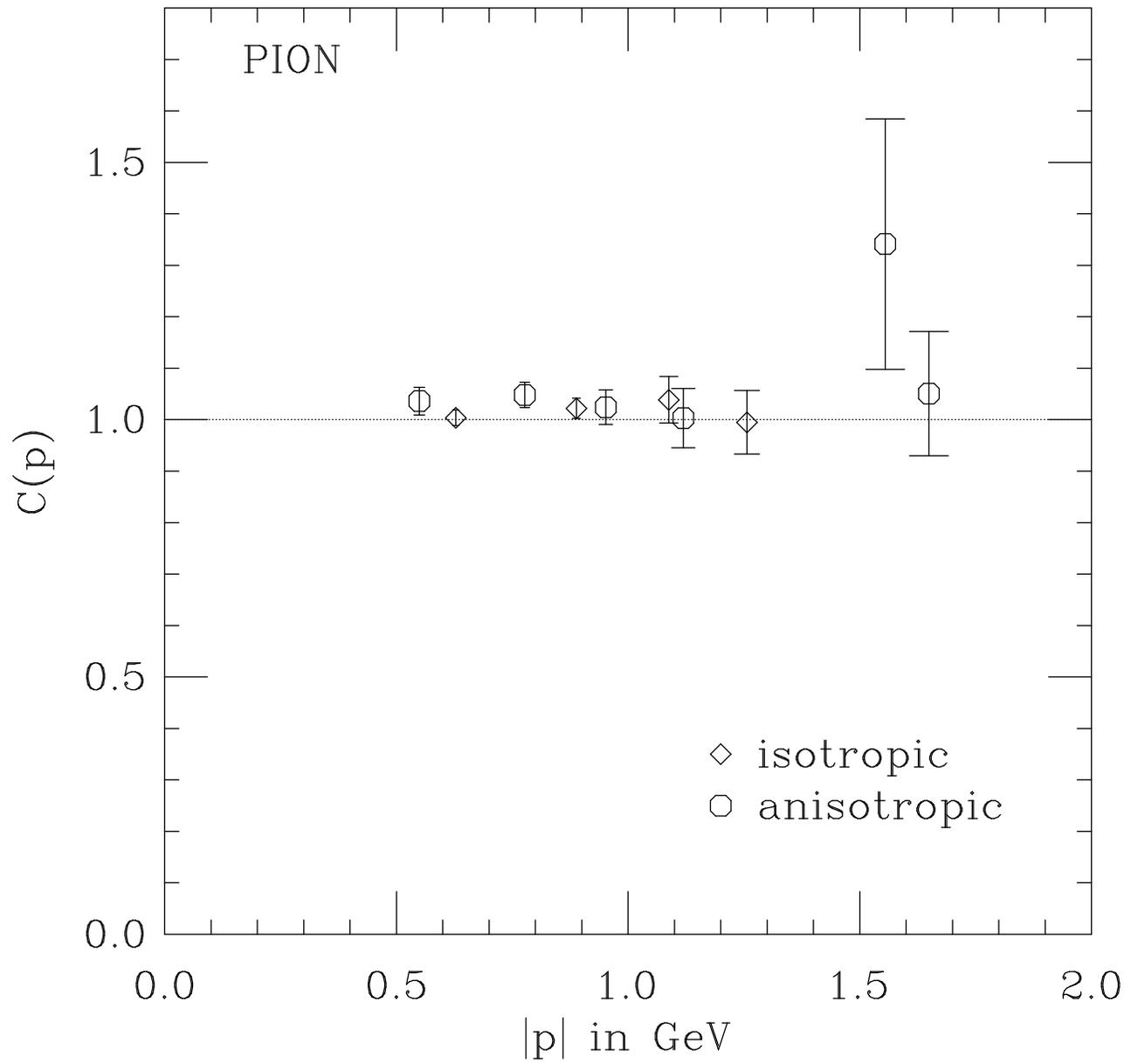}}
\end{center}
\caption{
$C(p)$ versus momentum for the pion.
  }
\end{figure}

\newpage
\begin{figure}
\begin{center}
\epsfysize=7.in
\centerline{\epsfbox{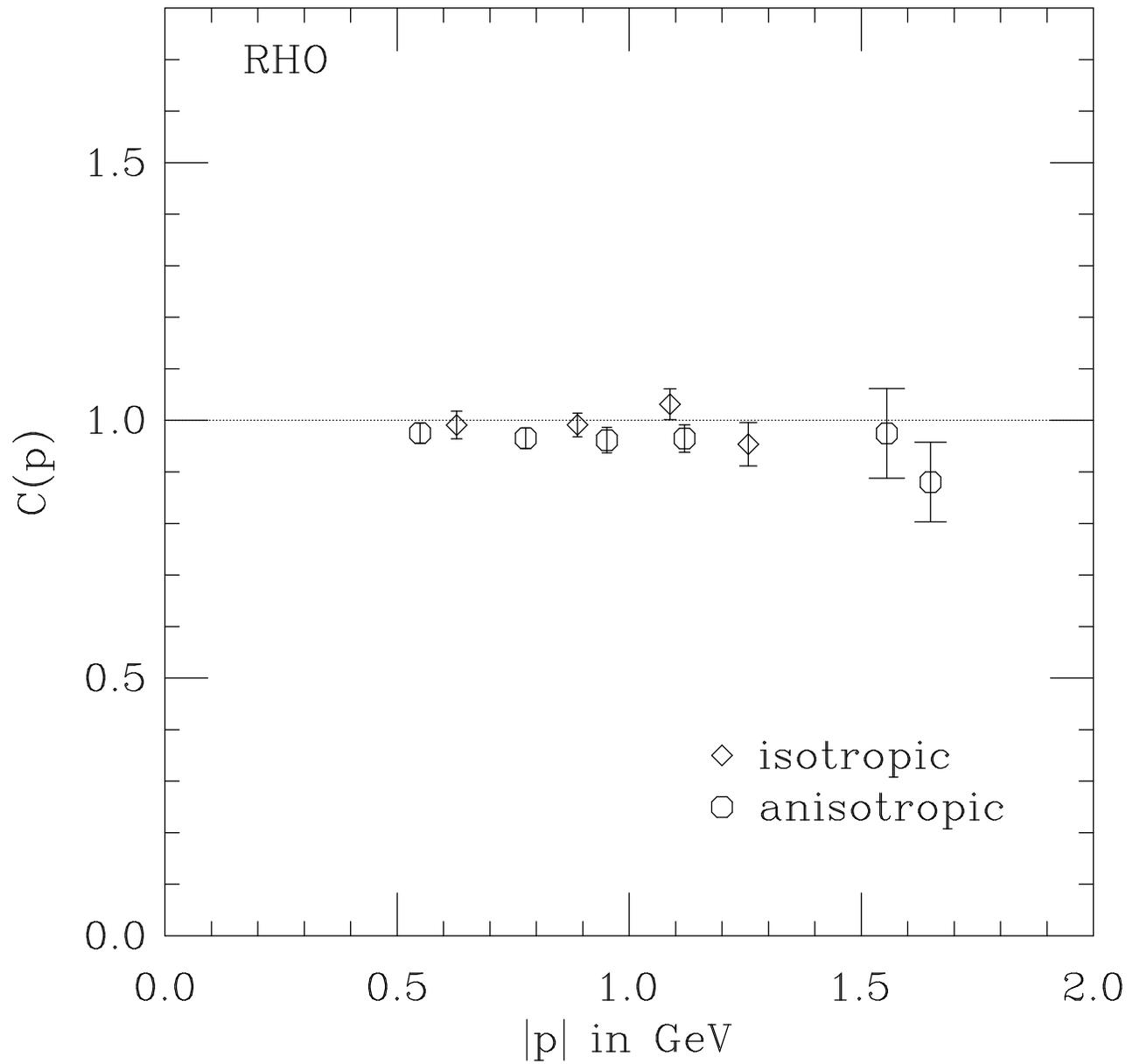}}
\end{center}
\caption{
$C(p)$ versus momentum for the rho.
  }
\end{figure}

\newpage
\begin{figure}
\centerline{
\ewxy{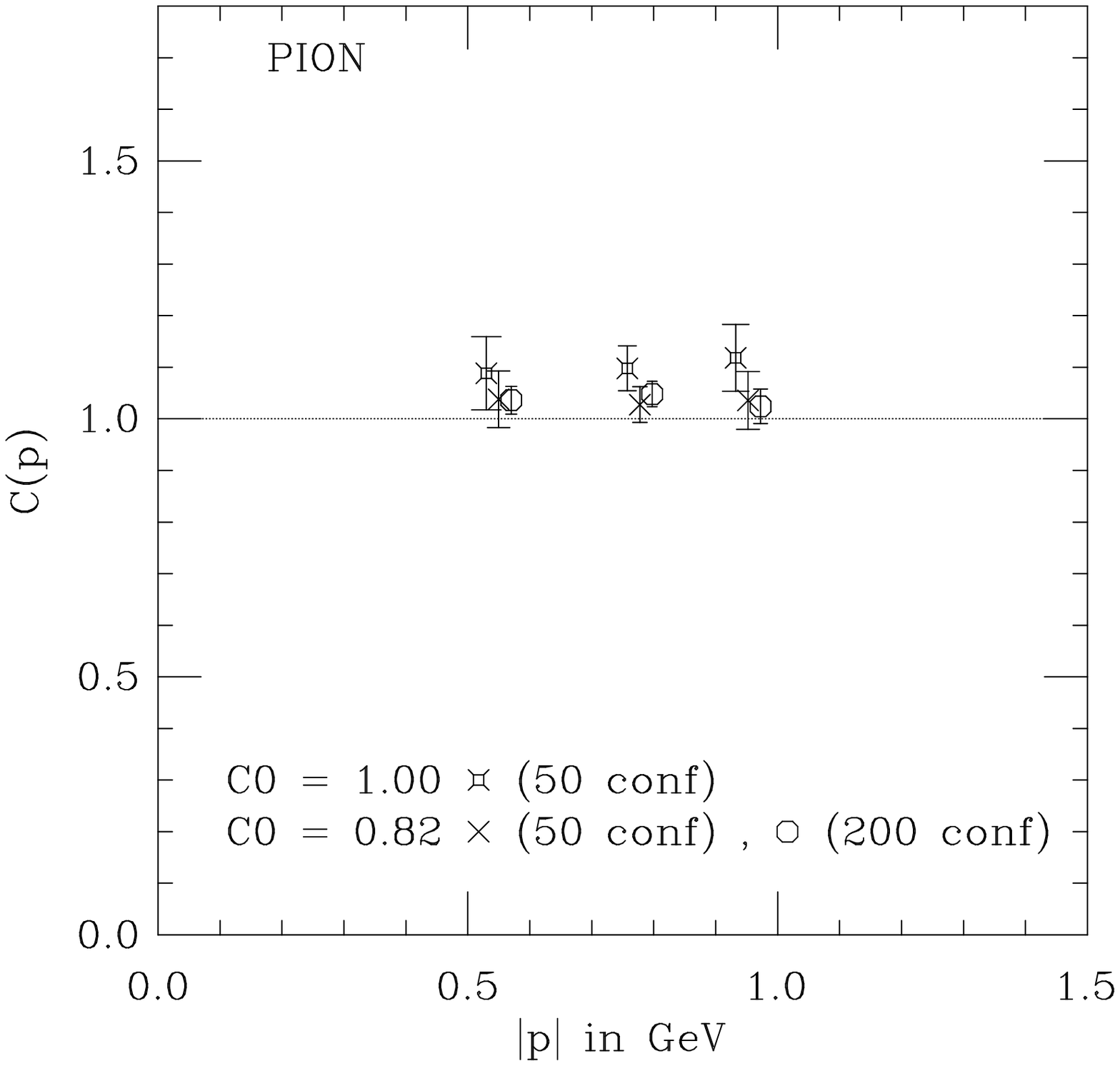}{90mm}
\ewxy{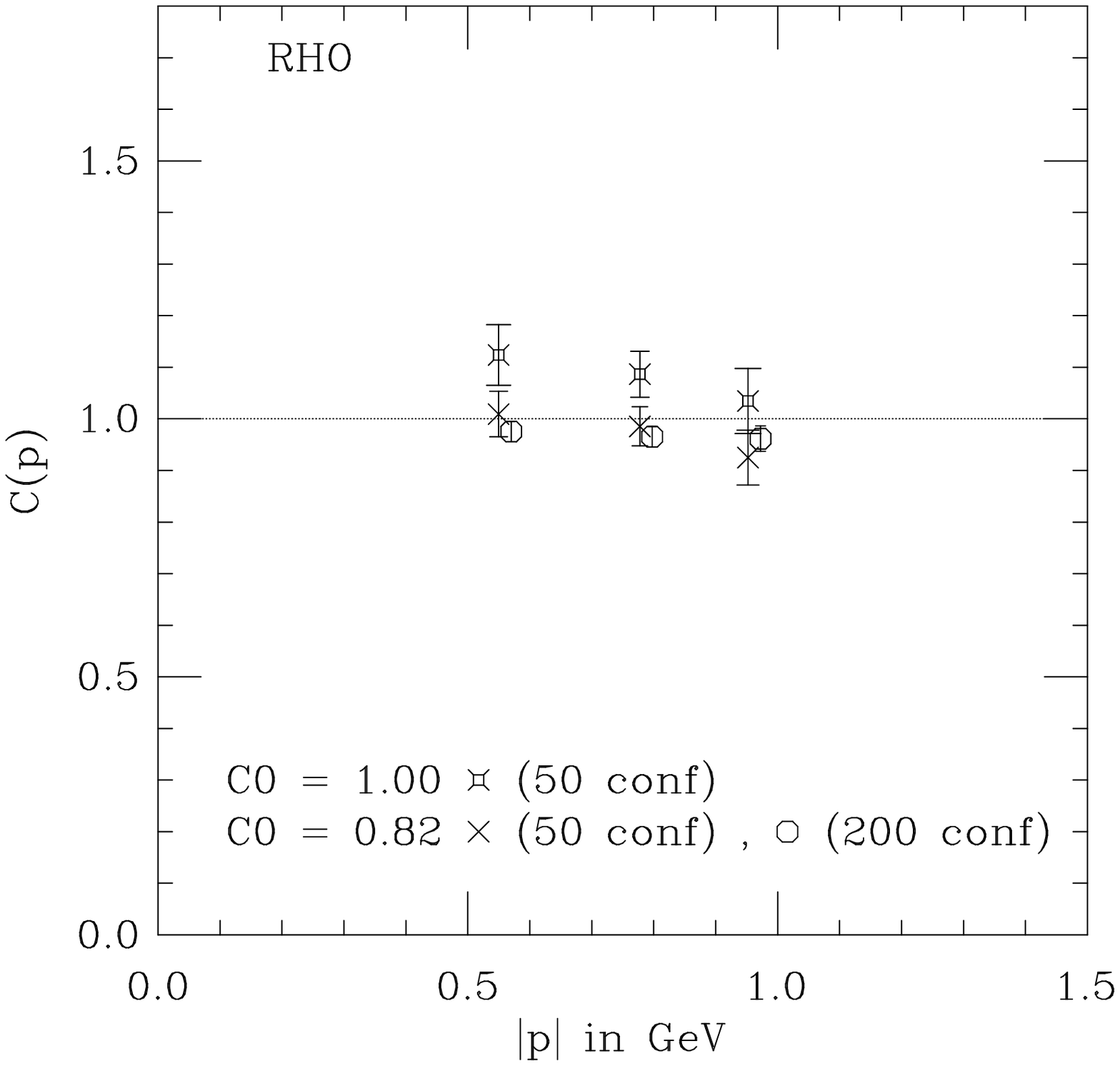}{90mm}
}
%\centerline{(i) \hspace{10cm} (ii)}
\caption{ 
Effect of the one-loop correction to $C_0$ on $C(p)$ for  anisotropic 
lattices.
}
\end{figure}

%\newpage
\begin{figure}
\centerline{
\ewxy{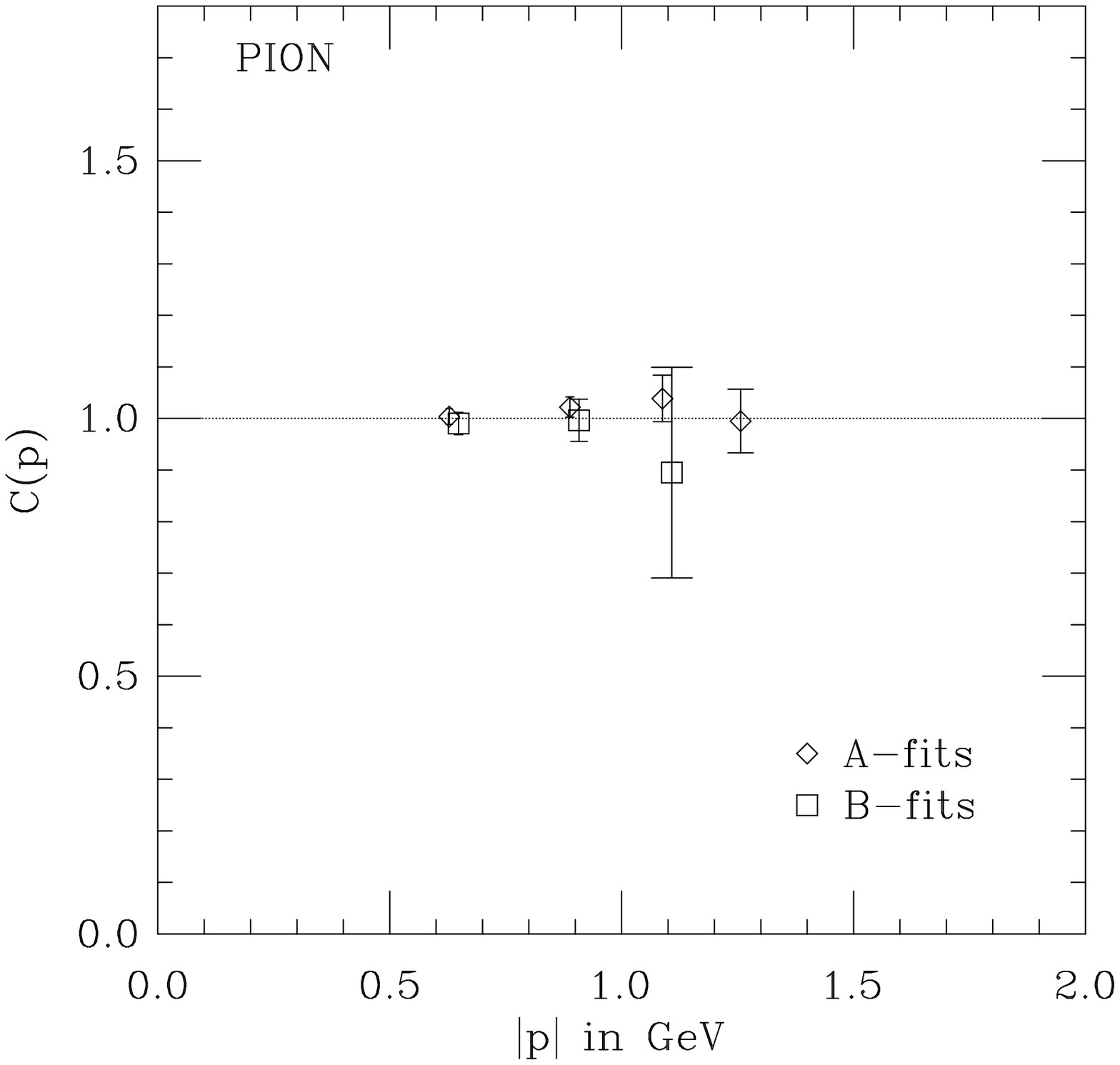}{90mm}
\ewxy{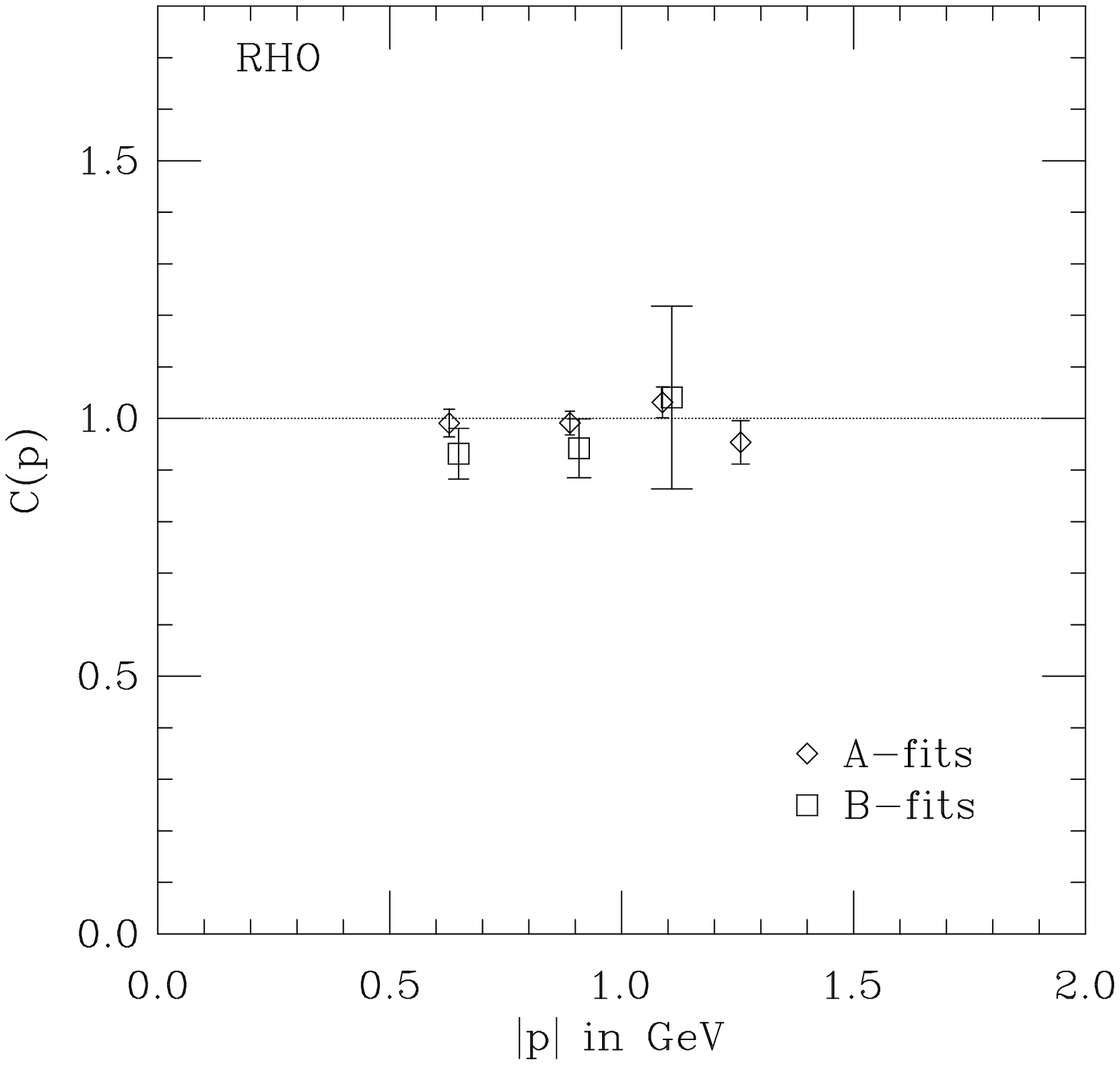}{90mm}
}
%\centerline{(i) \hspace{10cm} (ii)}
\caption{ 
Comparison of A-fit and B-fit (see text)
 results for $C(p)$ on isotropic lattices. 
% B-fit points have been shifted horizontally for clarity.
}
\end{figure}

%\newpage
%\begin{figure}
%\begin{center}
%\epsfysize=7.in
%\centerline{\epsfbox{mkin_mesps.ps}}
%\end{center}
%\caption{
%Kinetic mass versus momentum of correlator from which it was extracted.
%The horizontal line gives the experimental $B_s$ mass.
%  }
%\end{figure}

\newpage
\begin{figure}
\begin{center}
\epsfysize=7.in
\centerline{\epsfbox{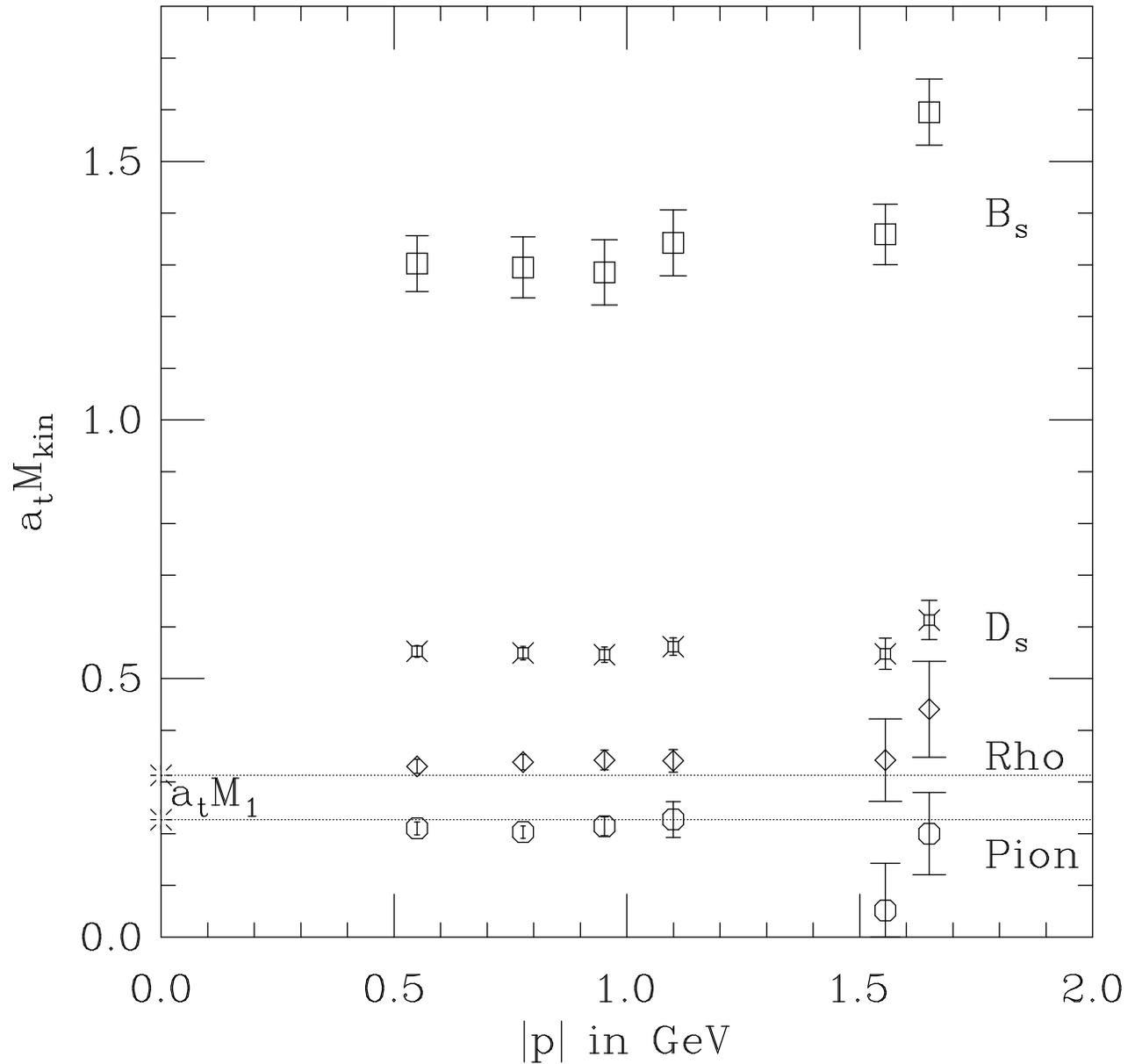}}
\end{center}
\caption{
Kinetic mass  in lattice units 
versus momentum of correlator from which it was extracted 
for the $B_s$, $D_s$, rho and pi mesons.  All results are from anisotropic 
lattices.  The horizontal lines show the rest masses $a_tM_1$ for the 
rho and the pi. 
  }
\end{figure}

\newpage
\begin{figure}
\begin{center}
\epsfysize=7.in
\centerline{\epsfbox{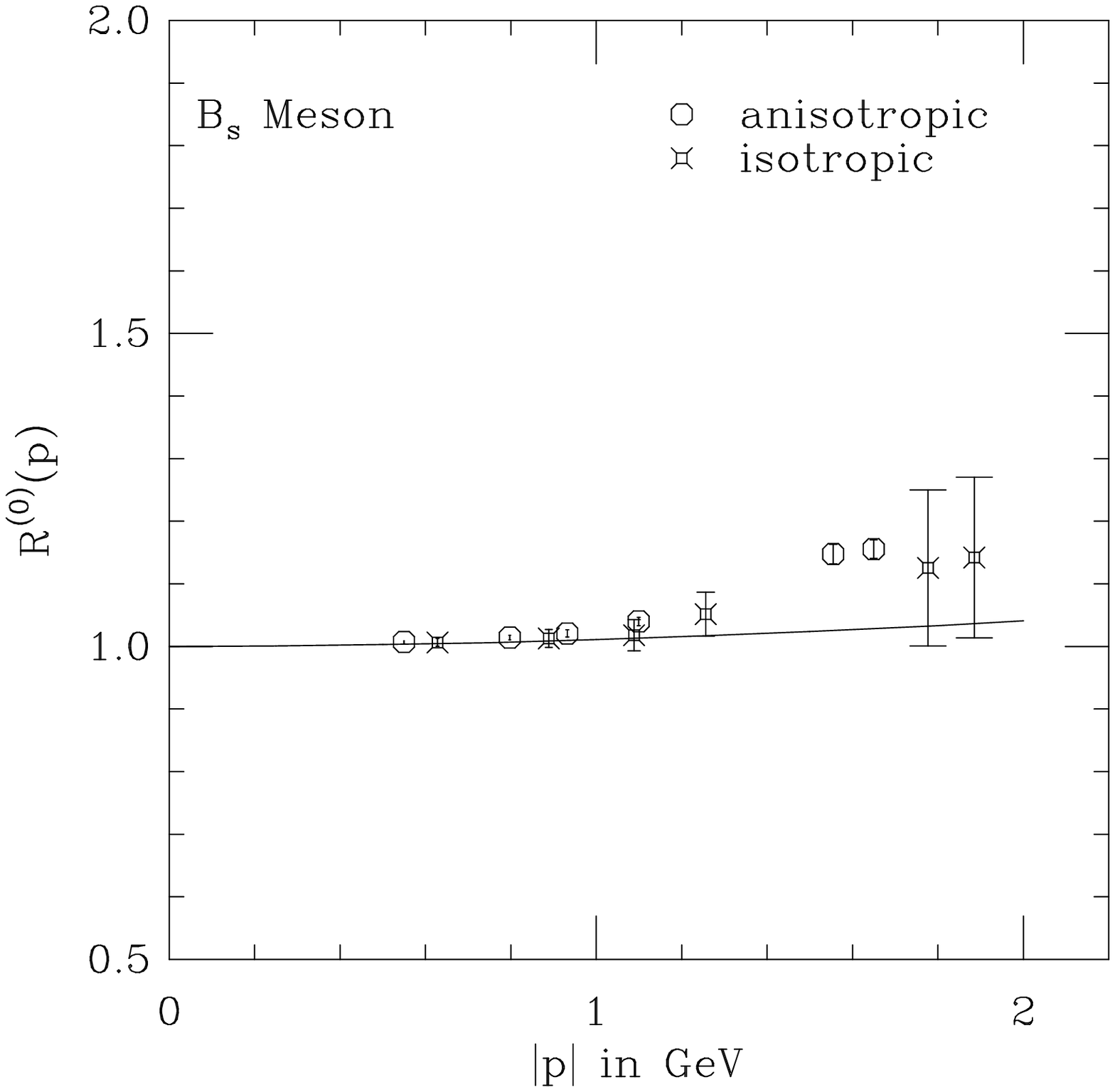}}
\end{center}
\caption{
$R^{(0)}(p)$ as defined in  eq.(\ref{ratio0}) for the $B_s$ meson. 
The full line shows $\sqrt{E(p)}/\sqrt{M_{PS}}$.
  }
\end{figure}

\newpage
\begin{figure}
\begin{center}
\epsfysize=7.in
\centerline{\epsfbox{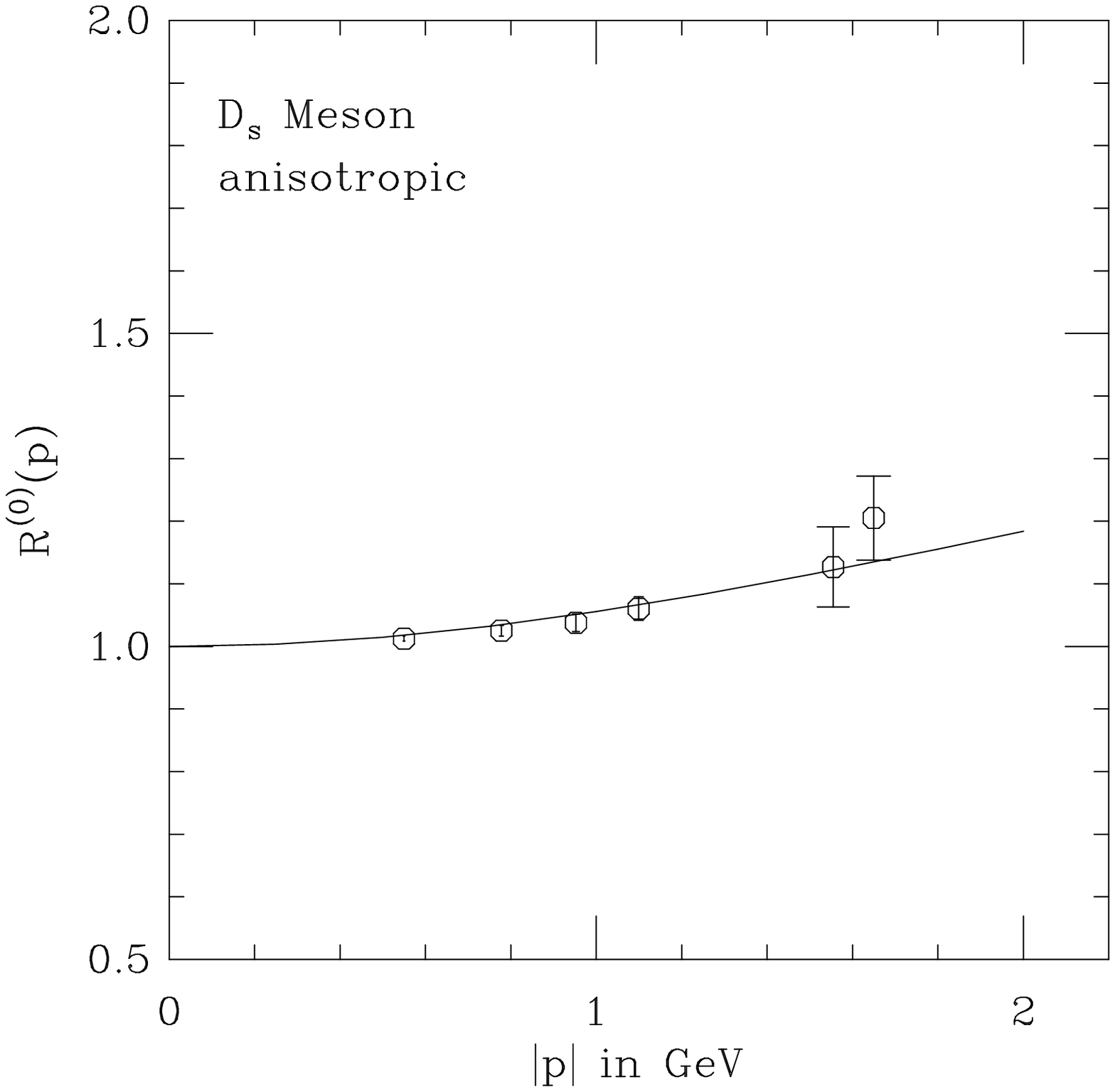}}
\end{center}
\caption{
$R^{(0)}(p)$ as defined in  eq.(\ref{ratio0}) for the $D_s$ meson. 
The full line shows $\sqrt{E(p)}/\sqrt{M_{PS}}$.
  }
\end{figure}

%\newpage
%\begin{figure}
%\begin{center}
%\epsfysize=7.in
%\centerline{\epsfbox{fbratcomp.ps}}
%\end{center}
%\caption{
%Comparison of $R^{(0)}(p)$ as defined in  eq.(\ref{ratio0}) for the $B_s$ meson
% from the coarse isotropic lattice of this work and results from 
%finer lattices using less improved actions studied in [].
%The full line shows $\sqrt{E(p)}/\sqrt{M_{PS}}$.
%  }
%\end{figure}

\newpage

\begin{figure}
\centerline{
\ewxy{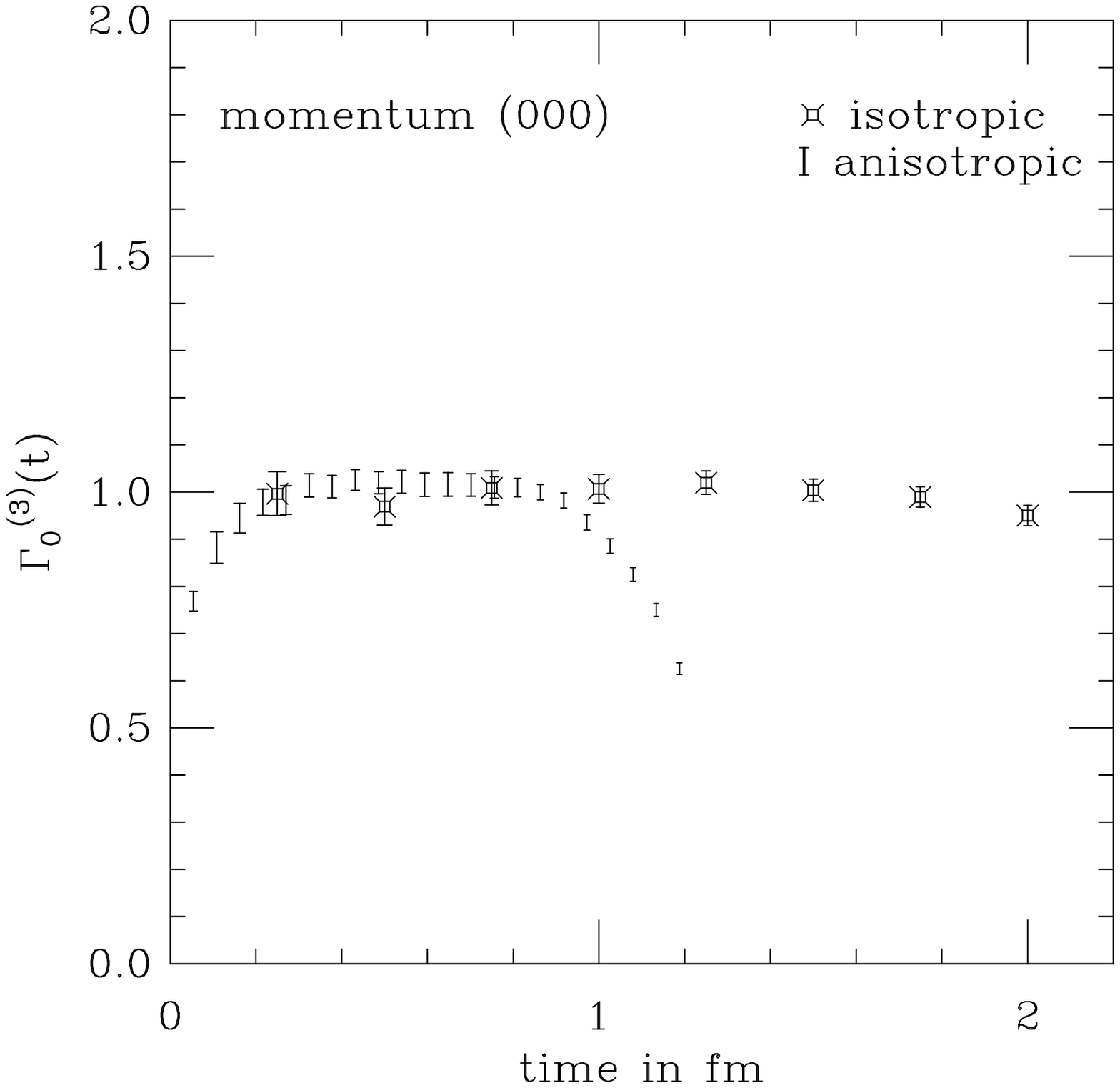}{90mm}
\ewxy{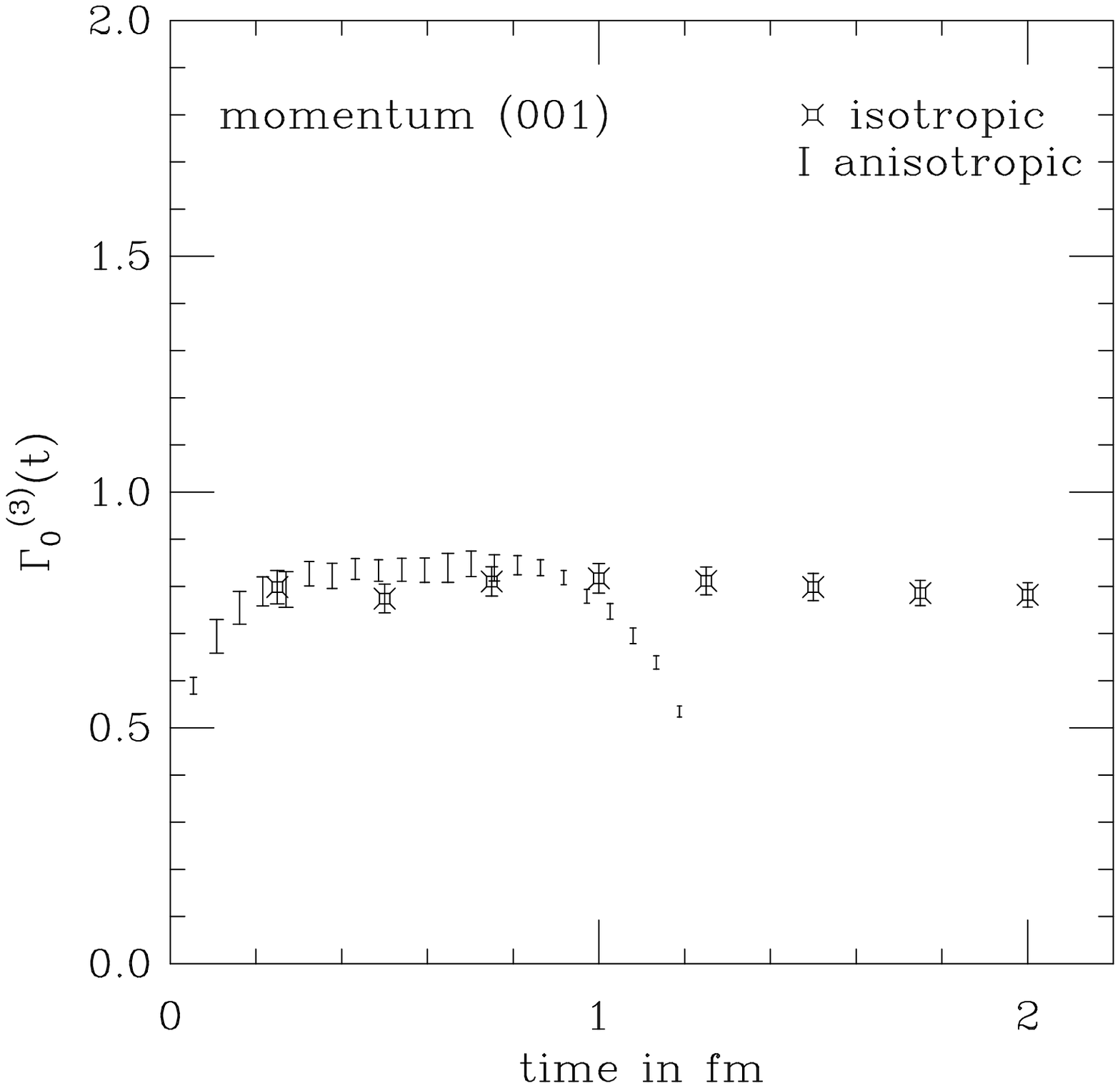}{90mm}
}
%\centerline{(i) \hspace{10cm} (ii)}
\caption{ $\Gamma^{(3)}_0(t)$ of eq.(\ref{c3}) for pion momentum  (0,0,0)
and (0,0,1).
}
\end{figure}

\begin{figure}
\centerline{
\ewxy{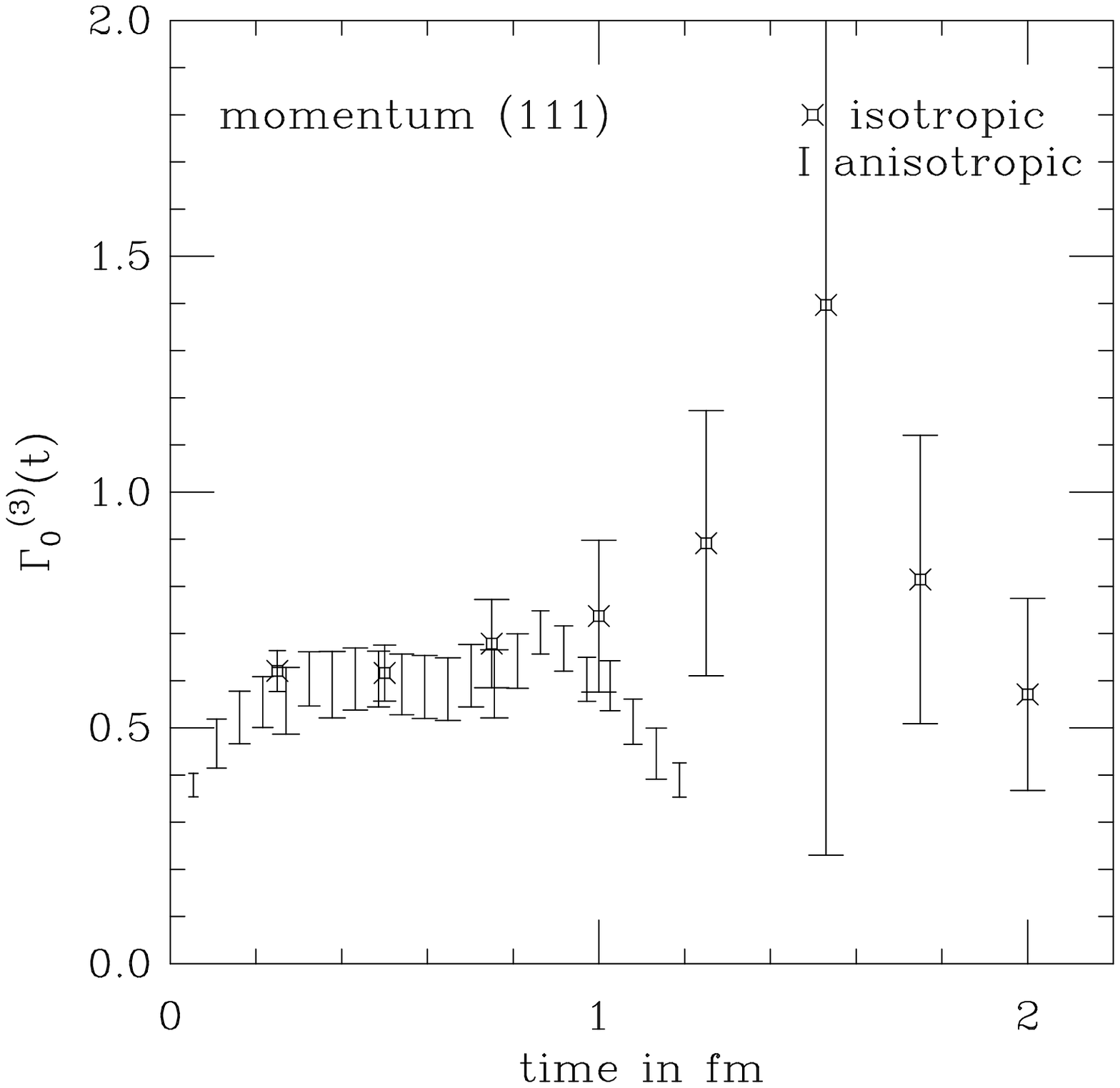}{90mm}
\ewxy{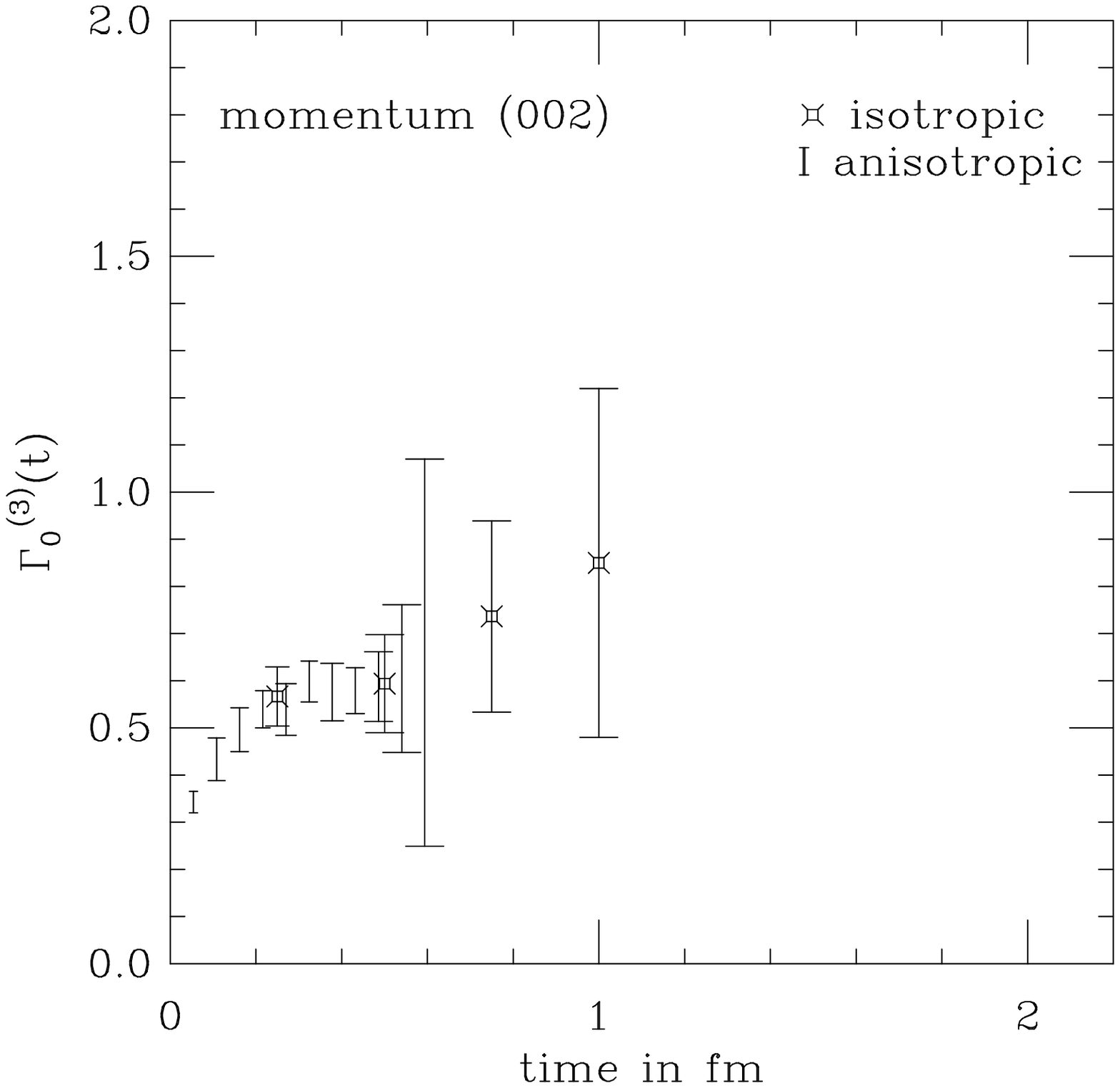}{90mm}
}
%\centerline{(i) \hspace{10cm} (ii)}
\caption{ $\Gamma^{(3)}_0(t)$ of eq.(\ref{c3}) for pion momentum  (1,1,1)
and  (0,0,2).
}
\end{figure}

\newpage

\begin{figure}
\centerline{
\ewxy{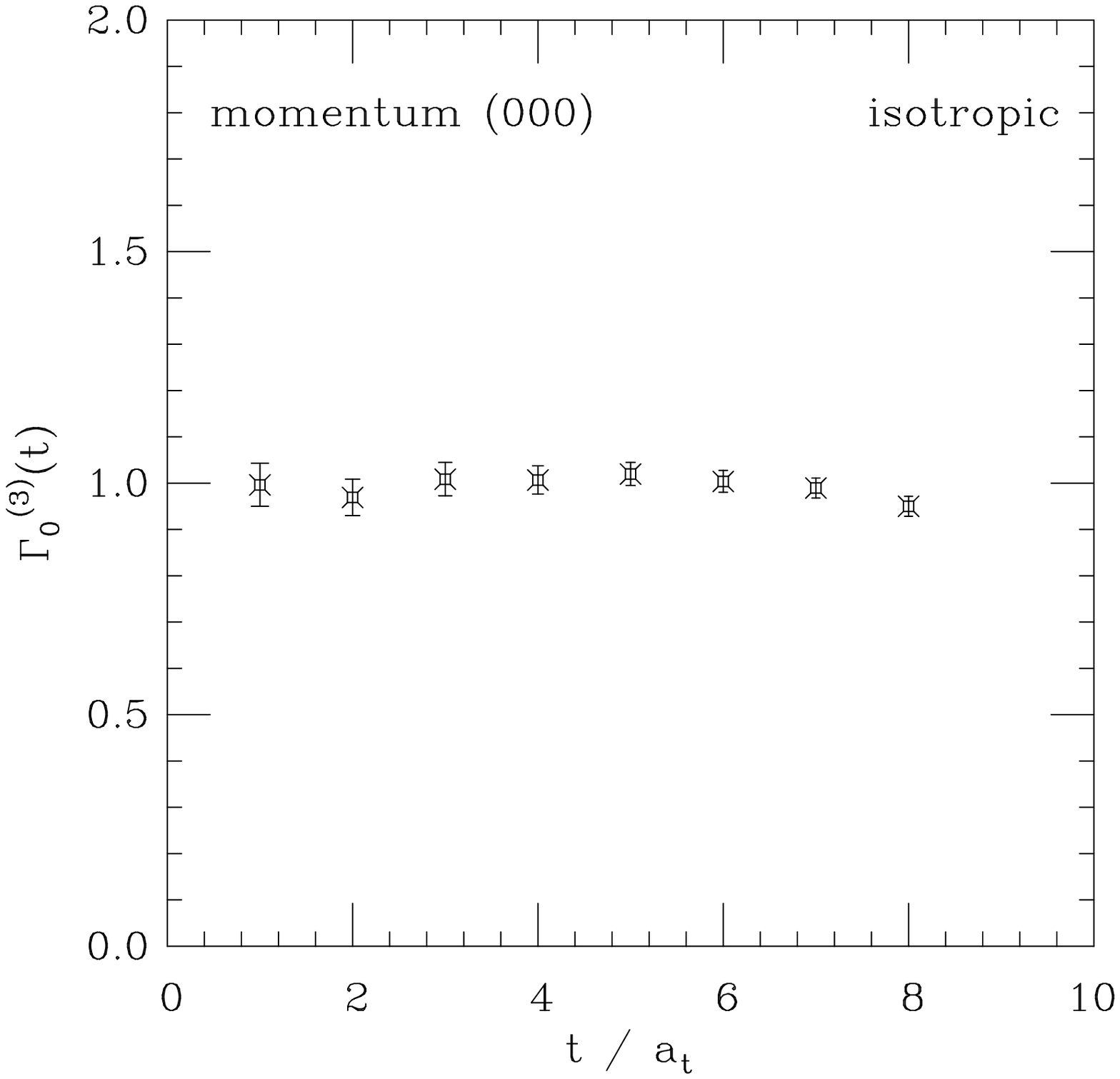}{90mm}
\ewxy{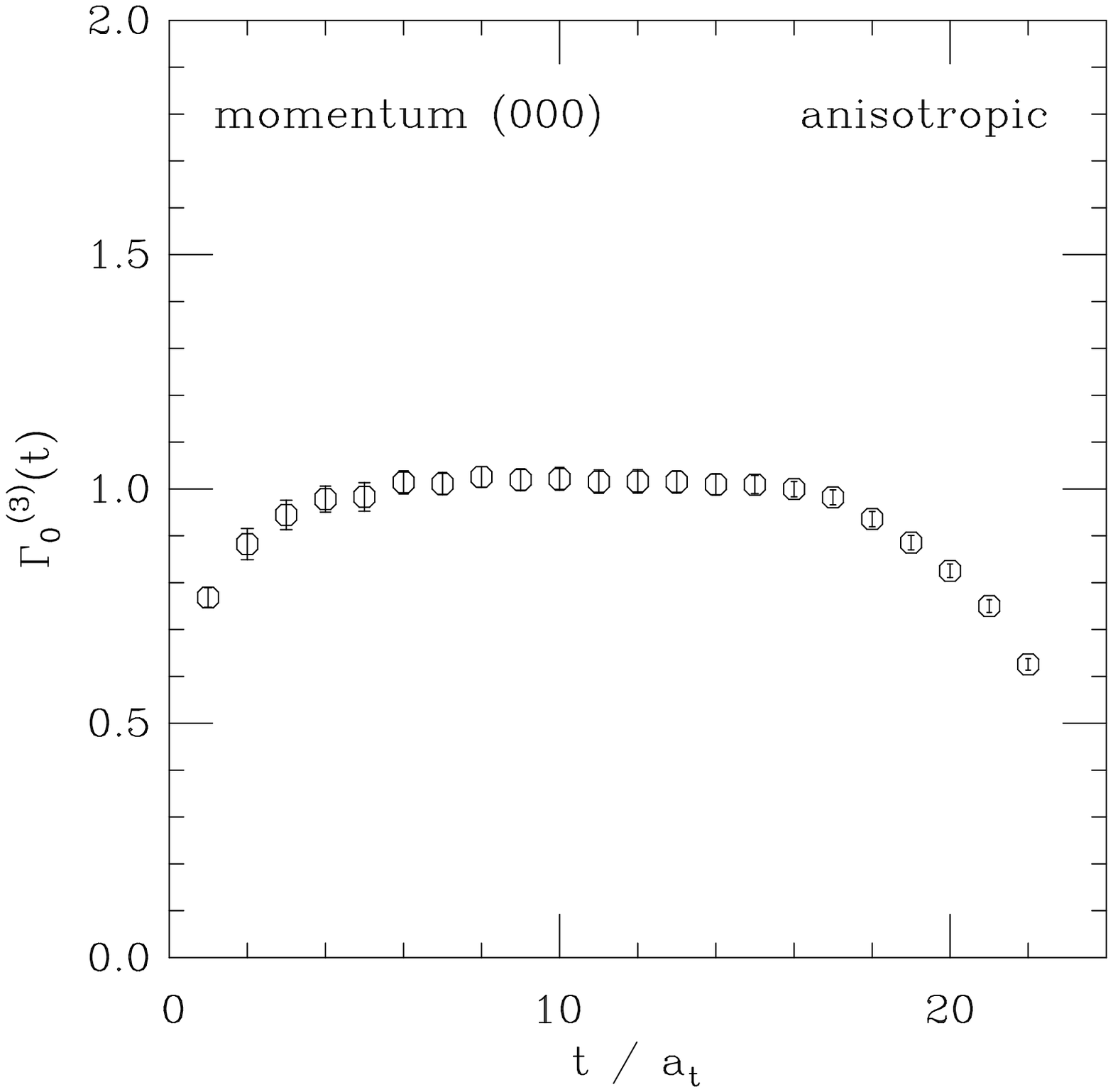}{90mm}
}
%\centerline{(i) \hspace{10cm} (ii)}
\caption{ $\Gamma^{(3)}_0(t)$ of eq.(\ref{c3}) for pion momentum  (0,0,0)
versus time in lattice units.
}
\end{figure}

\begin{figure}
\centerline{
\ewxy{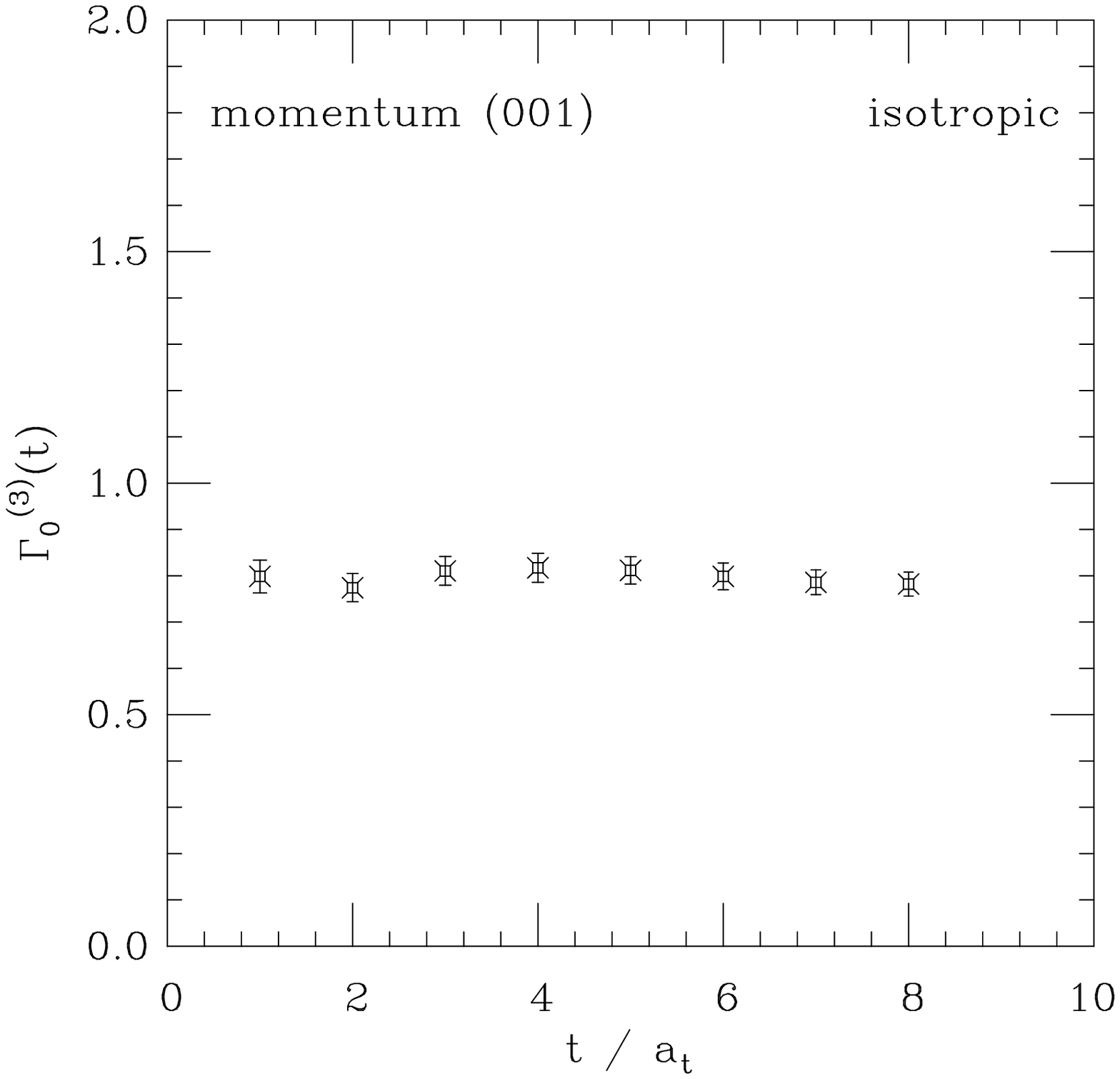}{90mm}
\ewxy{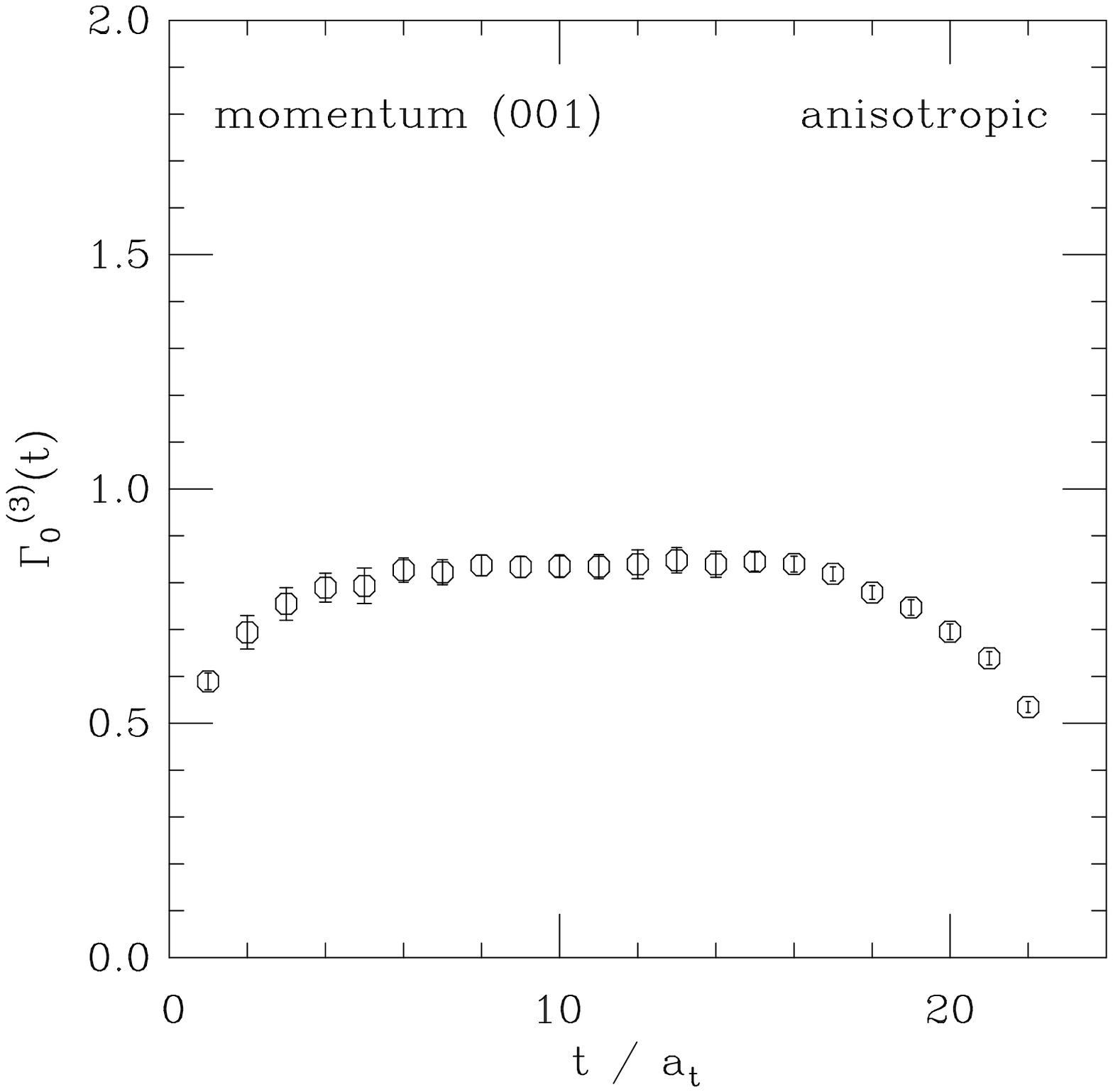}{90mm}
}
%\centerline{(i) \hspace{10cm} (ii)}
\caption{ $\Gamma^{(3)}_0(t)$ of eq.(\ref{c3}) for pion momentum  (0,0,1)
versus time in lattice units.
}
\end{figure}

\newpage

\begin{figure}
\centerline{
\ewxy{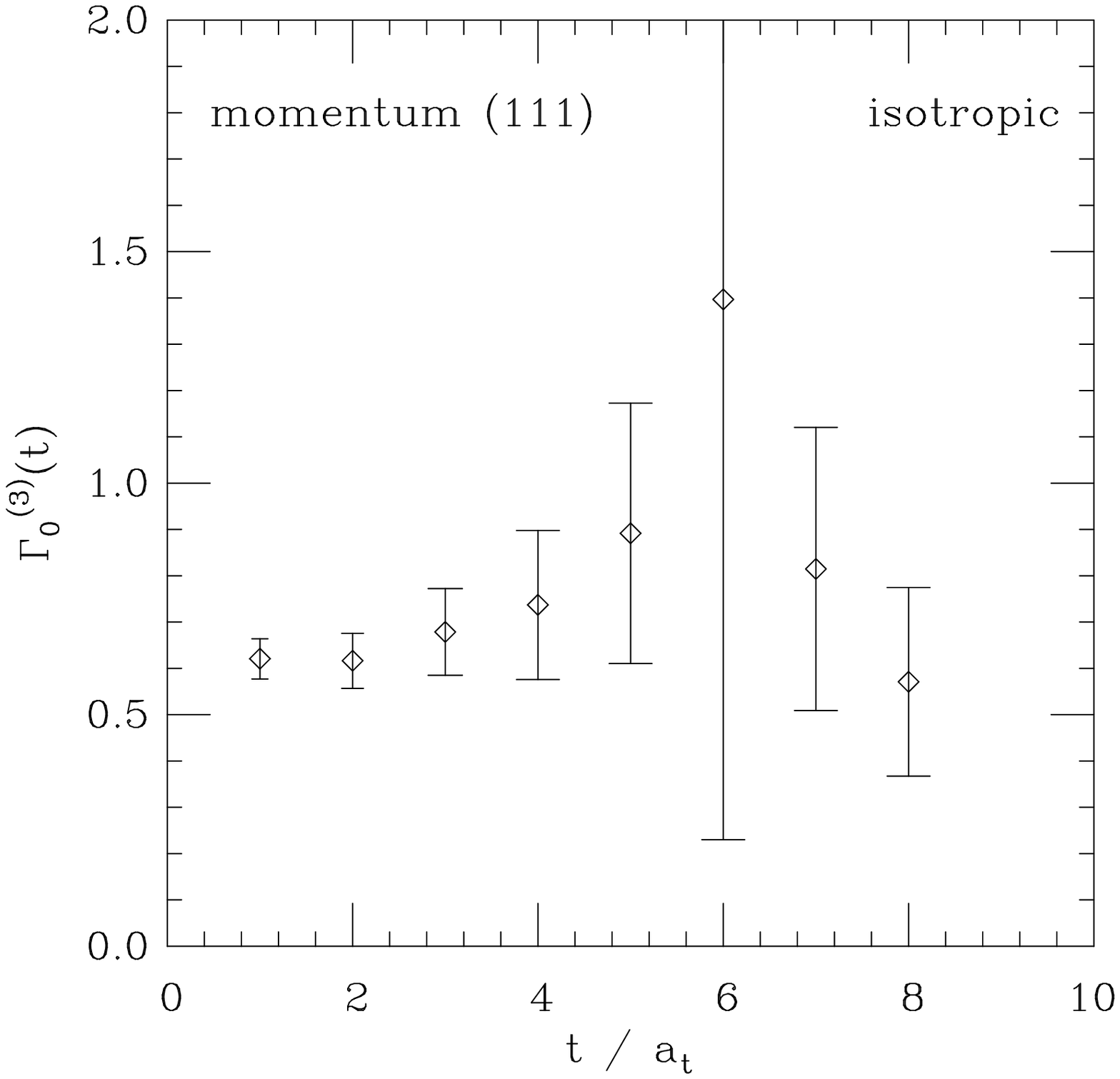}{90mm}
\ewxy{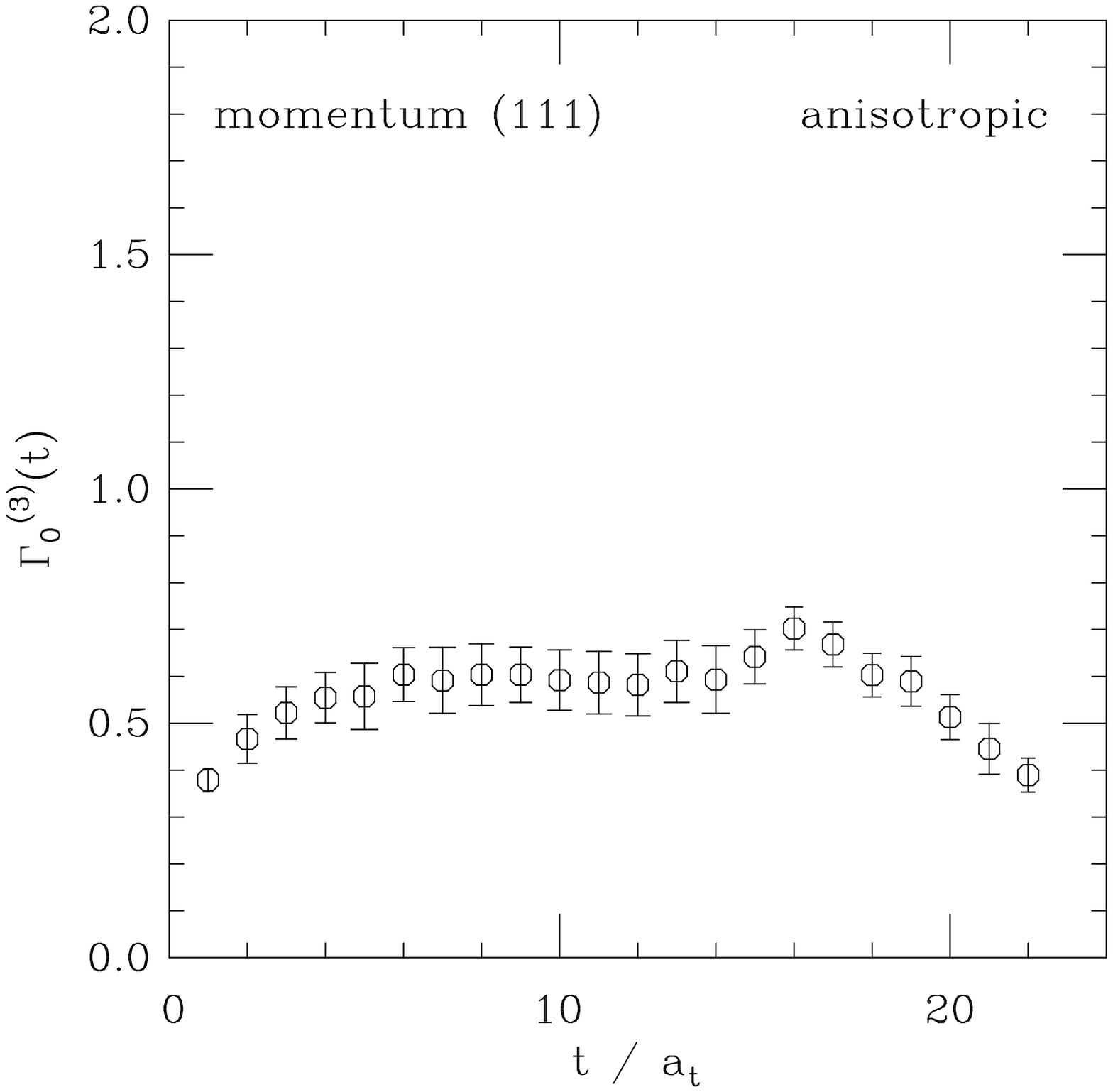}{90mm}
}
%\centerline{(i) \hspace{10cm} (ii)}
\caption{ $\Gamma^{(3)}_0(t)$ of eq.(\ref{c3}) for pion momentum  (1,1,1)
versus time in lattice units.
}
\end{figure}

\begin{figure}
\centerline{
\ewxy{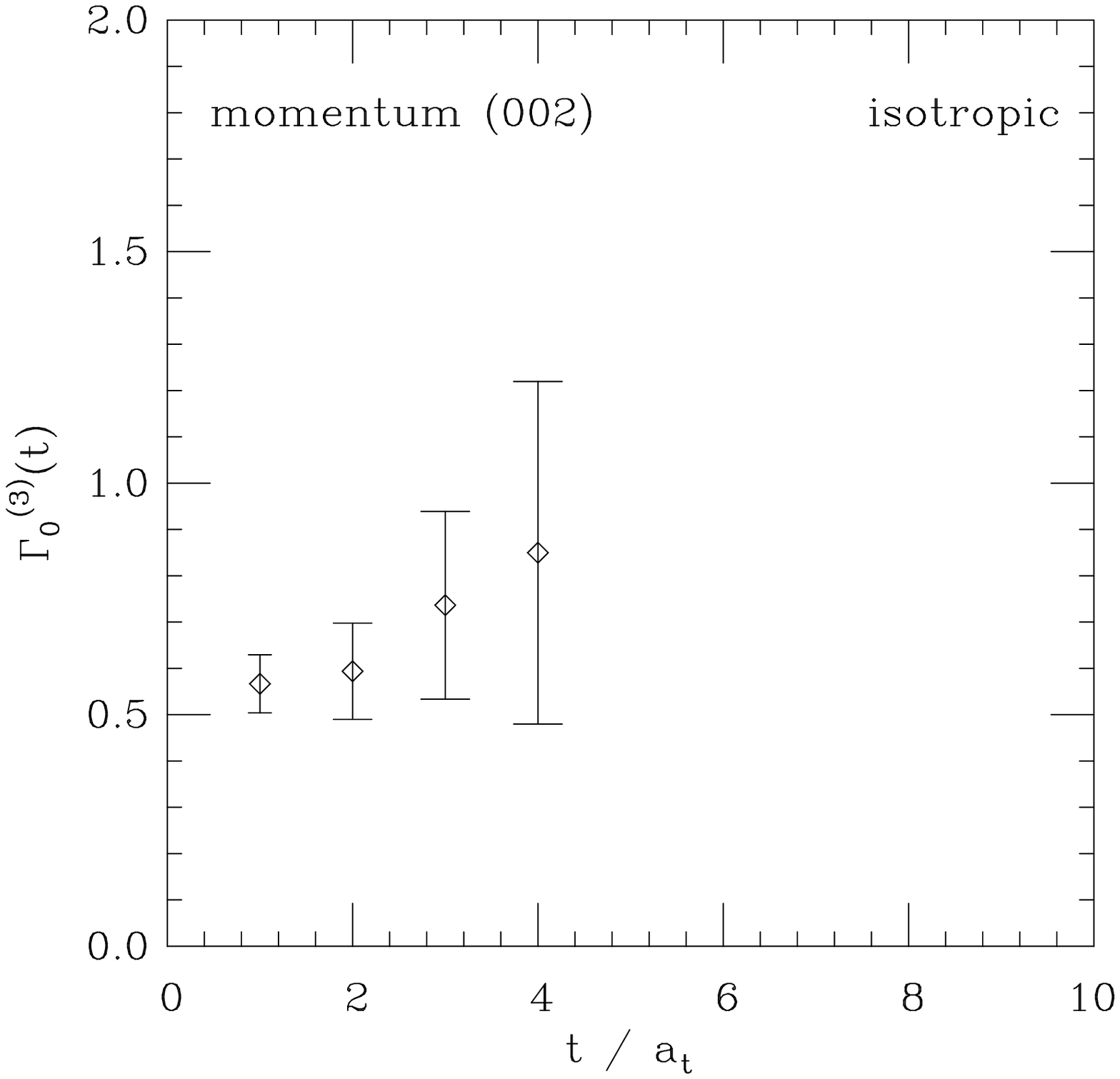}{90mm}
\ewxy{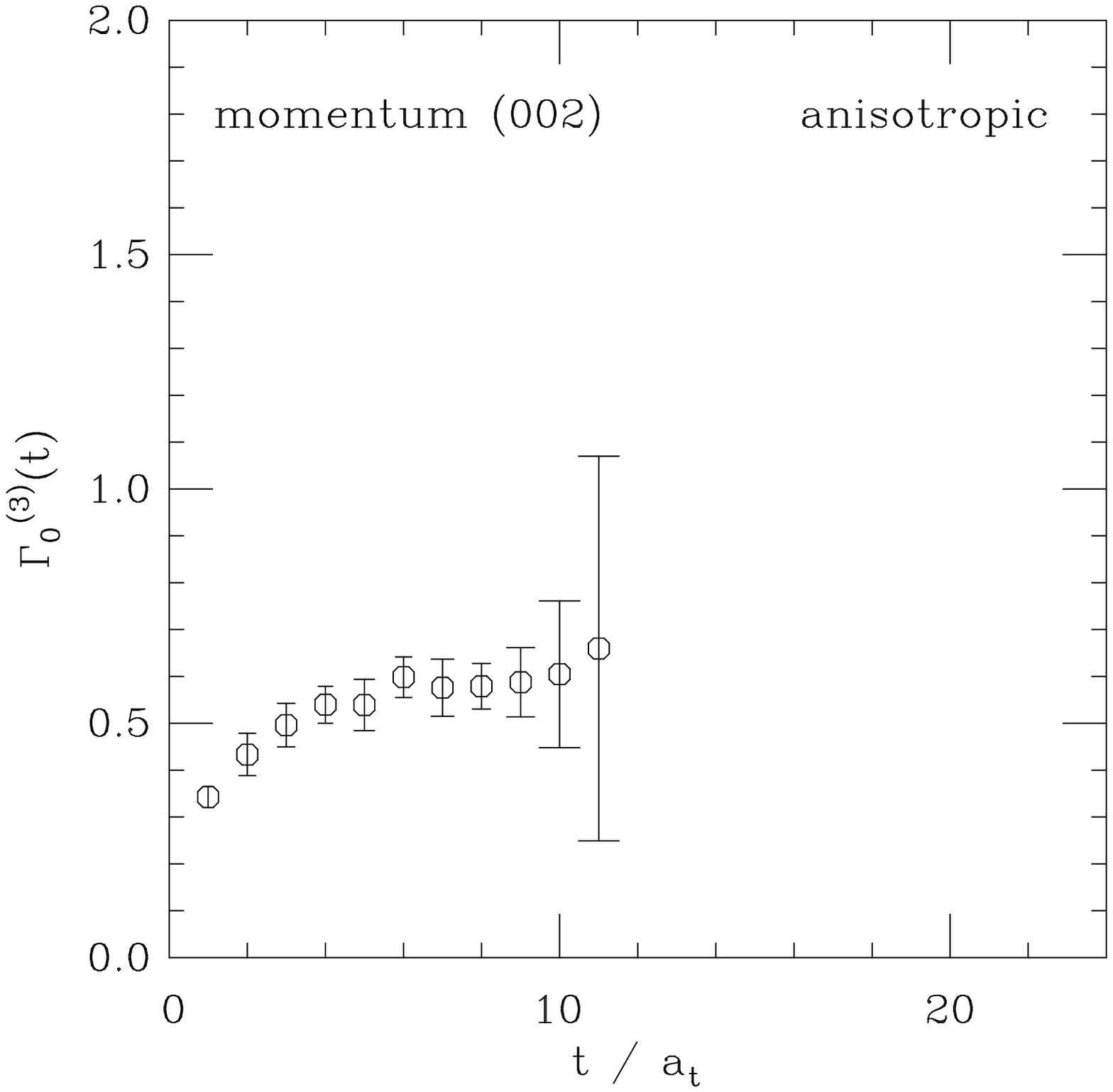}{90mm}
}
%\centerline{(i) \hspace{10cm} (ii)}
\caption{ $\Gamma^{(3)}_0(t)$ of eq.(\ref{c3}) for pion momentum  (0,0,2)
versus time in lattice units.
}
\end{figure}

\newpage

\begin{figure}
\centerline{
\ewxy{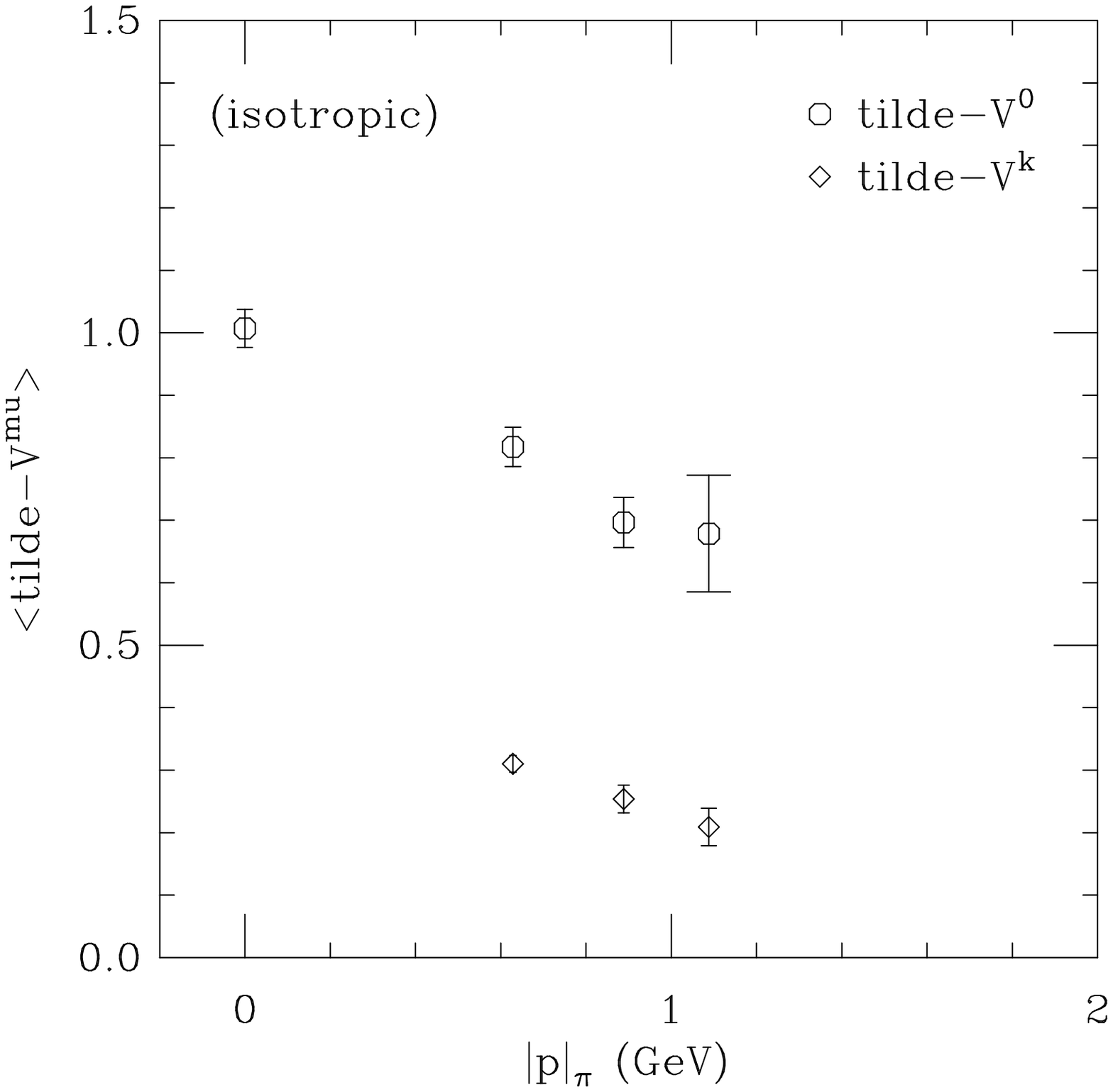}{90mm}
\ewxy{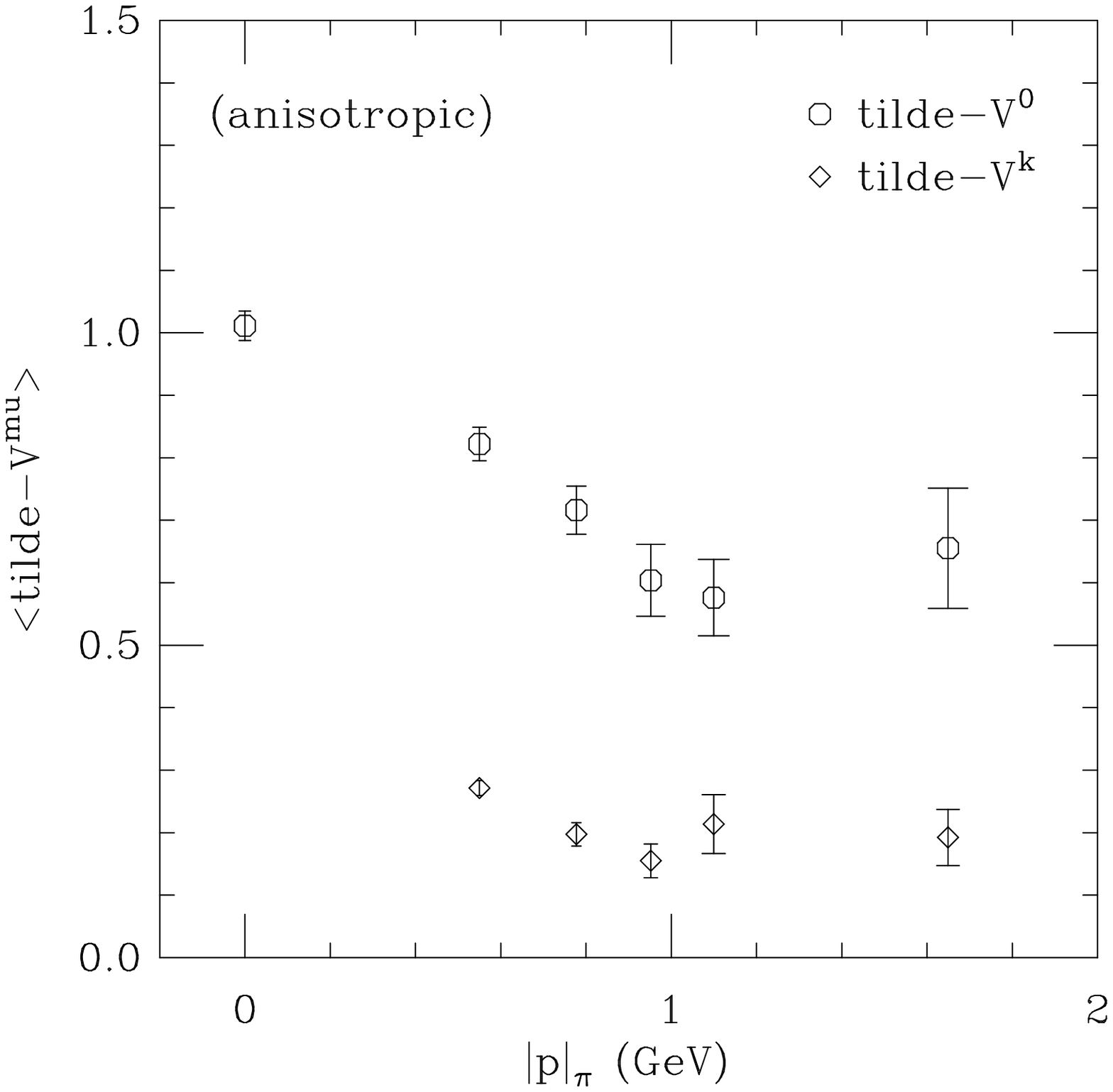}{90mm}
}
%\centerline{(i) \hspace{10cm} (ii)}
\caption{ $\vev{\tilde{V}^\mu}$ of eq.(\ref{vtilde}) from isotropic 
and anisotropic lattices versus the pion momentum.
}
\end{figure}

\newpage
\begin{figure}
\begin{center}
\epsfysize=7.in
\centerline{\epsfbox{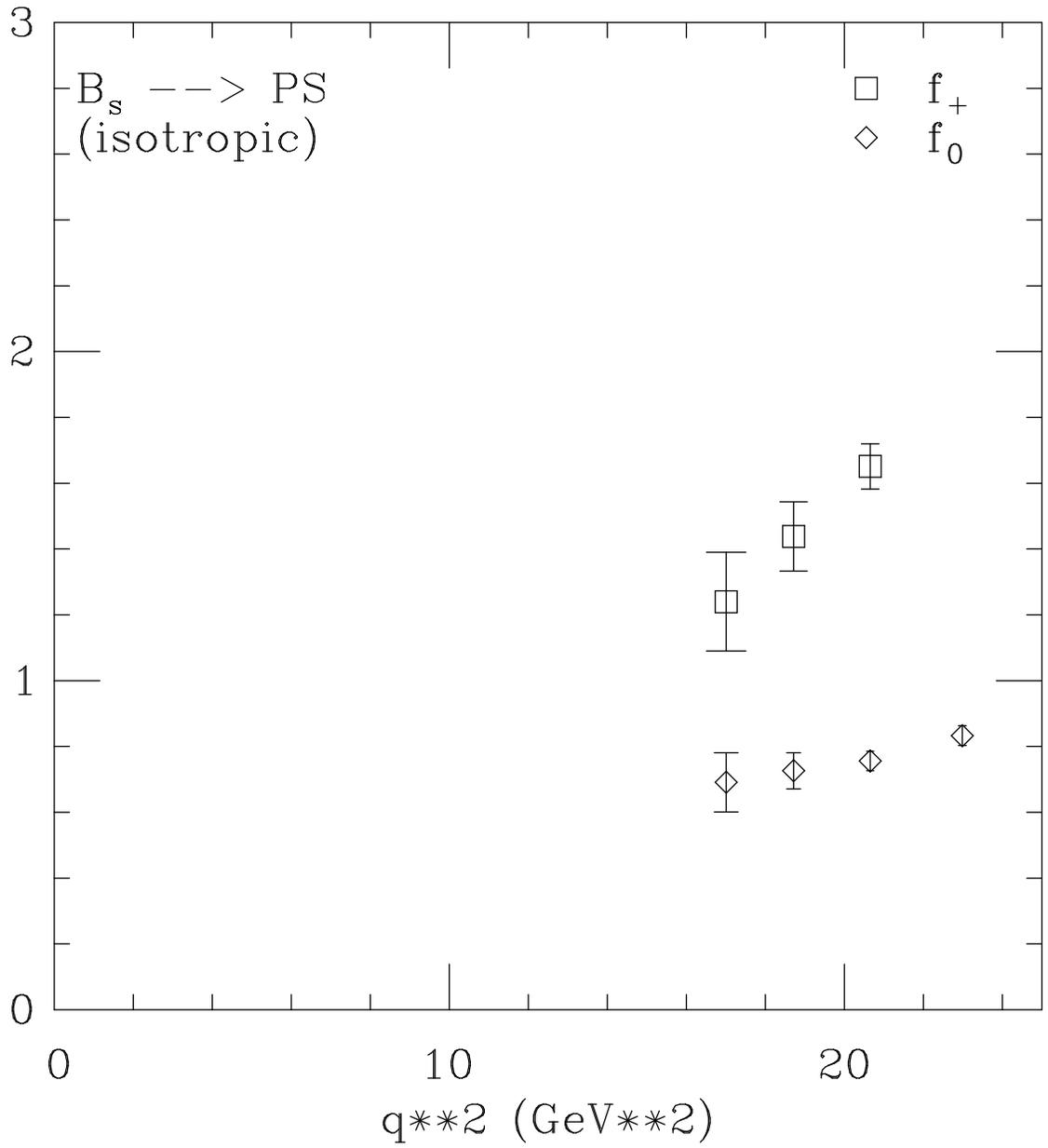}}
\end{center}
\caption{
The form factors $f_0(q^2)$ and $f_+(q^2)$ for $B_s$ decays 
from isotropic lattices.
See Table II for actual values of 
the decaying heavy meson and daughter meson masses.
  }
\end{figure}

\newpage
\begin{figure}
\begin{center}
\epsfysize=7.in
\centerline{\epsfbox{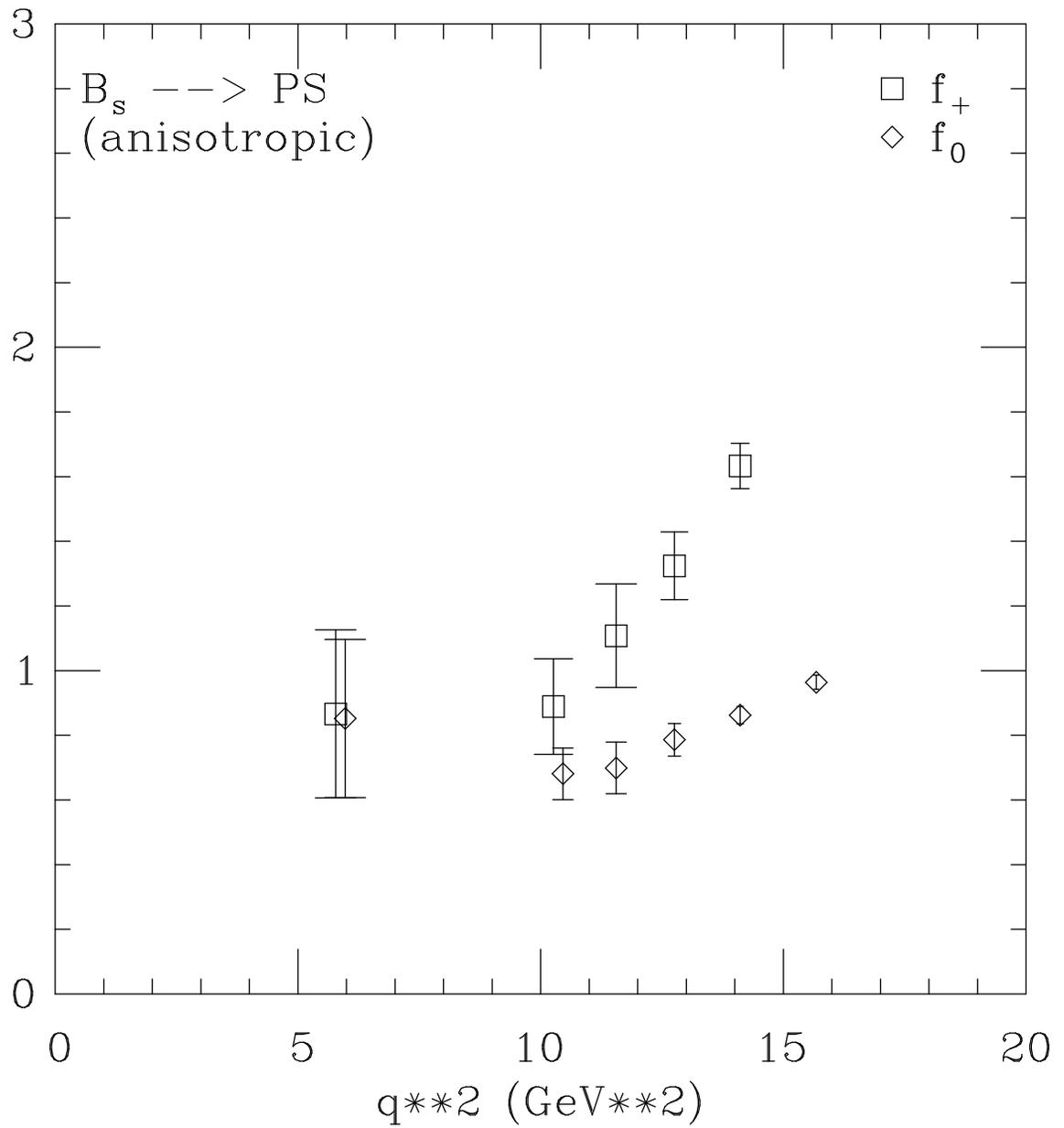}}
\end{center}
\caption{
The form factors $f_0(q^2)$ and $f_+(q^2)$ for $B_s$ decays 
from anisotropic lattices.  
See Table II for actual values of 
the decaying heavy meson and daughter meson masses.  Some points have 
been shifted horizontally for clarity.
  }
\end{figure}

\newpage
\begin{figure}
\begin{center}
\epsfysize=7.in
\centerline{\epsfbox{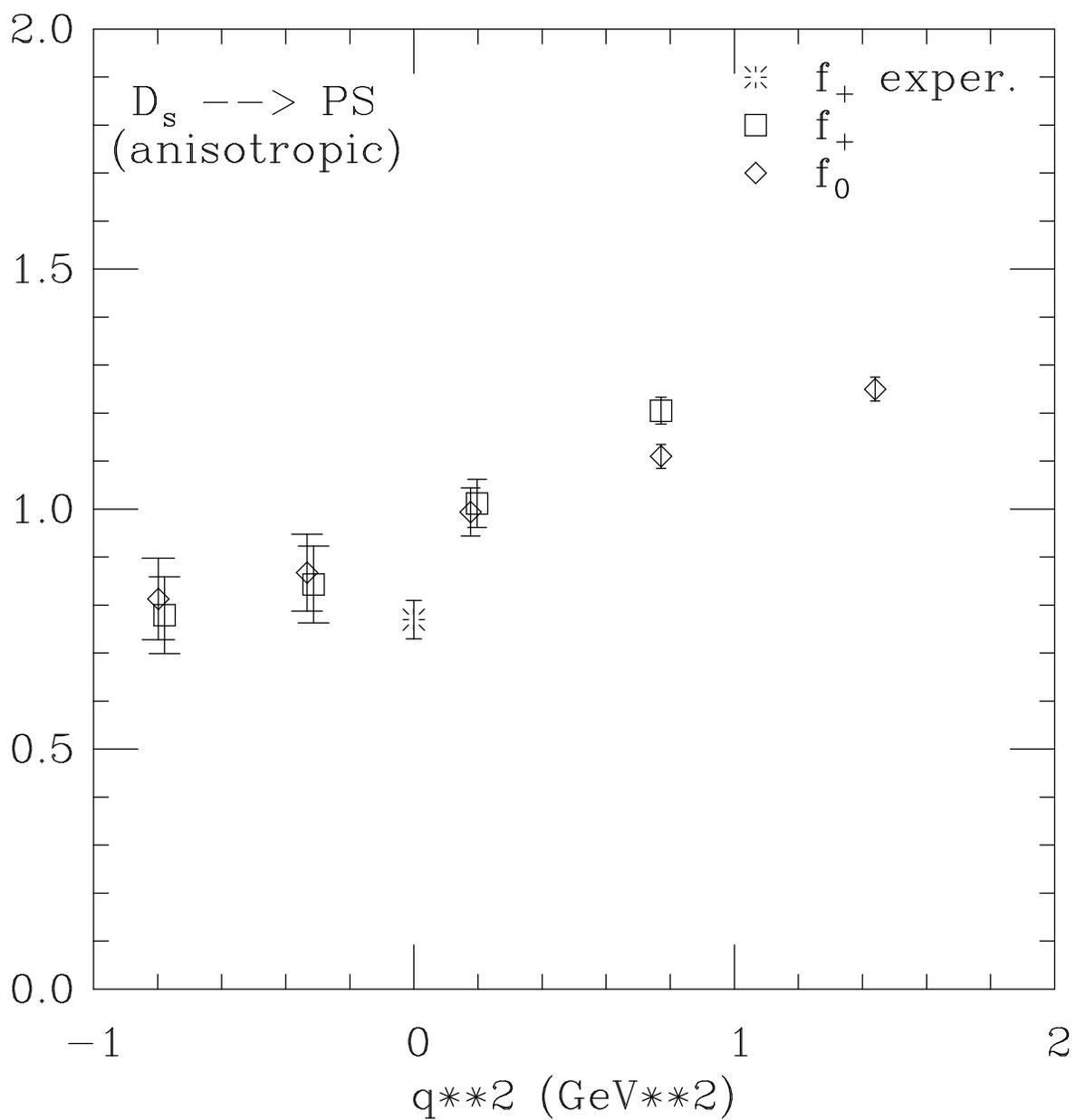}}
\end{center}
\caption{
The form factors $f_0(q^2)$ and $f_+(q^2)$ for $D_s$ decays 
from anisotropic lattices. The ``burst'' shows an experimentally
determined value of  $f_+(q^2=0)$ 
for the decay $D^0 \rightarrow K^-l^+\nu$ \protect\cite{cleo}.
  }
\end{figure}

\end{document}